\documentclass{article}

\usepackage{PRIMEarxiv}

\usepackage[utf8]{inputenc} 
\usepackage[T1]{fontenc}    
\usepackage{hyperref}       
\usepackage{url}            
\usepackage{booktabs}       
\usepackage{amsfonts}       
\usepackage{nicefrac}       
\usepackage{microtype}      
\usepackage{lipsum}
\usepackage{fancyhdr}       
\usepackage{graphicx}       

\usepackage[dvipsnames]{xcolor}
\usepackage{subcaption}
\usepackage{caption}
\usepackage{amsmath}
\usepackage{mathtools}
\usepackage{amssymb}
\usepackage{multirow}
\usepackage{adjustbox}
\usepackage{stackengine}
\usepackage{tikz}
\usepackage{color}
\usepackage{bbold}
\usetikzlibrary{angles,quotes}
\usepackage{rotating}
\usepackage{pifont}

\newcommand{\cmark}{\ding{51}}%
\newcommand{\xmark}{\ding{55}}%
\newcommand\omicron{o}
\DeclareMathOperator{\arctantwo}{arctan2}

\title{Stochastic Geometry Models for Texture Synthesis\\of Machined Metallic Surfaces:\\Sandblasting and Milling
}

\author{
  Natascha Jeziorski \\
  Fraunhofer ITWM Kaiserslautern\\
  University of Kaiserslautern-Landau\\
   \And
  Claudia Redenbach \\
  University of Kaiserslautern-Landau
}

\begin{document}
\maketitle

\begin{abstract}
Training defect detection algorithms for visual surface inspection systems requires a large and representative set of training data. Often there is not enough real data available which additionally cannot cover the variety of possible defects.
Synthetic data generated by a synthetic visual surface inspection environment can overcome this problem.
Therefore, a digital twin of the object is needed, whose micro-scale surface topography is modeled by texture synthesis models.
We develop stochastic texture models for sandblasted and milled surfaces based on topography measurements of such surfaces.
As the surface patterns differ significantly, we use separate modeling approaches for the two cases.
Sandblasted surfaces are modeled by a combination of data-based texture synthesis methods that rely entirely on the measurements.
In contrast, the model for milled surfaces is procedural and includes all process-related parameters known from the machine settings.
\end{abstract}


\section{Introduction}
\label{intro}
\begin{figure*}
    \centering
    \begin{subfigure}[t]{0.4\columnwidth}
    \centering
        \includegraphics[height=3.5cm]{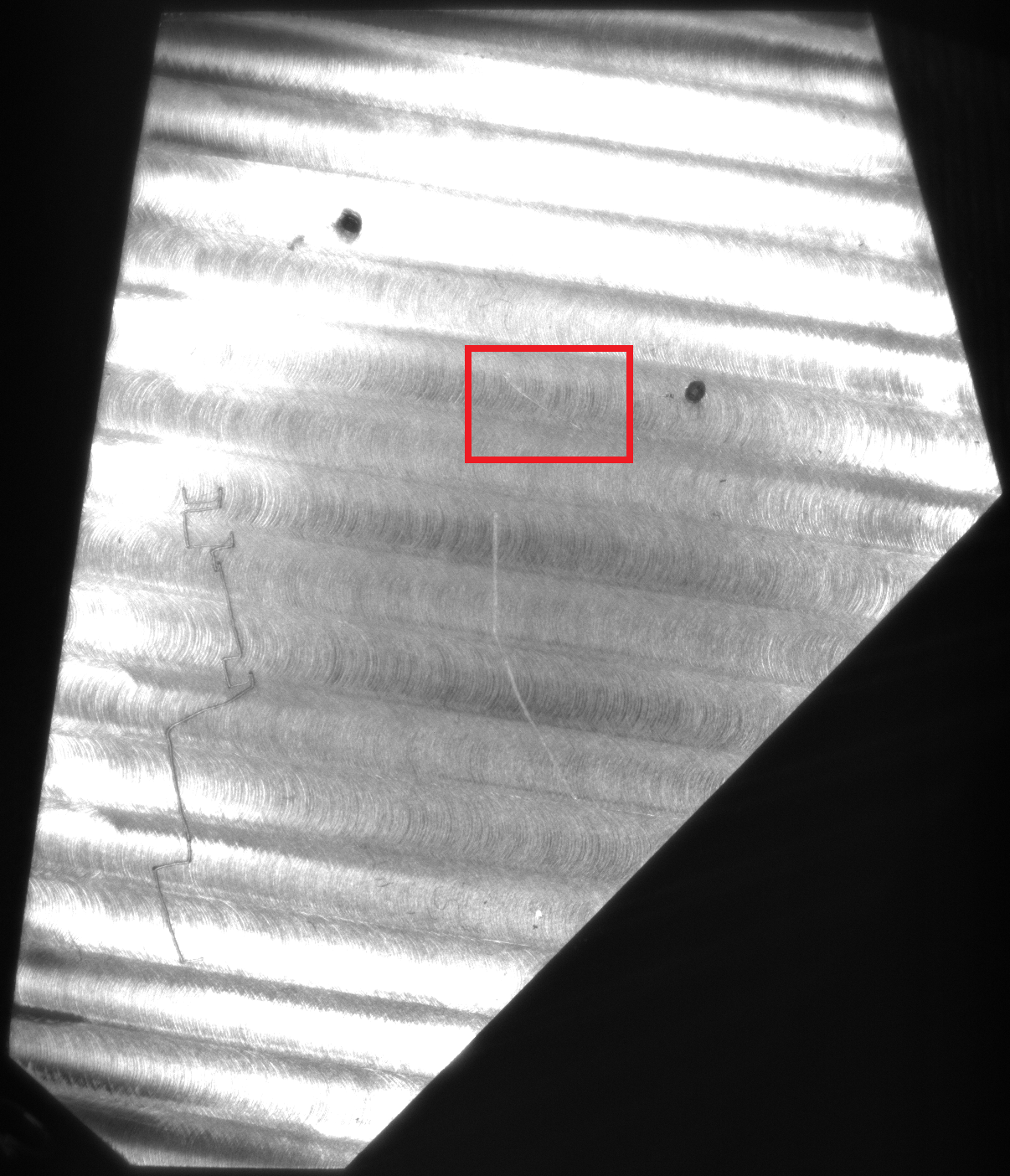}
        \includegraphics[height=3.5cm]{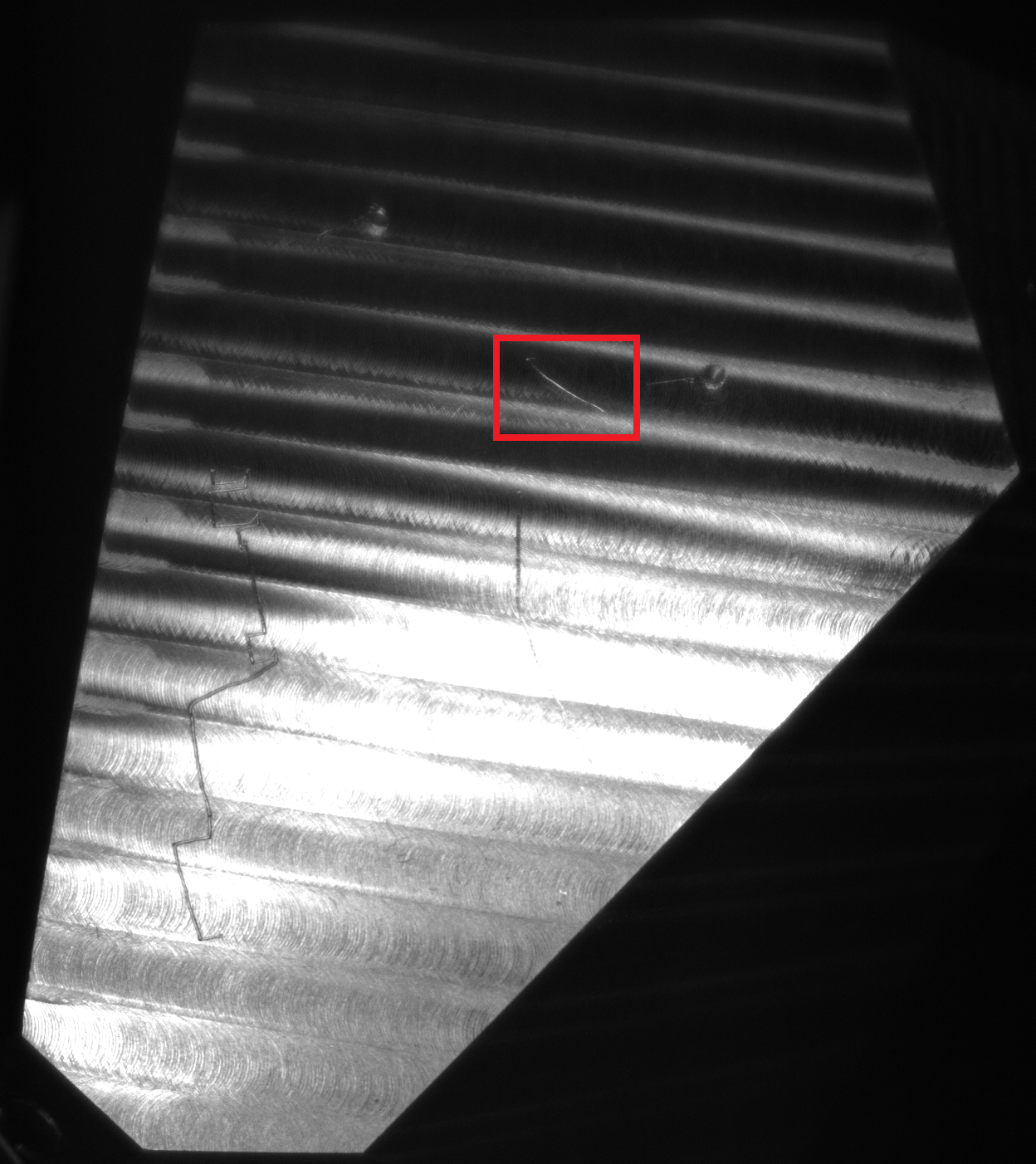}
        \caption{Acquisition images of a surface that contains defects (dents and scratches) taken from different viewing angles. A scratch with changing visibility is marked.}
        \label{fig:intro:application:acquistion}
    \end{subfigure}
    \hspace{3pt}
    \begin{subfigure}[t]{0.2\columnwidth}
        \centering
        \includegraphics[height=3.5cm]{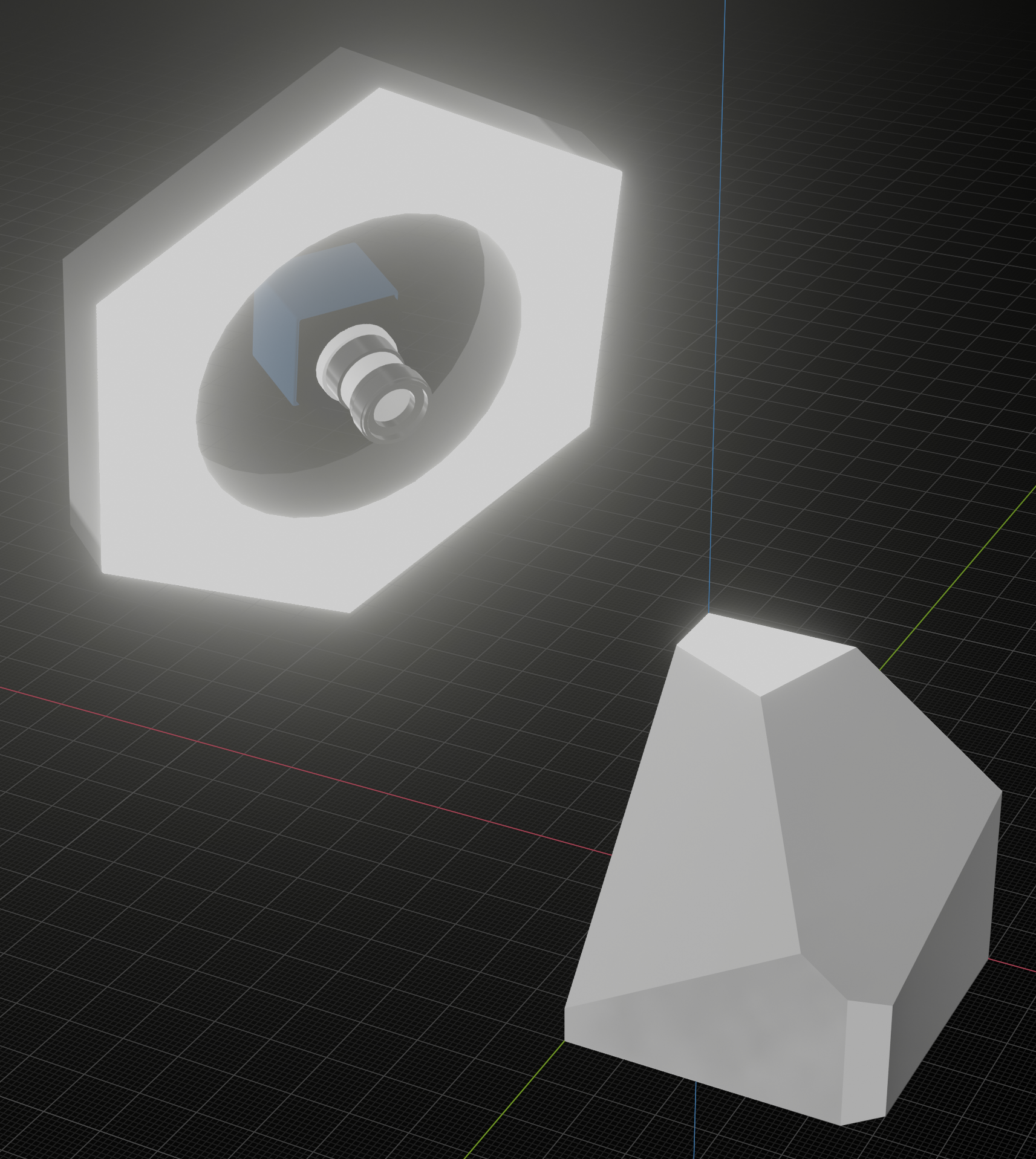}
        \caption{Illustration of the synthetic visual surface inspection system created in \textsc{Blender}.}
        \label{fig:intro:application:scene}
    \end{subfigure}
    \hspace{3pt}
    \begin{subfigure}[t]{0.2\columnwidth}
        \centering
        \includegraphics[height=3.5cm]{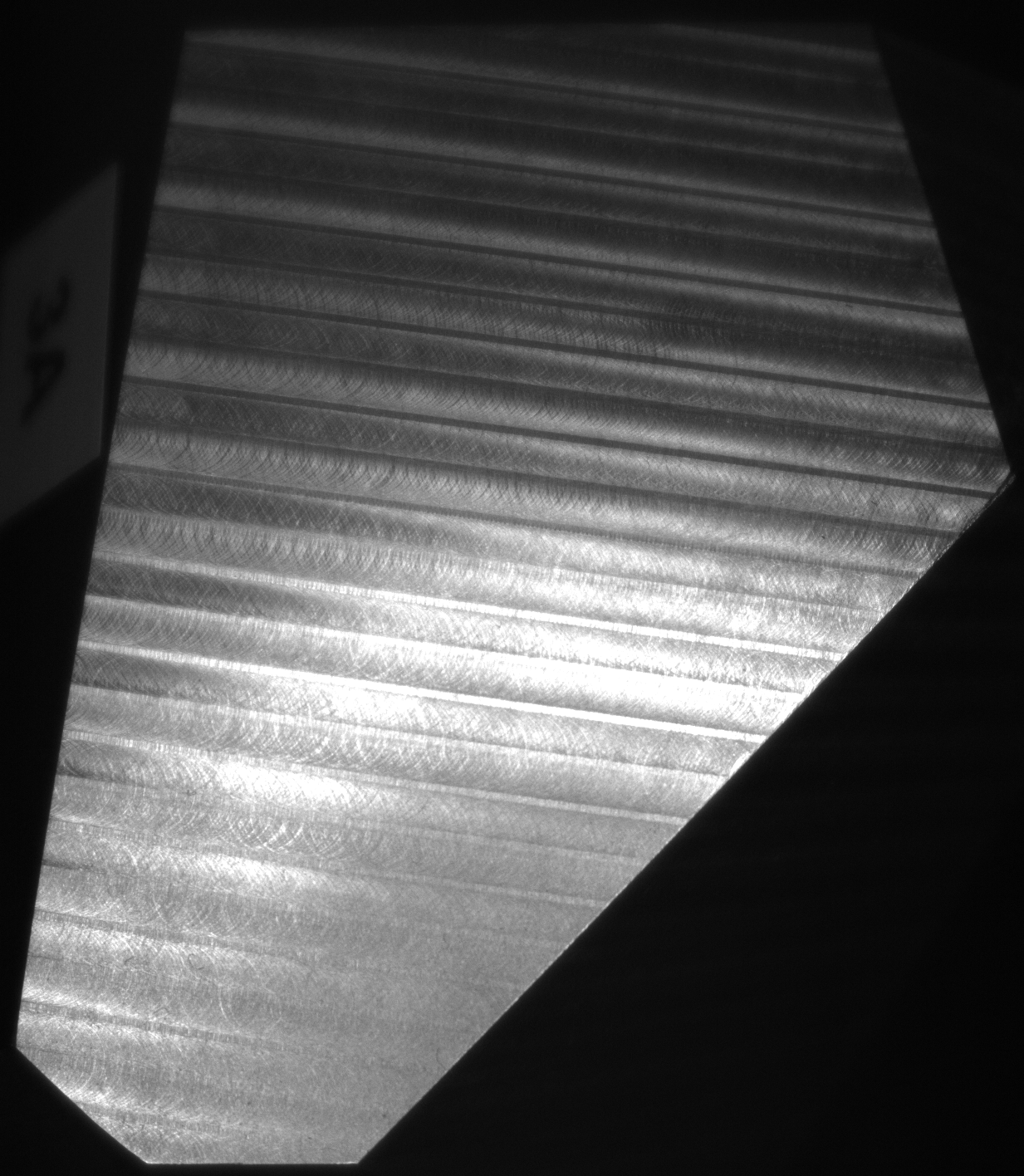}
        \caption{Rendered milled surface without defects.}
        \label{fig:intro:application:render}
    \end{subfigure}
    \caption{Illustration of the synthetic visual surface inspection system and comparison between acquisition and rendered images of a milled surface. Images are provided by Fraunhofer ITWM.}
    \label{fig:intro:application}
\end{figure*}

Automatic defect detection by visual inspection systems becomes increasingly important in industry. 
To facilitate the development of inspection systems, defect detection methods should be optimized in terms of robustness, speed, variability of detected defects and adaptability to objects that differ significantly in shape and appearance.
The main components of a visual inspection system are the camera, the light source and the object to be analyzed \cite{Lovro2020ImageSynthesis}.
Machine learning models are commonly used for automatic defect detection \cite{Juraj2023Kupplung}.
The quality of those models highly depends on the availability of a sufficiently large amount of training data.
Data here refers to images of objects captured during the acquisition step in the inspection system accompanied by a ground truth for the defect detection. 
Examples are given in Figure \ref{fig:intro:application:acquistion}.
In practice, it can be very expensive to acquire and annotate the required amount of real data. 
Additionally, it may be impossible to capture the variety of defects that can occur. 
The use of synthetic data promises solutions to both problems.

Synthetic image data can be generated by the pipeline described in \cite{SynosIs_in_preparation} using a synthetic visual surface inspection system.
Therefore, a digital twin of the object to be analyzed is needed.
Assume that the geometry and the material of the object are known.
Then, the surface topography and defect shapes need to be modeled. 
Surface topography strongly influences the light reflection of the object surface as shown in Figure \ref{fig:intro:application:acquistion}.
We model surface topographies by textures which are functions specifying the small-scale changes of a surface \cite{Pharr_bookCG}.
We concentrate on texture synthesis methods for the surface topography of metallic objects with planar surfaces.
Defects are then geometrically imprinted into the object.
Simulations of particular defect shapes have been provided in the literature. 
For instance, Bosnar and Gospodenti\'c \cite{Lovro2023Defects} developed approaches to simulate scratches and bumps, while Jung and Redenbach \cite{Christian2022Cracks} establish models for crack simulation. 
However, a much larger variability of defects may occur in practice. 

To complete the synthetic visual surface inspection system, also a synthetic camera and a synthetic light source are needed, see Figure \ref{fig:intro:application:scene} for an exemplary 3d scene.
The image acquisition is simulated by rendering the virtual 3d scene into a 2d image.
Similar images as in the real acquisition are generated, see Figure \ref{fig:intro:application:render} for an example.

\begin{figure*}
    \centering
    \includegraphics[height=3.5cm]{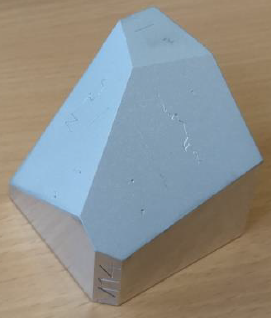}
    \includegraphics[height=3.5cm]{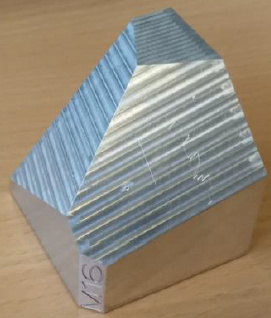}
    \includegraphics[height=3.5cm]{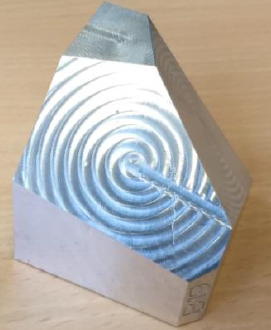}
    \caption{Test objects made of aluminum having a base about $5\text{ cm}$ in size and differently machined surfaces: sandblasted (left), parallel milled (middle) and spiral milled (right). The objects were produced by Fraunhofer IOF.}
    \label{fig:intro:testbody}
\end{figure*}

The surface topography of metallic products is determined by the surface processing methods used during their manufacturing.
Here, we focus on two methods resulting in completely different surface patterns, namely sandblasting and milling.
See Figure \ref{fig:intro:testbody} for test objects treated by those machining methods.
In sandblasting, a strong air-jet mixed with an abrasive such as sand is used to mat a smooth surface, to smooth a rough surface or to remove contaminants such as rust or dirt. The process results in a non-periodic homogeneous random pattern.
The process parameters are the pressure of the air-jet, the stand-off distance (distance between the tool and the surface), the grain size distribution, the blasting duration and the blasting angle.

Contrary, milling, more precisely face-milling, creates periodically arranged deterministic ring-shaped structures on the surface.
It is a well-known machining method used to form work pieces in high precision.
A milling tool equipped with an end mill moves over the surface following a prescribed path. 
While moving, the milling head rotates and its blades leave ring-shaped scratches.
Process parameters include the diameter of the milling head, the width of the blades, the axial depth of cut, the radial width of cut, the feed rate and others \cite{2011LacalleMilling}. 
Both the toolpath and the milling head diameter are deterministic while the variation within the rings is random.

Various different approaches for synthesis of the texture of such surfaces exist.
\textbf{Process-based} approaches generate textures by simulating the physical process of the surface treatment in high precision.
Contrary, \textbf{phenomenological} approaches create new surface textures based on a given realization thereof.
In general, those simulation methods can be classified into statistics based, patch rearrangement and machine learning methods, see \cite{Raad2017Overview} for an overview.
\textbf{Statistics based} methods generate random output textures that mimic selected statistics of the input. 
\textbf{Patch rearrangement} methods take given realizations in form of images as input, copy and randomly rearrange smaller patches of the image to generate a new texture image.
\textbf{Machine learning} methods use generative models to generate output images resembling the input texture \cite{Gatys2015CNN,Ulyanov2017TextureNet,Wang2021GenConvNet,Xie2016CoopNet,Xie2016gCNN}.

In computer graphics, texture synthesis methods are often divided into procedural and data-based.
The first group includes parameterized methods that define new textures on a continuous domain.
Considering machined surfaces, process parameters can be directly included as model parameters.
However, they are restricted and specialized to a specific surface processing method.
Contrary, for data-based models, the whole information to describe a pattern is delivered by an image.
Since less assumptions on the imaged textures are required, data-based models are typically more widely applicable than procedural methods.

For sandblasting, Oranli et al. \cite{2023OranliSandParticles} developed a process-based method describing the surface topography to assess the resulting surface roughness virtually. 
Varying air pressures and stand-off distances are considered.
Aluminum particles serve as grains to sandblast polymeric objects.
The model describes the surface deformation realistically when differently shaped grains hit the surface.
The particle specific velocity is computed and taken into account.
Yu et al. \cite{2022YuSandParticlesSpheres} simulate the sandblasting process on wind turbine blades to estimate the best parameter choices. 
The grain shape is assumed to be spherical while different grain radii, air pressures and stand-off distances are considered.

A class of parametric models that can be used for stochastic simulation of sandblasted surfaces are Gaussian random fields.
However, little is known about the relation of process and model parameters.
We decided to use data-based methods as they are applicable to random stationary patterns with short-range correlations and are thus suitable for sandblasted surfaces. Details on the models are discussed in Section \ref{sec:sandblasting}.

In the literature, there exist several process-based models for milling. 
Felh\H{o} et al. \cite{Felho2015SurfaceRoughnessFaceMilling} consider various shapes of inserts which approximate the blades. 
Theoretical line profile roughness values of simulated surfaces are compared to measurements from identically machined real surfaces made of steel.
This research was extended to analyze the influence of the feed rate, the speed at which the tool moves over the surface \cite{2018FelhoFaceMillingRoughnessErrors}.
Besides the line profile roughness values, the surface profile roughness values are compared as well. 
However, modeling is restricted to a small central part of the ring-shaped milling pattern.
Kundr\'ak et al. \cite{Kundrak2022RoughnessFaceMilling} also investigated the influence of different inserts and feed rates, comparing two more lateral positions and identifying differences to the central part in the height values. 
The re-cutting effect resulting from multiple passages of the tool over the same spot on the surface is not taken into account in the surface models.
Hadad and Ramezani \cite{Hadad2016FaceMillingImages} included the re-cutting effect by giving an in-depth description of the cutting edge path.

Moreover, Tosello et al. \cite{2023ToselloMachiningModelingSummary} summarize models to simulate surface topographies produced by different machining methods such as milling, grinding and laser machining to enable process optimization such that the resulting surfaces fulfill the requested functional properties.
For milling, they mostly restrict to ball end milling and present various, mainly process-based methods.

Bosnar et al. \cite{Lovro2022TextureSynthesis} developed a stochastic model for circular, parallel and knurling texture synthesis without taking an explicit machining method into account.
Parts of that model can be taken as basis for stochastic models for other milling types.
To the best of our knowledge, there is no literature in which milled surface textures were considered as examples of data-based models.
Since most of these methods reproduce spatially homogeneous structures they are less suitable for milled surfaces because of their ring-shaped components.

This paper introduces texture synthesis models for sandblasted and milled surfaces.
Model fitting is based on topography measurements of such machined surfaces which are introduced in Section \ref{sec:data}.
Section \ref{sec:sandblasting} describes the use of phenomenological methods to synthesize texture images of sandblasted surfaces.
A stochastic texture synthesis model is developed for milled surfaces which is introduced in Section \ref{sec:milling}.
Section \ref{sec:conclusion} serves as conclusion.

\section{Data}
\label{sec:data}

The models for sandblasted and milled textures are developed based on measurements of the surface topography of machined work pieces with planar surfaces. 
Both machining methods were performed using several different sets of processing parameters. 
For sandblasting, we modify only the pressure. 
For milling, there are three parameters.
The tool-path defines whether parallel or spiral milling is applied.
Then, the milling head diameter and the radial with of cut ($a_e$ \cite{2011LacalleMilling}) are chosen.
The latter defines the step size of the milling tool proportionally to the milling head diameter.
Table \ref{tab:data:overviewSurfaces} gives an overview of all available samples.

\begin{table*}[h]
\caption{Overview of both manufacturing processes with the parameters used in each case. Optical 3d images obtained by focus-variation microscopy provided by Fraunhofer IOF. The imaged region for sandblasting is $7.3 \text{ mm}\times 5.7 \text{ mm}$ and for milling $14.6 \text{ mm}\times 11.5 \text{ mm}$.}
\label{tab:data:overviewSurfaces}
\centering
\resizebox{\textwidth}{!}{
\begin{tabular}{|c@{\hskip 1.5pt}c|@{\hskip 3pt}c@{\hskip 3pt}||c@{\hskip 1.5pt}c|@{\hskip 3pt}c@{\hskip 3pt}c@{\hskip 3pt}c@{\hskip 3pt}|@{\hskip 3pt}c@{\hskip 3pt}c@{\hskip 3pt}c@{\hskip 1pt}|}
	\hline
	\multicolumn{3}{|c||}{\textbf{Sandblasting}} & \multicolumn{2}{|}{}&\multicolumn{6}{c|}{\textbf{Milling}}\\
	\cline{4-11}
	\multicolumn{3}{|c||}{} &\multicolumn{2}{|c@{\hskip 3pt}}{\small \textbf{path}}& \multicolumn{3}{c|@{\hskip 3pt}}{\small parallel}& \multicolumn{3}{c|}{\small spiral}\\
	\multicolumn{3}{|c||}{} &\multicolumn{2}{|c@{\hskip 3pt}}{$\mathbf{a_e}$}& \small $0.2$ & \small $0.5$ & \small $0.8$ & \small $0.2$ & \small $0.5$ & \small $0.8$\\
	\hline
    \multirow{2}*[-12pt]{\rotatebox[origin=c]{90}{\small \textbf{pressure}}}
	&\rotatebox[origin=c]{90}{\small $2.5$ bar}&
	\raisebox{-0.425\height}{\includegraphics[height=1.5cm]{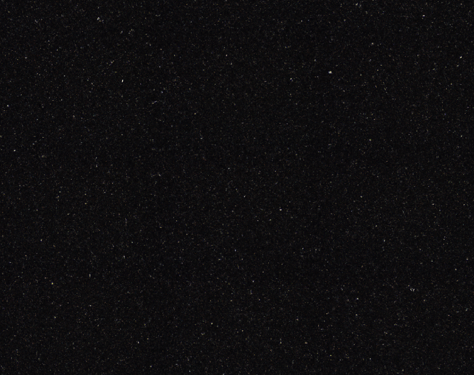}}&
    \multirow{2}*[5pt]{\rotatebox[origin=c]{90}{\small \textbf{head diameter}}}&
	\rotatebox[origin=c]{90}{\small $4$ mm}&
	\raisebox{-0.425\height}{\includegraphics[height=1.5cm]{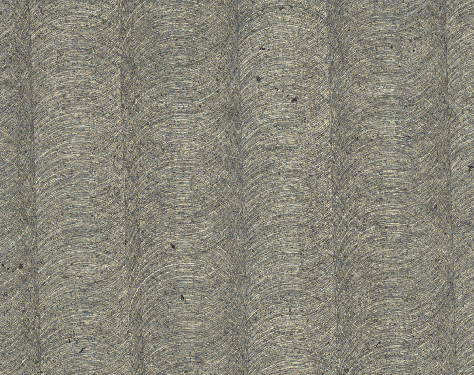}}&
	\raisebox{-0.425\height}{\includegraphics[height=1.5cm]{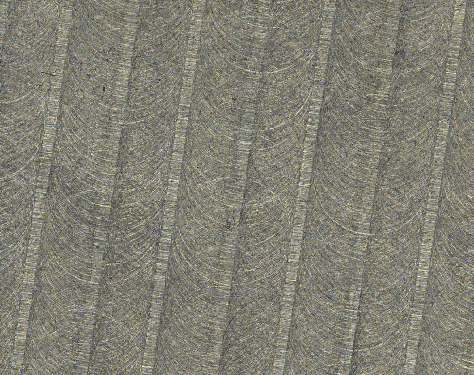}}&
	\raisebox{-0.425\height}{\includegraphics[height=1.5cm]{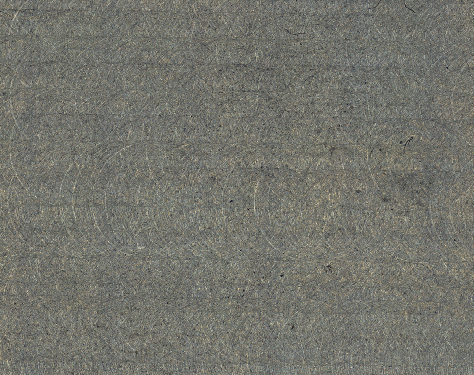}}&
	\raisebox{-0.425\height}{\includegraphics[height=1.5cm]{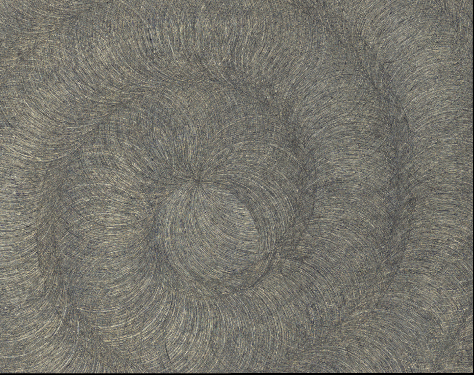}}&
	\raisebox{-0.425\height}{\includegraphics[height=1.5cm]{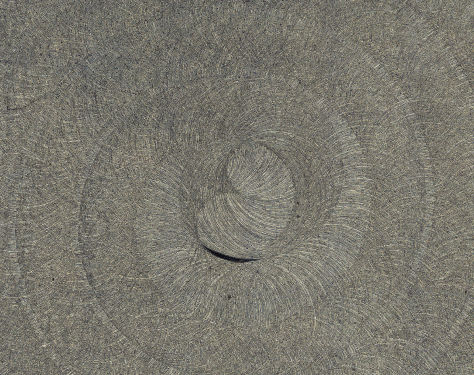}}&
	\raisebox{-0.425\height}{\includegraphics[height=1.5cm]{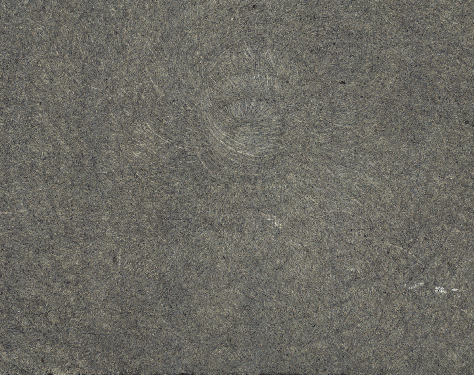}} \rule{0pt}{27pt} \\[18pt]
    &\rotatebox[origin=c]{90}{\small $6$ bar}&
	\raisebox{-0.425\height}{\includegraphics[height=1.5cm]{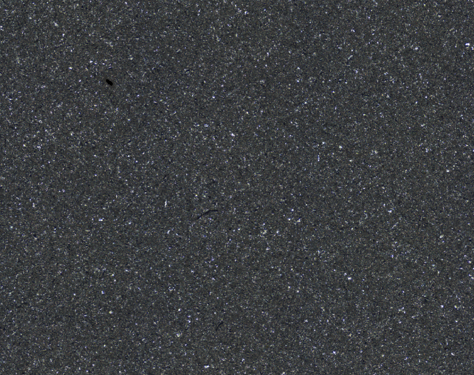}}&&
	\rotatebox[origin=c]{90}{\small $8$ mm}&
	\raisebox{-0.425\height}{\includegraphics[height=1.5cm]{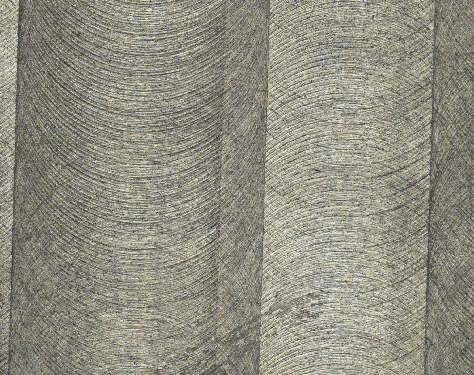}}&
	\raisebox{-0.425\height}{\includegraphics[height=1.5cm]{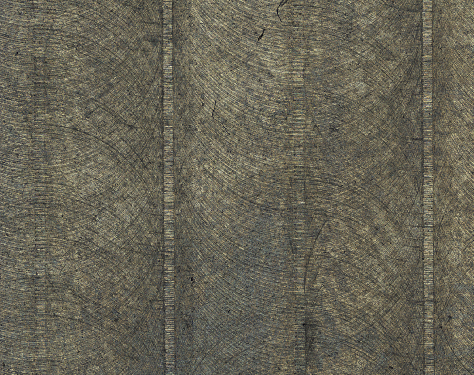}}&
	\raisebox{-0.425\height}{\includegraphics[height=1.5cm]{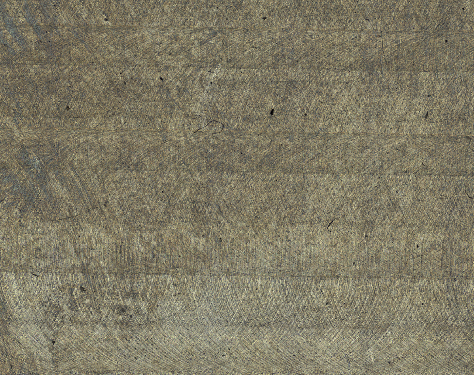}}&
	\raisebox{-0.425\height}{\includegraphics[height=1.5cm]{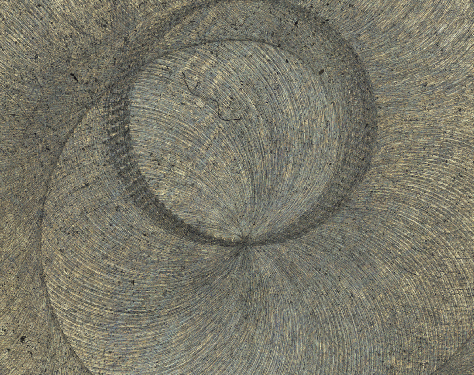}}&
	\raisebox{-0.425\height}{\includegraphics[height=1.5cm]{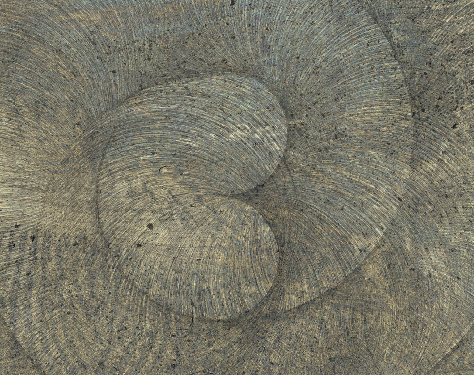}}&
	\raisebox{-0.425\height}{\includegraphics[height=1.5cm]{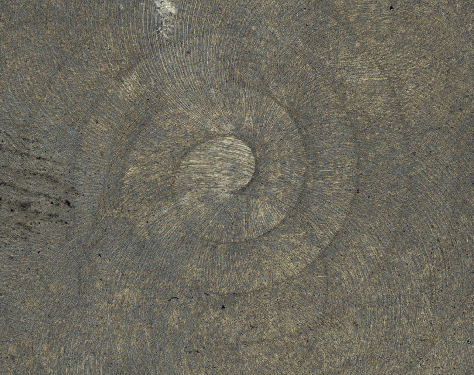}} \rule[-21pt]{0pt}{0pt} \\
	\hline						
\end{tabular}}
\end{table*}

\begin{figure*}[h]
	\centering
    \includegraphics[height=4cm]{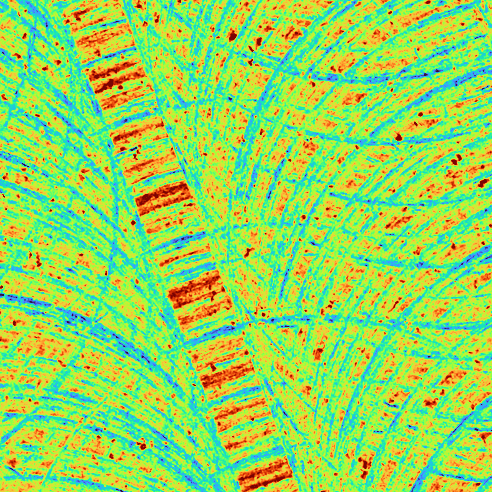}
    \includegraphics[height=3cm]{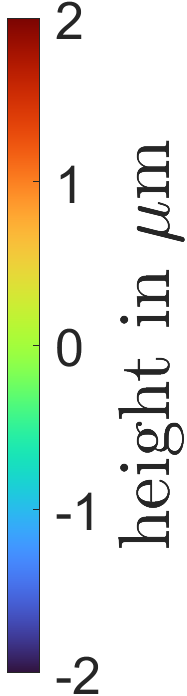}
    \hspace{8pt}
	\includegraphics[height=4cm]{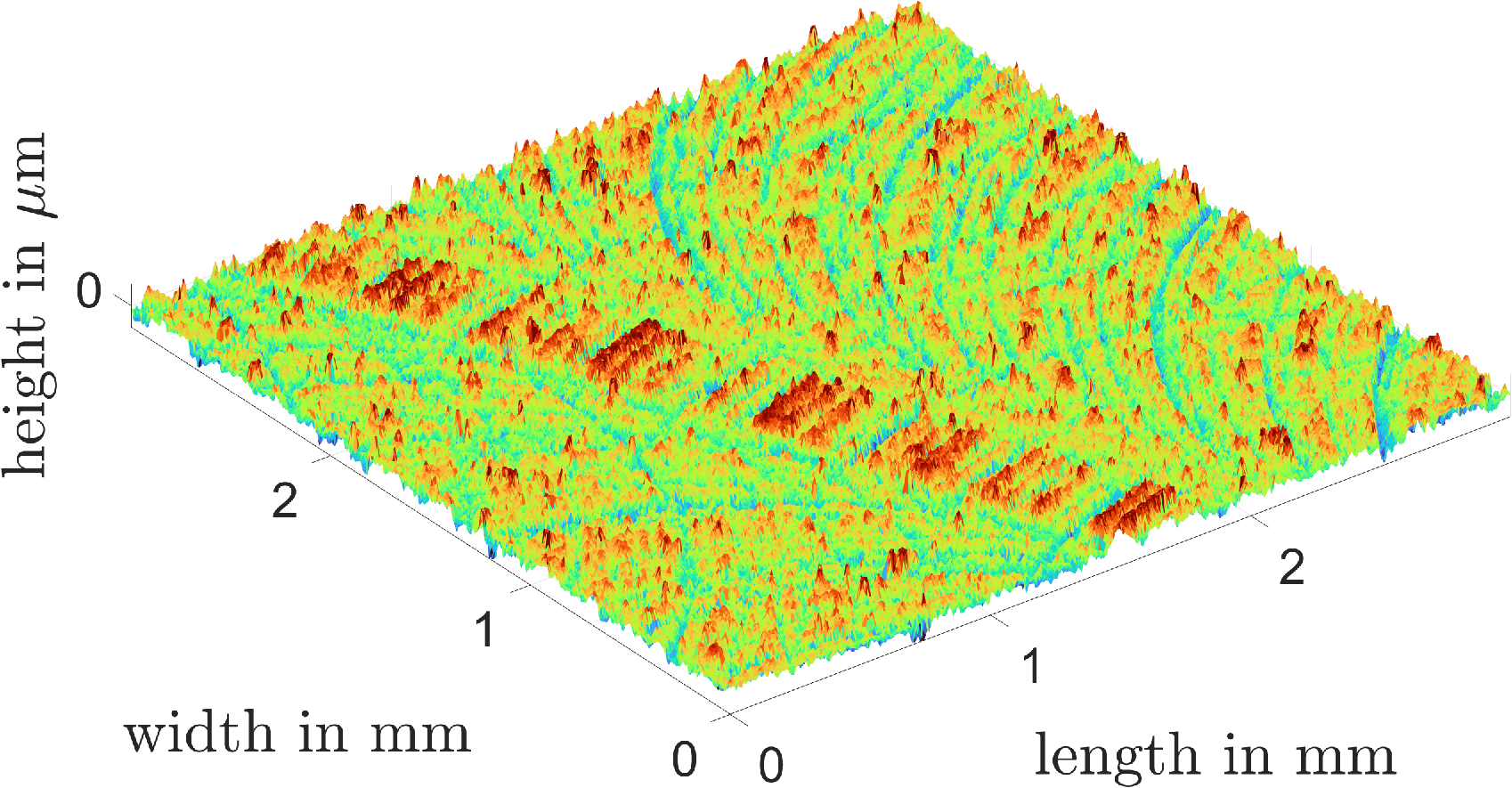}
    \caption{Measurement of a milled surface provided by Fraunhofer IOF. 2d height image (left) and its representation as a surface in $\mathbb{R}^3$ (right). Imaged region is $3\text{ mm}\times 3\text{ mm}$.}
	\label{fig:data:heightimage}
\end{figure*}

The measurements were performed using the focus-variation microscope Bruker alicona InfiniteFocus G4 with a vertical resolution of $420$ nm. 
Data are provided as 2d height images $\mathcal{H}$ of size $M\times N$ and homogeneous pixel spacing $\nu$.
The gray value in a pixel with coordinates $(x,y)$ represents the height value $\mathcal{H}(x,y)$ at the given surface point. 
Hence, $(x,y,\mathcal{H}(x,y))$ can be interpreted as a 2d surface in $\mathbb{R}^3$, see Figure \ref{fig:data:heightimage}. 
In our measurements, we have $\nu=1.75\,\mu\text{m}$ and  image sizes varying between $4301\times 3175$ and $4440\times3288$.  

\section{Model for sandblasted surfaces}
\label{sec:sandblasting}

\begin{figure*}
	\centering
	\includegraphics[height=4cm]{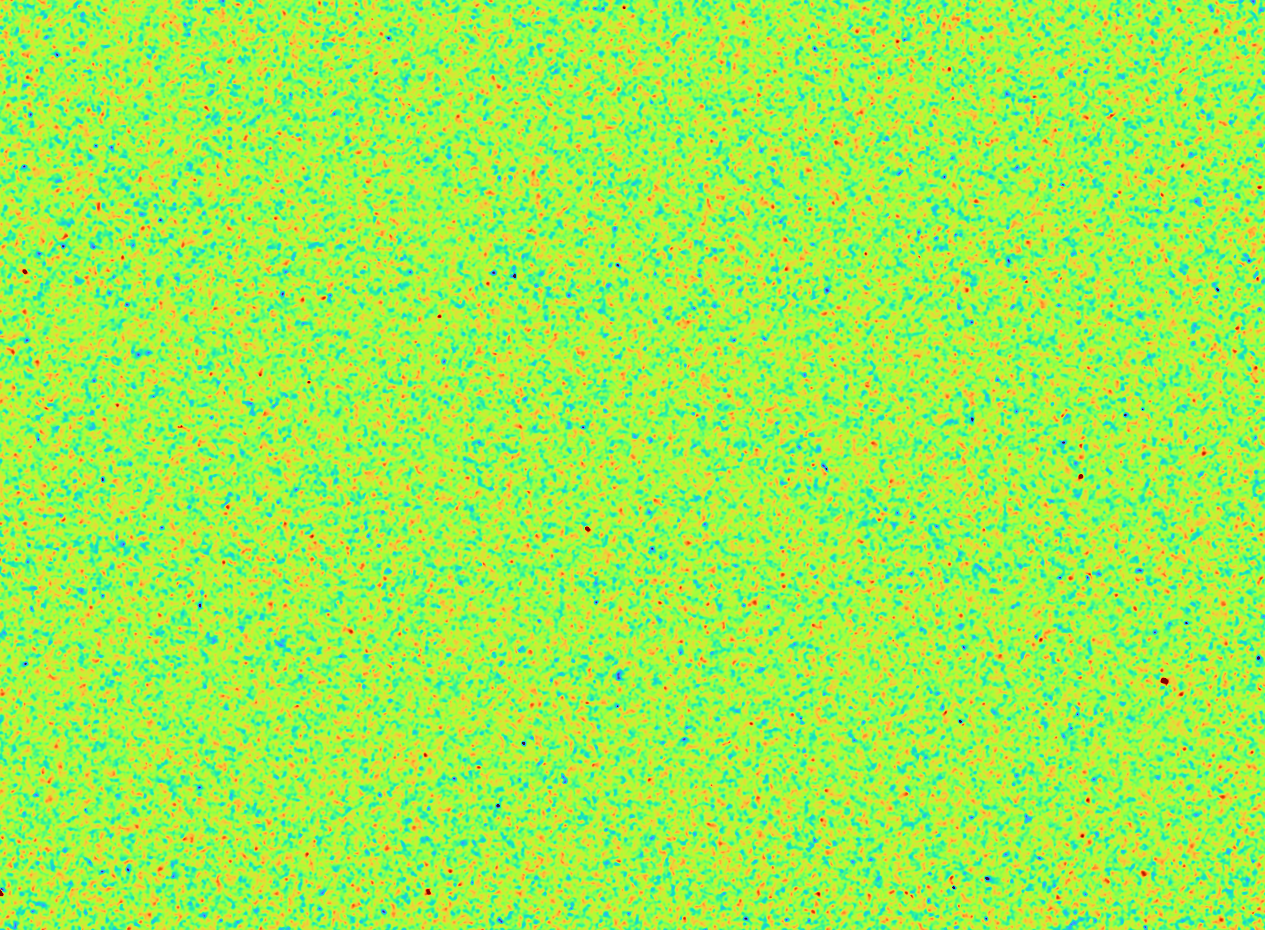}
	\includegraphics[height=4cm]{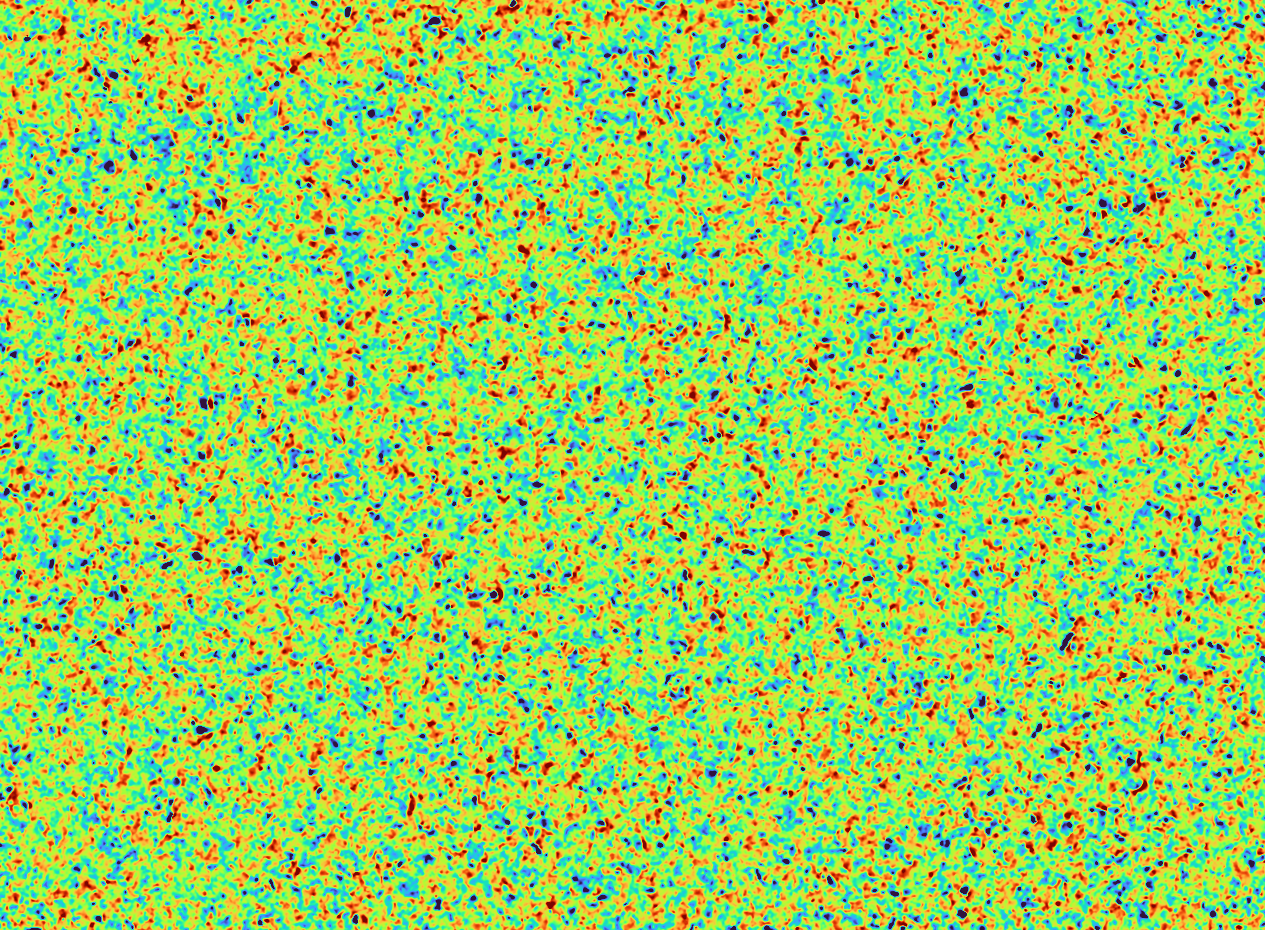}
	\includegraphics[height=3cm]{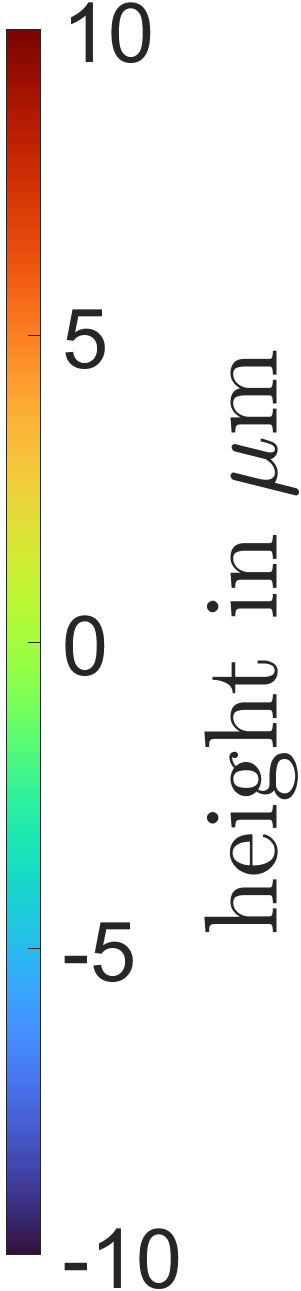}
	\caption{Measurements of sandblasted surfaces using an air-jet with pressure $2.5$ bar (left) and $6$ bar (right) provided by Fraunhofer IOF. The imaged region is $5.6\text{ mm}\times 7.7\text{ mm}$.}
	\label{fig:sand:measurements}
\end{figure*}

Sandblasting produces a homogeneous surface topography that can be interpreted as a stationary random field.
Almost no deterministic components exist. 
Depending on the air-jet's pressure, the resulting patterns differ in depth and their degree of roughness, see Figure \ref{fig:sand:measurements}. 
We conjecture that the distance between the tool and the surface as well as the grain size distribution of the sand also influence the textures \cite{2020BechikhSandParameter1,2023OranliSandParticles}.

Fitting Gaussian random fields to the measurements did not result in visually similar texture images.
Additionally, it is not obvious how to define a parametric model that takes process-induced parameters into account.
Thus, we use data-based texture generating methods that take the measured texture images as input and generate similar outputs.
Here, we limit ourselves to statistic based and patch rearrangement methods, as these are comprehensible and provide sufficiently good results. 
Although many statistical constraints are fitted within the presented statistic based methods, they are still few compared to the amount of features in deep learning methods for texture generation. 
In particular, the method of Portilla and Simoncelli is still considered as state-of-the-art and used to compare to recent methods.

Statistics based methods are used to generate texture images of the same size and pixel spacing as the measurements.
Several methods are summarized and compared in Section \ref{sec:sand:methods}.
Then, patch rearrangement methods are used to produce images of arbitrary size.
This procedure is detailed in Section \ref{sec:sand:SpacingScaling}.

\subsection{Texture generation methods}
\label{sec:sand:methods}

Four different statistical data-based texture generating methods are considered: the asymptotic discrete spot noise (ADSN) \cite{Galerne2011AdsnRpnTheoretical,Leclaire2015ADSNexact}, the random phase noise (RPN) \cite{Galerne2011AdsnRpnTheoretical}, and the methods by Heeger and Bergen (HB) \cite{Heeger1995HBfirst,Vacher2014HeegerBergen} and by Portilla and Simoncelli (PS) \cite{Portilla2000PSold,Vacher2021PortillaSimoncelli}.
All methods are divided into two steps.
First, certain statistics of the input image are computed.
Then, the output image is calculated by first creating a Gaussian white noise image $\mathcal{G}$ with $\mathcal{G}(p)\sim\mathcal{N}(0,1)$ i.i.d. for $p\in\Omega$ and then adjusting it so that the desired statistics match.
Here, let $\mathcal{I}$ be the input image of size $M\times N$ defined on the grid $\Omega=\left\{1,\dots,M\right\}\times\left\{1,\dots,N\right\}$.
Since all methods are based on the Fourier transform, we assume $\mathcal{I}$ to be periodic to avoid artifacts.

\subsubsection{Asymptotic discrete spot noise}

Galerne et al. \cite{Galerne2011AdsnRpnTheoretical} introduced the ADSN as limit of the normalized discrete spot noise (DSN).
The DSN of order $n$ associated with input image $\mathcal{I}$ is then given as sum of random shifts $\mathcal{I}_k(p)=\mathcal{I}(p-p_k)$ of $\mathcal{I}$ for $p_k$ independent and identically uniformly distributed in $\Omega$, namely $DSN_n(\mathcal{I})=\sum_{k=1}^n \mathcal{I}_k$.
By the central limit theorem, the DSN asymptotically follows a Gaussian distribution $$\sqrt{n}\left(\frac{1}{n}DSN_n(\mathcal{I})-\hat{\mu}_\mathcal{I}\right)\xrightarrow{d}ADSN\left(\mathcal{I}\right)\sim\mathcal{N}\left(0,\widehat{C}_\mathcal{I}\right),$$
where $\hat{\mu}_\mathcal{I}$ is the arithmetic mean and $\widehat{C}_\mathcal{I}$ the covariance matrix of $\mathcal{I}$.
The ADSN can thus be simulated by sampling from the multivariate normal distribution $\mathcal{N}\left(\hat{\mu}_\mathcal{I},\widehat{C}_\mathcal{I}\right)$.

ADSN is a non-iterative and parameter-free method. 
Figure \ref{fig:sand:methods} shows an exemplary output.
The Fourier modulus of a newly generated texture image is the Fourier modulus of the input image 
multiplied with Rayleigh noise \cite{Galerne2011AdsnRpnTheoretical}.
Thus, minor differences are visible when comparing their sample auto-correlations, see Figure \ref{fig:sand:methodsComp:autocorr}, which are obtained by the inverse Fourier transformation of the squared Fourier modulus.

\subsubsection{Random phase noise}
The random phase noise associated with an image $\mathcal{I}$ generates new texture images having the same Fourier modulus as $\mathcal{I}$ but a random phase, see \cite{Galerne2011RpnImages,Galerne2011AdsnRpnTheoretical,Holten2006RPNfirst} for more details.
This method is also non-iterative and parameter-free.

Assume $M$ and $N$ to be odd.
Then, the RPN is defined in the Fourier domain $\widehat{\Omega}=\left\{-\frac{M-1}{2},\dots,\frac{M-1}{2}\right\}\times\left\{-\frac{N-1}{2},\dots,\frac{N-1}{2}\right\}$ by $\widehat{RPN}(\mathcal{I})(\xi) = \widehat{\mathcal{I}}(\xi)\cdot\exp\left(\text{i}\theta(\xi)\right)$ for $\xi\in\widehat{\Omega}$.
Here, $\theta$ is a uniform random phase which is defined as function $\theta:\widehat{\Omega}\rightarrow(-\pi,\pi]$ fulfilling the following properties
\begin{enumerate}
	\item $\theta(-\xi)=-\theta(\xi)$ for all $\xi\in\widehat{\Omega}$
    \item $\theta(\xi)\sim\mathcal{U}((-\pi,\pi])$ for $\xi\in\widehat{\Omega}\setminus\left\{(0,0),\left(0,\frac{N-1}{2}\right),\left(\frac{M-1}{2},0\right),\left(\frac{M-1}{2},\pm\frac{N-1}{2}\right)\right\}$
	\item $\theta(\xi)\sim\mathcal{U}(\{0,\pi\})$ for $\xi\in\left\{(0,0),\left(0,\frac{N-1}{2}\right),\left(\frac{M-1}{2},0\right),\left(\frac{M-1}{2},\pm\frac{N-1}{2}\right)\right\}$
	\item For any subset $\widehat{\omega}\subset\widehat{\Omega}$ not containing symmetric points ($\xi\in\widehat{\omega}\Rightarrow-\xi\notin\widehat{\omega}$), the random variables $\left\{\theta(\xi):\xi\in\widehat{\omega}\right\}$ are independent.
\end{enumerate}
A Gaussian white noise image has a uniform random phase.
In spite of assumption 3. above, we assume additionally $\theta((0,0))=0$ such that $RPN(\mathcal{I})$ maintains the sample mean of $\mathcal{I}$. Note that the method is also valid for even $M$ or $N$. 
See Figure \ref{fig:sand:methods} for a realization of the RPN.
Since $\mathcal{I}$ and $RPN(\mathcal{I})$ have the same Fourier modulus, their sample auto-correlations coincide as can be seen in Figure \ref{fig:sand:methodsComp:autocorr}.

\subsubsection{Method of Heeger and Bergen}
This method was primarily introduced by Heeger and Bergen \cite{Heeger1995HBfirst}.
Briand et al. \cite{Vacher2014HeegerBergen} published a more detailed version including an in-depth description of the method and possible extensions.
In addition, an implementation is provided.

The HB-method iteratively modifies an initial image $HB_0=\mathcal{G}$.
Consider $K$ iterations resulting in updated texture images $HB_k,\,k=1,\dots,K$.
In each iteration, the histogram of $HB_{k-1}$ is first matched to that of $I$.
Then, the image is decomposed by using the steerable pyramid introduced by Simoncelli and others \cite{Simoncelli1992SteerablePyramid,Simoncelli1995SteerablePyramid}.
Histogram matching is applied on all sub-bands to adapt the image pyramid to that of $\mathcal{I}$.
Finally, the updated texture image $HB_{k}$ is reconstructed from the adapted image pyramid.

The histogram matching algorithm first sorts the gray values of both images and thus gives each pixel a rank. 
If ties occur, the rank is assigned according to the scanning order \cite{Vacher2014HeegerBergen}.
Then, each pixel in the image to be matched is assigned the gray value of that pixel in the reference image having the same rank.

The steerable pyramid is used for image decomposition. 
The image is linearly decomposed into multiple scaled and oriented sub-bands.
Therefore, split the image first into its high and low frequency part.
Then, oriented symmetric band-pass filters are applied to the low frequency image followed by down-sampling of factor two.
The latter is done in the Fourier domain by cropping out the central part of the image.
That procedure is repeated $P$ times considering $Q$ orientations each.
In total, the pyramid consist of $PQ+2$ images of different sizes.
To enable down-sampling, the image size must be a multiple of $2^P$.
The steerable image decomposition is a pseudo-invertible operator \cite{Simoncelli1995SteerablePyramid} which allows to  reconstruct the image from the pyramid.
The convolutions for the filters are computed by multiplication in the Fourier domain.
Since numerical artifacts cause complex-valued sub-bands having a negligible imaginary part just their real parts are used for histogram matching.

The output of the HB-method is denoted by $HB_K^{P,Q}(\mathcal{I})$.
Briand et al. \cite{Vacher2014HeegerBergen} recommended $P=Q=4$ and showed that convergence is fast such that $K=5$ is sufficient. Additionally, they provided \textsc{matlab}-code which we used for the computation.
Figure \ref{fig:sand:methods} shows that edge effects occur in form of elongated structures. 
In order to remove that artifact, the output's boundary region can be cropped.

\subsubsection{ Method of Portilla and Simoncelli}

The method of Portilla and Simoncelli \cite{Portilla2000PSold} is an extension of the HB-method. 
There are two major differences to the HB-method.

First, the complex steerable pyramid is taken.
Instead of real-valued oriented band-pass filters, complex-valued filters are used which are obtained by applying the respective Hilbert filters \cite{Vacher2021PortillaSimoncelli}.
The real part of each filter coincides with the corresponding real-valued filter used for the HB-method.

Second, instead of using image-wise histogram matching, several summary statistics are matched to those estimated from the input image.
An algorithm for doing so is given in \cite{Portilla2000PSold,Vacher2021PortillaSimoncelli}.
A recommendation for the choice of relevant summary statistics is based on experimental evaluation done by Portilla and Simoncelli \cite{Portilla2000PSold}. 
Marginal statistics (e.g. mean, variance, skewness) and auto-correlations on the low-frequency bands at each scale and auto- and pairwise cross-correlations of the modulus of all oriented sub-bands are considered as relevant.
The necessity of the selected statistics is confirmed when showing the effect of removing some of them \cite{Portilla2000PSold,Vacher2021PortillaSimoncelli}.

The output of the iterative PS-method is denoted by $PS_K^{P,Q,S}(\mathcal{I})$. 
Compared to the HB-method, just one additional parameter is needed: the adjustment of auto-correlations is limited to its central area of size $S\times S$ for odd $S\in\mathbb{N}$.
Vacher and Briand \cite{Vacher2021PortillaSimoncelli} discuss the influence of the parameter choices. 
In particular, they recommend to choose $S=7$. 
Further, they provide code which we used for the computations.
Figure \ref{fig:sand:methods} shows a realization of the PS-method.

\subsubsection{Comparison of methods}
\label{sec:sand:methods:comparison}

The input image is assumed to be periodic which is usually not the case for arbitrary images.
Moisan \cite{Moisan2011PEriodicSmoothDecomposition} describes how to obtain periodic boundary conditions on non-periodic images by decomposing the image into its periodic component $I_\text{per}$ and a smooth component.
We will use $I_\text{per}$ as input image.

\begin{figure*}
	\centering
	\footnotesize
	\stackunder[5pt]{\includegraphics[height=3cm]{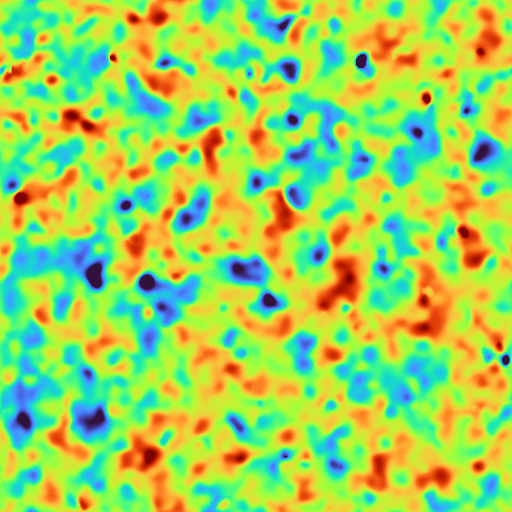}}{$\mathcal{I}_\text{per}$}
	\stackunder[5pt]{\includegraphics[height=3cm]{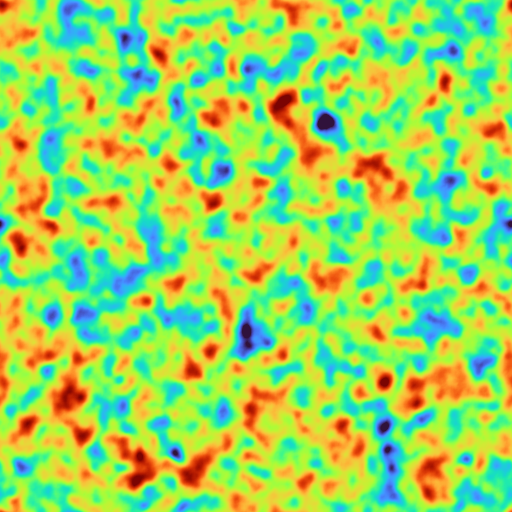}}{$ADSN\left(\mathcal{I}_\text{per}\right)$}
	\stackunder[5pt]{\includegraphics[height=3cm]{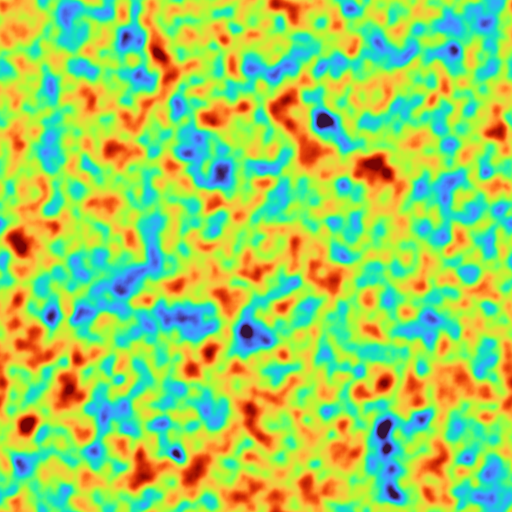}}{$RPN\left(\mathcal{I}_\text{per}\right)$}
	\stackunder[5pt]{\includegraphics[height=3cm]{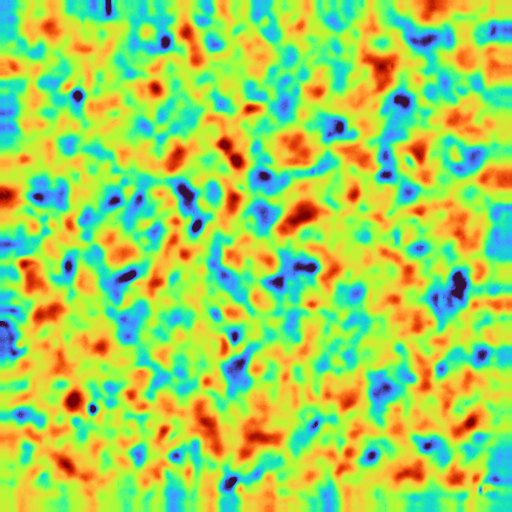}}{$HB_5^{4,4}\left(\mathcal{I}_\text{per}\right)$}
	\stackunder[5pt]{\includegraphics[height=3cm]{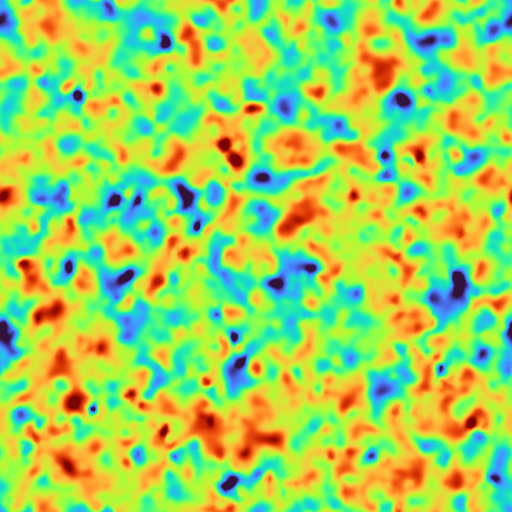}}{$PS_5^{4,4,7}\left(\mathcal{I}_\text{per}\right)$}
    \includegraphics[height=3cm]{images/sand/colorbar10um.png}
	\caption{Output of texture generation methods using the same initial Gaussian white noise image. Image size is $512\times 512$ and pixel spacing $1.75\,\mu$m which corresponds to an imaged region of $0.9\text{ mm}\times0.9\text{ mm}$.}
	\label{fig:sand:methods}
\end{figure*}

\begin{figure*}
	\centering
    \begin{subfigure}[b]{\columnwidth}
        \centering
        \footnotesize
        \stackunder[5pt]{\includegraphics[height=3cm]{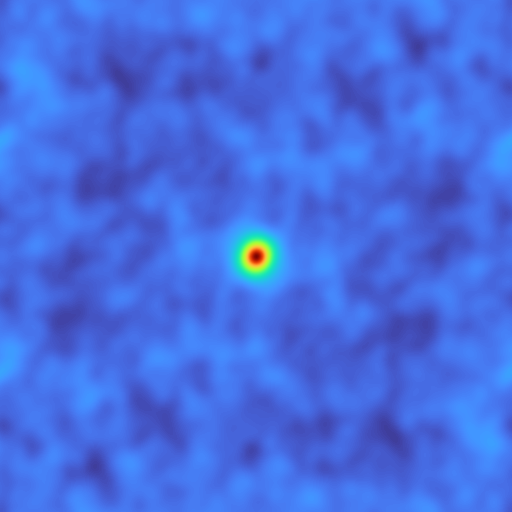}}{$\mathcal{I}_\text{per}$}
        \stackunder[5pt]{\includegraphics[height=3cm]{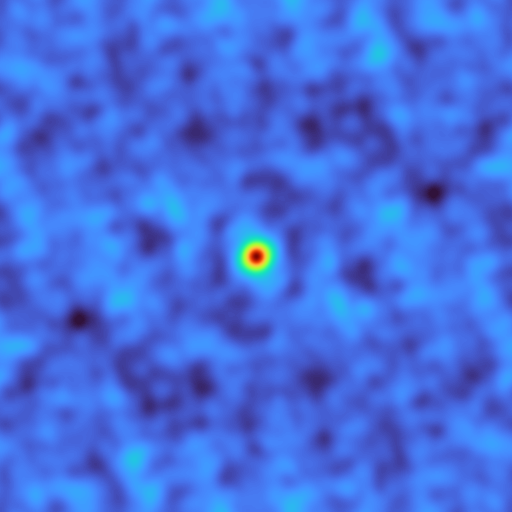}}{$ADSN\left(\mathcal{I}_\text{per}\right)$}
    	\stackunder[5pt]{\includegraphics[height=3cm]{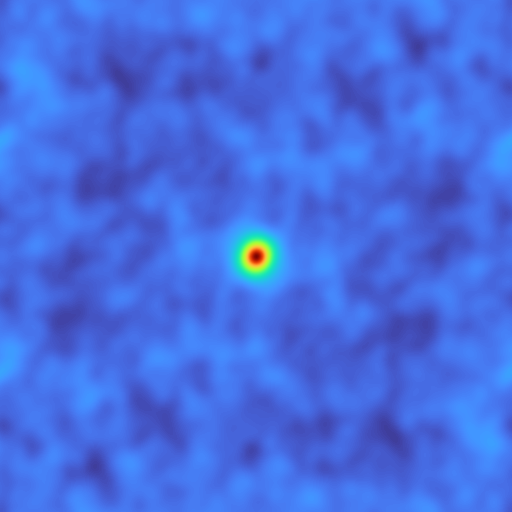}}{$RPN\left(\mathcal{I}_\text{per}\right)$}
    	\stackunder[5pt]{\includegraphics[height=3cm]{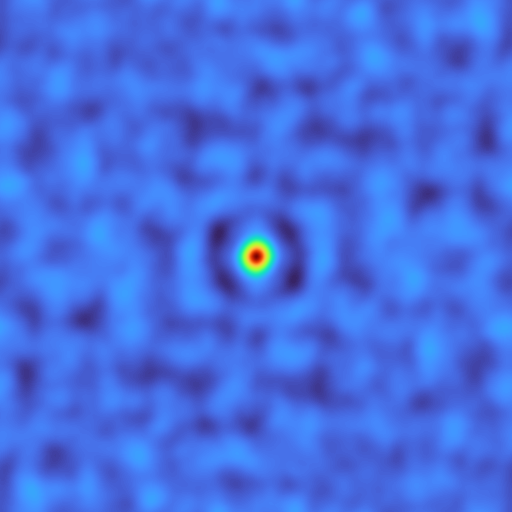}}{$HB_5^{4,4}\left(\mathcal{I}_\text{per}\right)$}
    	\stackunder[5pt]{\includegraphics[height=3cm]{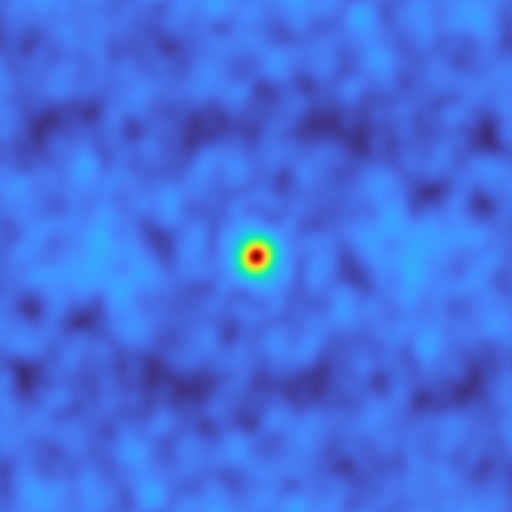}}{$PS_5^{4,4,7}\left(\mathcal{I}_\text{per}\right)$}
        \includegraphics[height=3cm]{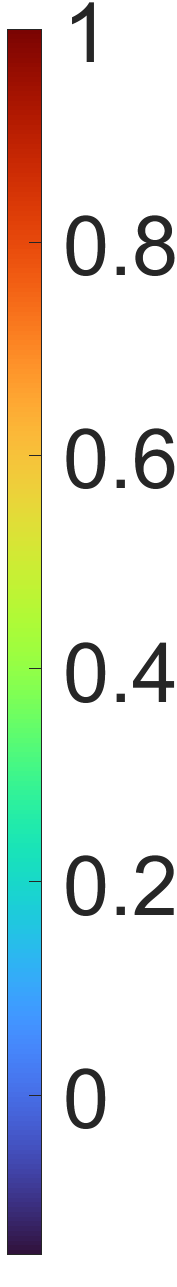}
        \caption{Auto-correlation matrices of images given in Figure \ref{fig:sand:methods}.}
    	\label{fig:sand:methodsComp:autocorr}
    \end{subfigure}\\[8pt]
    \begin{subfigure}[t]{0.48\columnwidth}
        \centering
        \includegraphics[height=4cm]{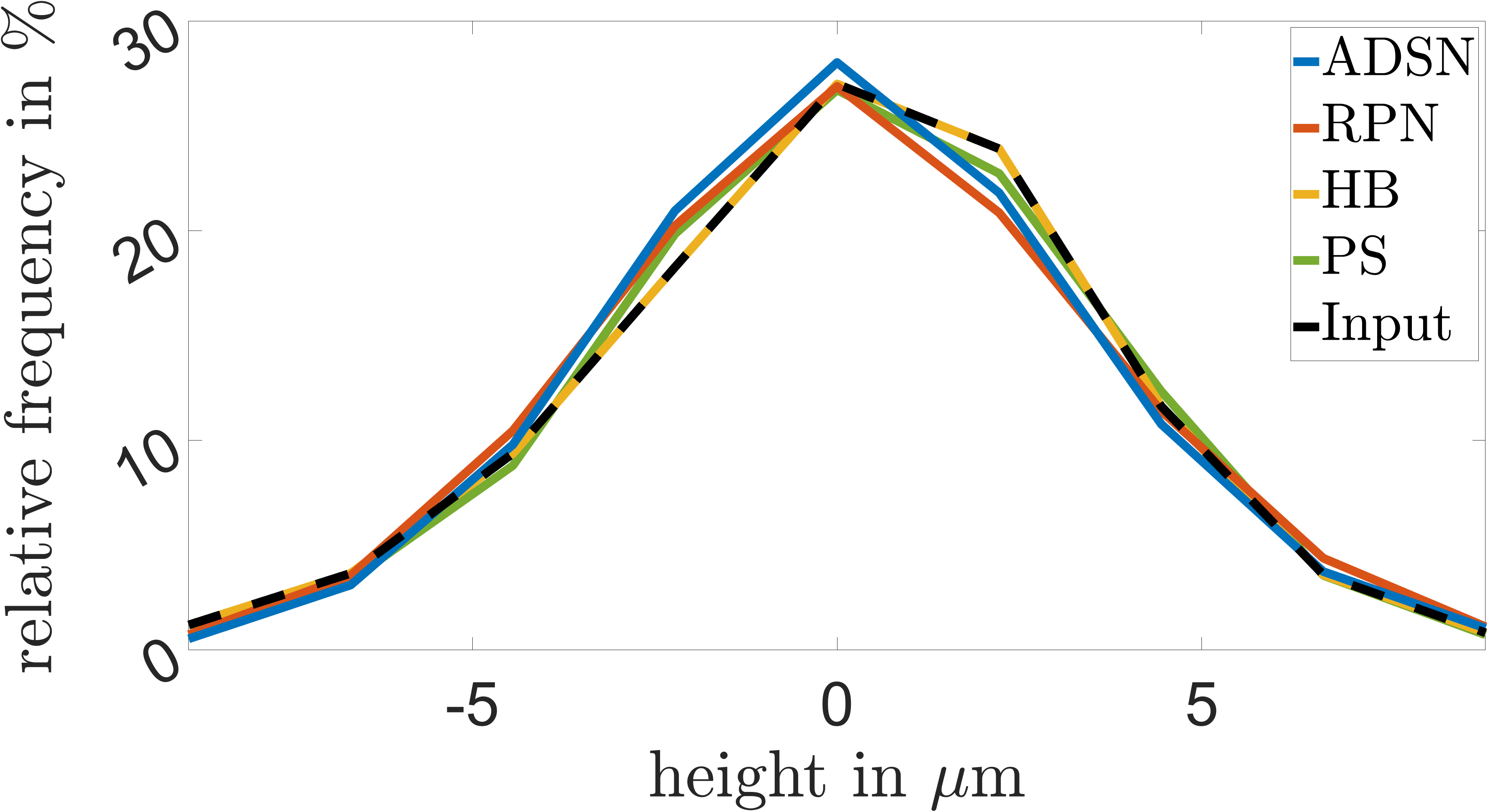}
        \caption{Height value distribution. }
        \label{fig:sand:methodsComp:histogram}
    \end{subfigure}
    \hspace{1em}
    \begin{subfigure}[t]{0.48\columnwidth}
        \centering
        \includegraphics[height=4cm]{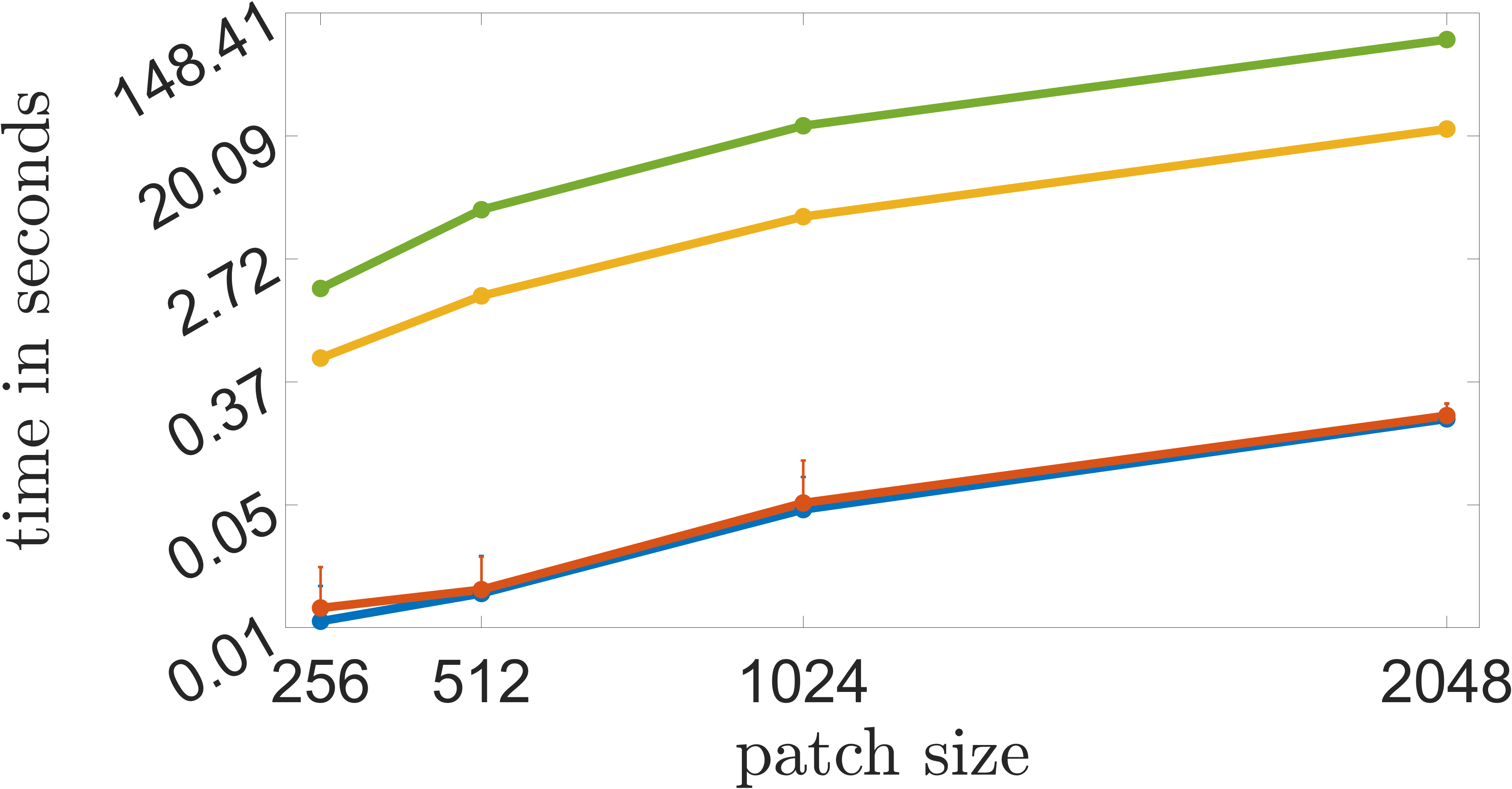}
        \caption{Mean run-time as means of $100$ repetitions of \textsc{matlab}-code of methods in logarithmic scale dependent on the patch size. Error bars indicate minimal and maximal values. Colors as in Figure \ref{fig:sand:methodsComp:histogram}.}
        \label{fig:sand:methodsComp:runtime}
    \end{subfigure}
    \caption{Comparison of texture generation methods. The run-time measurements were performed on a HP Z240 Tower Workstation equipped with Intel Core i7-3700 CPU at 3.40GHz $\times$ 8.}
\end{figure*}

Figure \ref{fig:sand:methods} summarizes all texture generating methods introduced previously. 
Visually, they all resemble the texture in the input image while being pixel-wise different.
The distributions of height values are similar as well, see Figure \ref{fig:sand:methodsComp:histogram}.
Information about spatial relations is given by the auto-correlation in Figure \ref{fig:sand:methodsComp:autocorr}.
It is noticeable that the auto-correlation of the RPN resembles the auto-correlation of the input image since both have the same Fourier modulus.
The result of the PS-method has a similar auto-correlation as well.
However, the auto-correlations of the other two methods show that smaller structures are erroneously present in their output images.
Thus, RPN and PS are the best methods with the first having a substantially faster run-time, see Figure \ref{fig:sand:methodsComp:runtime}.
Therefore, from now on we will use the RPN as texture generation method.

\subsection{Simulating texture images of arbitrary size and pixel spacing}
\label{sec:sand:SpacingScaling}

The methods presented in Section \ref{sec:sand:methods} generate images that maintain the image size and pixel spacing of the input image. 
Here, we discuss how to generate texture images $\mathcal{T}$ of arbitrary size $M_\mathcal{T}\times N_\mathcal{T}$ and pixel spacing $\nu_\mathcal{T}\geq\nu_{\mathcal{I}}$ larger than the input image. The case of higher resolved texture images ($\nu_\mathcal{T}<\nu_{\mathcal{I}}$) is not considered as the input $\mathcal{I}$ does not contain the required image information.

\begin{figure}[h]
    \centering
    \includegraphics[height=3cm]{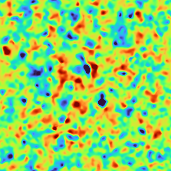}
    \includegraphics[height=3cm]{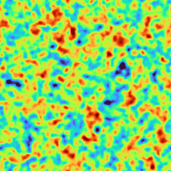}
    \includegraphics[height=3cm]{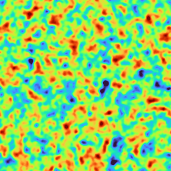}
    \includegraphics[height=3cm]{images/sand/colorbar10um.png}\\[2pt]
    \includegraphics[height=3cm]{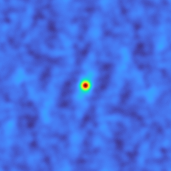}
    \includegraphics[height=3cm]{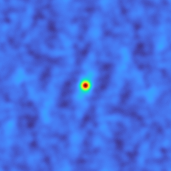}
    \includegraphics[height=3cm]{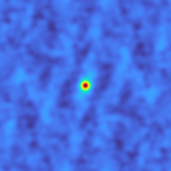}
    \includegraphics[height=3cm]{images/sand/SandMethodsColorbarAutocorr.png}
    \hspace{5.5pt}
    \caption{Adjustment of pixel spacing: Comparison between down-sampled reference image (left) and adjustments of pixel spacing either by down-sampling the synthetic texture image (center) or the input image (right). RPN is used for texture generation. Top: the textures' height image, bottom: their auto-correlations. Factor $3$ is used for down-sampling, so image size reduces to $171\times 171$ and pixel spacing is $5.25\,\mu\text{m}$. The covered area is $0.9\,\text{mm}\times0.9\,\text{mm}$.} 
    \label{fig:sand:spacing}
\end{figure}

There are two possibilities to create down-sampled texture images.
Either a texture image of the same pixel spacing as the input is generated and then down-sampled afterwards or the synthetic texture image is generated directly from a down-sampled version of the input.
Results of both options when using nearest-neighbor interpolation for down-sampling are shown in Figure \ref{fig:sand:spacing}.
Visual comparison does not reveal obvious differences to the down-sampled input image. 
Also, the related auto-correlation matrices look similar.
The main advantage of down-sampling prior to texture synthesis is the drastically reduced computation time. 
Thus, this procedure is used from now on.

Now, consider the adjustment to generate texture images of arbitrary size. 
A reduction of the image size is achieved by simple cropping of the input image or the output.
Note that HB and PS require input image sizes to be powers of $2$.

\begin{figure*}
    \centering
    \hspace{3cm}
    \includegraphics[height=3cm]{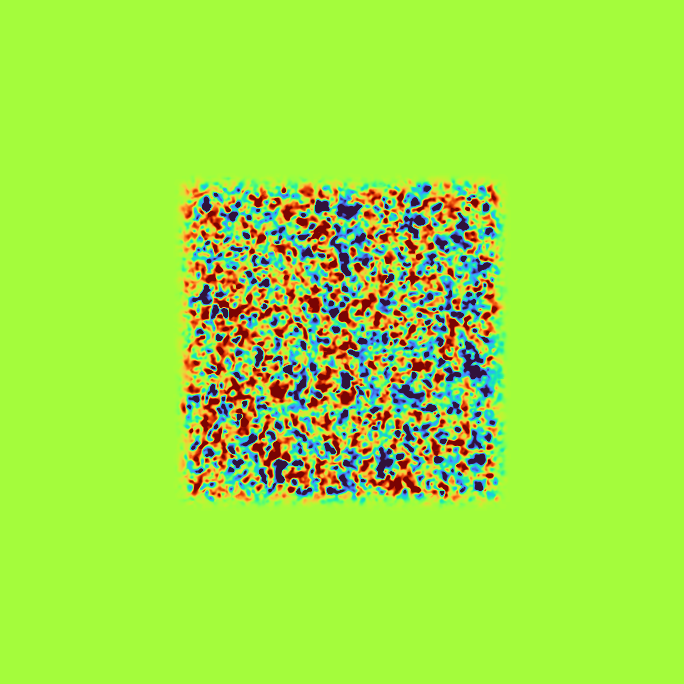}
    \includegraphics[height=3cm]{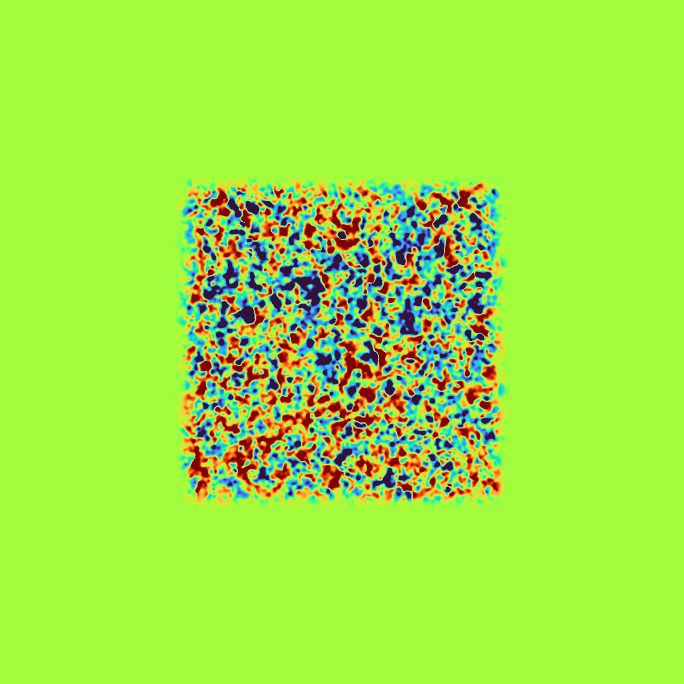}
    \phantom{\includegraphics[height=3cm]{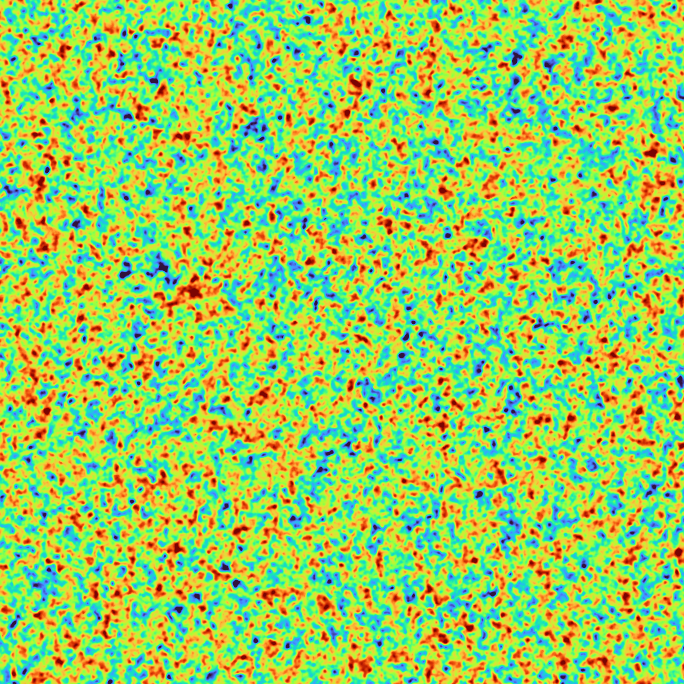}}
    \phantom{\includegraphics[height=3cm]{images/sand/colorbar10um.png}}\\[2pt]
    \includegraphics[height=3cm]{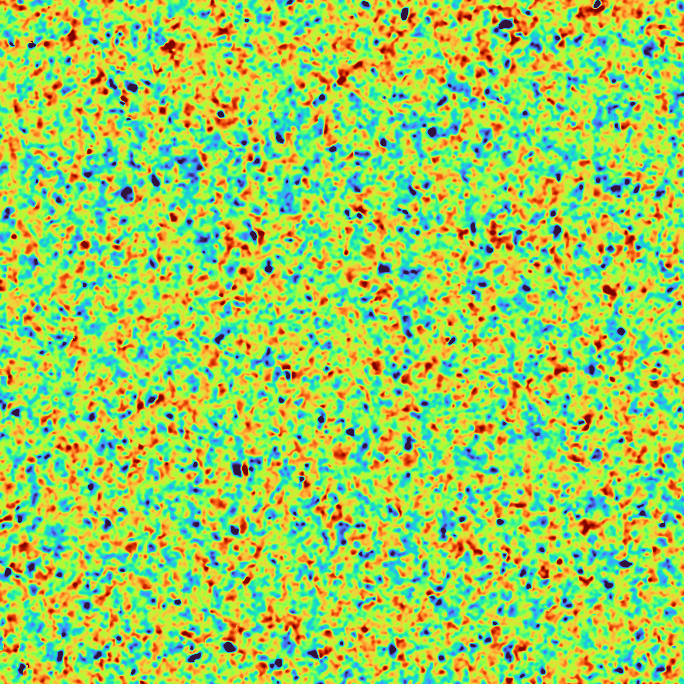}
    \includegraphics[height=3cm]{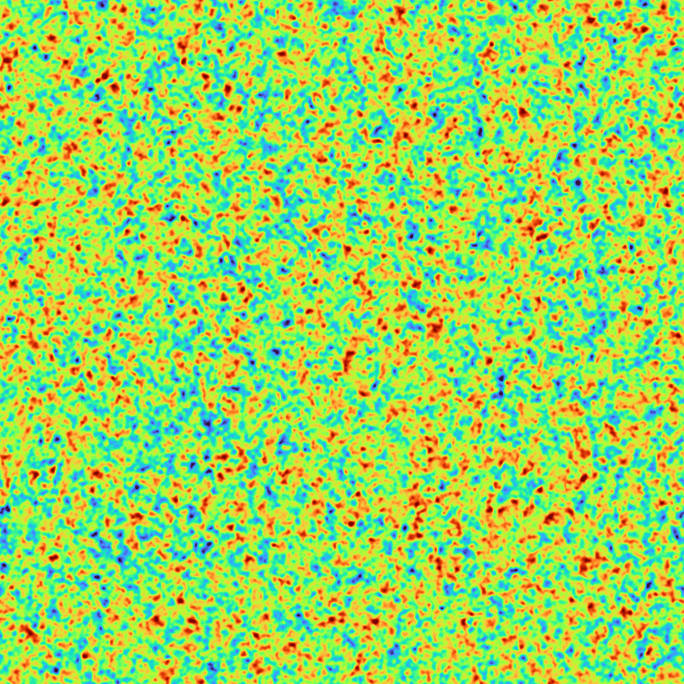}
    \includegraphics[height=3cm]{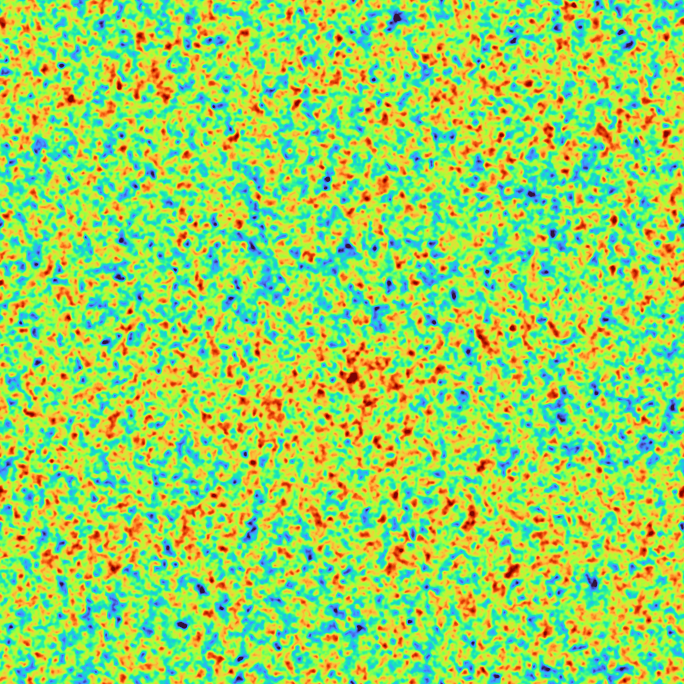}
    \includegraphics[height=3cm]{images/sand/SandMethodsEF.png}
    \includegraphics[height=3cm]{images/sand/colorbar10um.png}\\[2pt]
    \includegraphics[height=3cm]{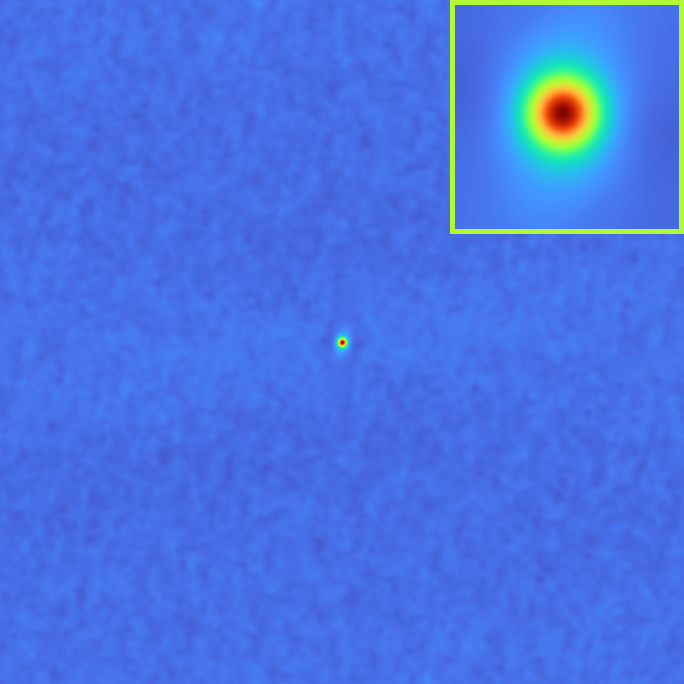}
    \includegraphics[height=3cm]{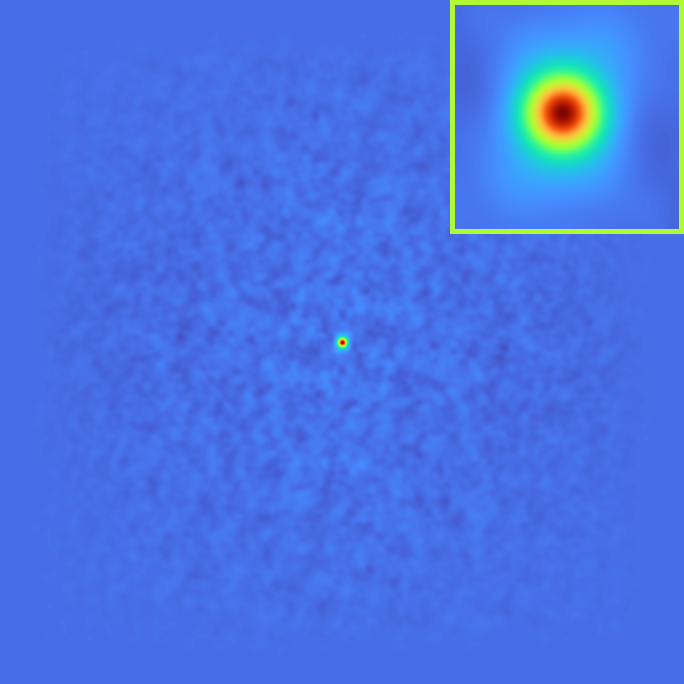}
    \includegraphics[height=3cm]{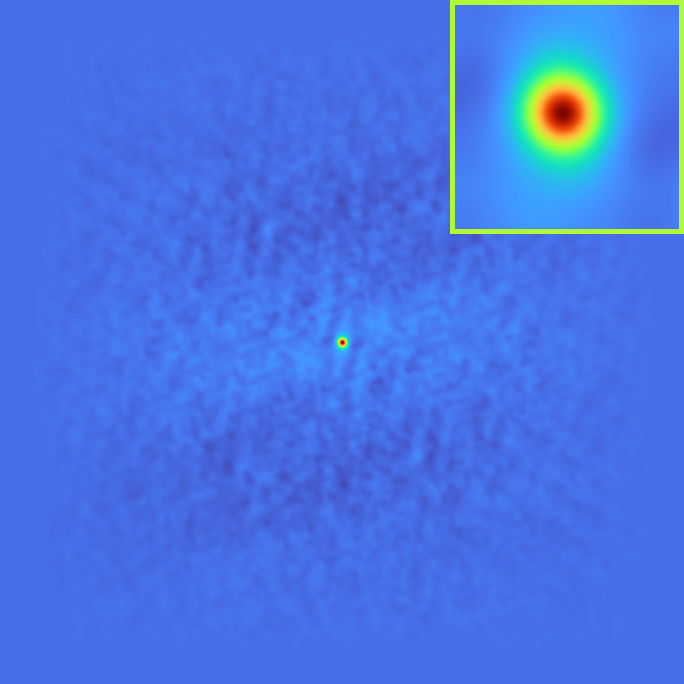}
    \includegraphics[height=3cm]{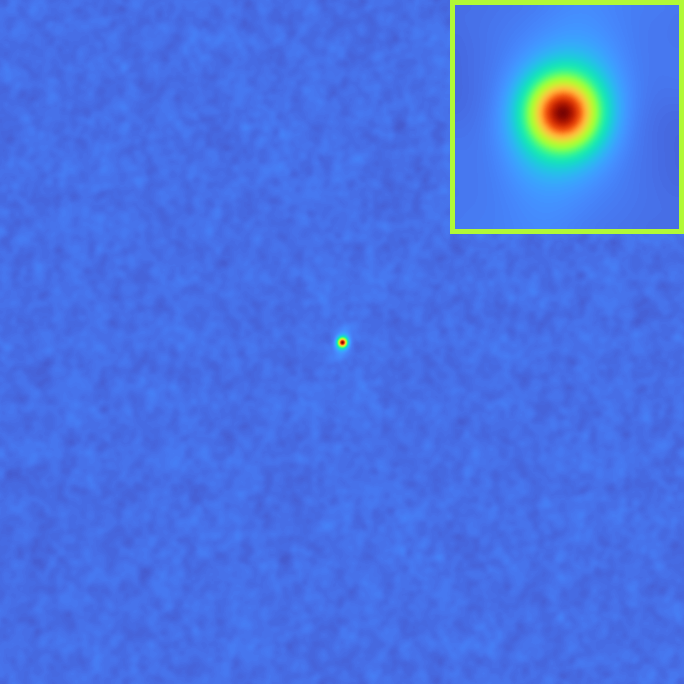}
    \includegraphics[height=3cm]{images/sand/SandMethodsColorbarAutocorr.png}
    \hspace{5.5pt}
    \caption{Comparison between methods to obtain texture images larger than the input image. Height images in the middle row: reference image (left), two realizations of the RPN (center) and a texture image generated using EF-stitching (right). EF-stitching is applied for patches of size $256\times 256$ and overlap width $128$. Input images for the RPN in the top row: extensions of different normalized cut-outs of size $342\times 342$ of the reference image. Corresponding auto-correlations with zoom-in of the central part in the bottom row. Images are of size $684\times 684$ and have pixel spacing $5.25\,\mu\text{m}$ such that they cover an area of size $3.6\,\text{mm}\times3.6\,\text{mm}$.}
    \label{fig:sand:scaling}
\end{figure*}

When attempting to increase the output image size compared to the input, the adjustment becomes more difficult. 
Under certain conditions on $M_\mathcal{T}$ and $N_\mathcal{T}$, HB and PS can produce output images that are larger than the input image \cite{Vacher2021PortillaSimoncelli,Vacher2014HeegerBergen}. 
When using ADSN and RPN, the input image is extended to match the desired output size \cite{Galerne2011AdsnRpnTheoretical}.
Therefore, pad $\nicefrac{\sqrt{M_\mathcal{T}N_\mathcal{T}}}{\sqrt{M_\mathcal{I}N_\mathcal{I}}}\left(\mathcal{I}-\hat{\mu}_\mathcal{I}\right)+\hat{\mu}_\mathcal{I}$ with $\hat{\mu}_\mathcal{I}$ to obtain an image of size $M_\mathcal{T}\times N_\mathcal{T}$ having the same mean and variance as $\mathcal{I}$, see Figure \ref{fig:sand:scaling} top row, and use that as input image. 
The border of the texture image in the central part is multiplied with a sigmoid function to obtain a smooth transition to the padding.

Using an extended input image may result in artifacts. 
For example, the auto-correlation of a synthetically generated texture image using a patch of the given measurement as input (Figure \ref{fig:sand:scaling} center right) indicates an anisotropic behavior along the $x$-axis which is not observed in the reference measurement (Figure \ref{fig:sand:scaling} left).
Especially when the input image is much smaller than the desired output ($M_\mathcal{I},N_\mathcal{I}\ll M_\mathcal{T},N_\mathcal{T}$), the resulting pattern looses randomness since structures are repeated periodically.

This effect is avoided when simulating smaller texture patches first which are then stitched together (EF-stitching).
The procedure is explained in Section \ref{sec:sand:SpacingScaling:EFstitching} and is considered the preferred method for generating larger texture images.

In summary, the procedure to generate texture images of arbitrary size and pixel spacing is as follows:
\begin{enumerate}
    \item \textbf{Adaption of pixel spacing:} Down-sample the input image by a factor $\nicefrac{\nu_\mathcal{T}}{\nu_\mathcal{M}}$.
    \item \textbf{Adaption of image size:} If the down-sampled input image has at least the size of $\mathcal{T}$, crop the image and apply a texture generation method.
    Otherwise, take patches of the down-sampled input image at random positions as input for a texture generation method and stitch those simulations using EF-stitching. 
    Note that the run-time of the texture generation method depends on the chosen patch size, see Figure \ref{fig:sand:methodsComp:runtime}.
\end{enumerate}

\subsubsection{Stitching method by Efros and Freeman}
\label{sec:sand:SpacingScaling:EFstitching} 

EF-stitching \cite{Efros2001EFOld} is a non-parametric phenomenological texture generation method based on stitching together small patches taken from the input image. 
It is an extension of the method of Efros and Leung \cite{Efros1999EfrosLeung}, in which a new image is synthesized pixel by pixel.
Raad and Galerne \cite{Raad2017EfrosFreeman} have taken up Efros and Freeman's method and published an in-depth description of it. 
Additionally, they describe a possible partial parallelization of the computation.

When applying EF-stitching, patches taken from the input image are placed equidistantly in the output image, taking into account an overlap region.
Within this region, a cutting edge for stitching is determined as a path where pixel values in the images are most similar.
To smooth the transition of both patches, the mean filter of size $3\times3$ is applied to the seam pixels.
The method inserts new patches in raster scan order (from top left to bottom right).
We use a modification of that method since newly generated patches are taken instead of cut-outs of the input image.

\begin{figure*}
	\centering
    \begin{subfigure}[t]{0.19\columnwidth}
        \centering
        \includegraphics[width=\columnwidth]{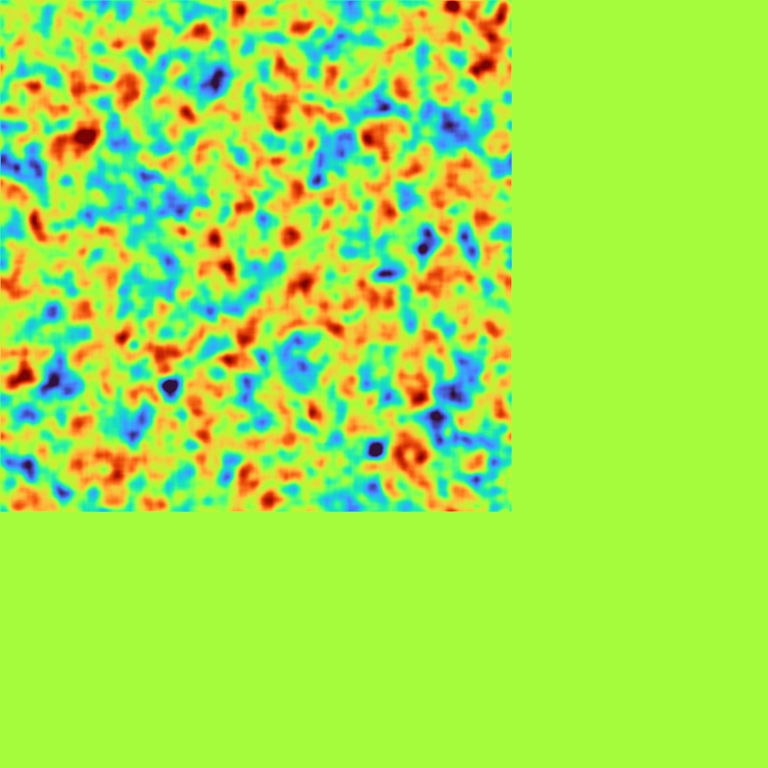}
        \caption*{Start.}
    \end{subfigure}
    \begin{subfigure}[t]{0.19\columnwidth}
        \centering
        \captionsetup{justification=centering}
        \includegraphics[width=\columnwidth]{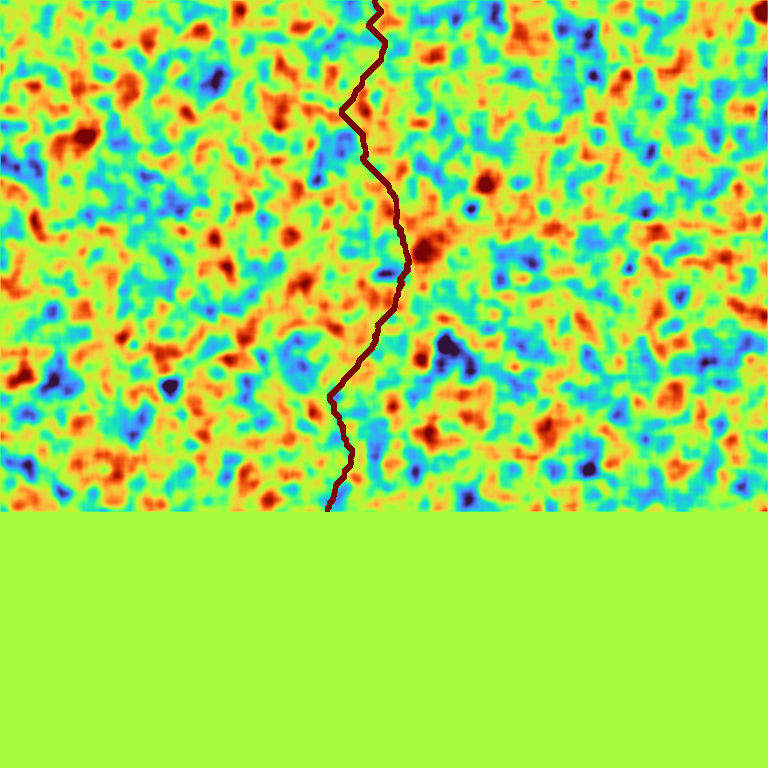}
        \caption*{Horizontal\\alignment.}
    \end{subfigure}
    \begin{subfigure}[t]{0.19\columnwidth}
        \centering
        \captionsetup{justification=centering}
        \includegraphics[width=\columnwidth]{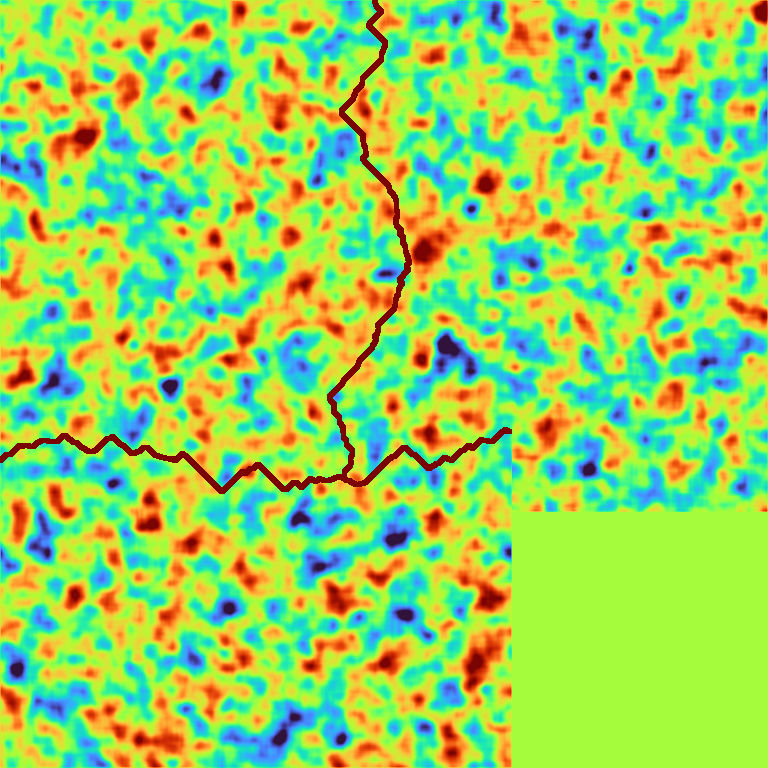}
        \caption*{Vertical\\alignment.}
    \end{subfigure}
    \begin{subfigure}[t]{0.19\columnwidth}
        \centering
        \captionsetup{justification=centering}
        \includegraphics[width=\columnwidth]{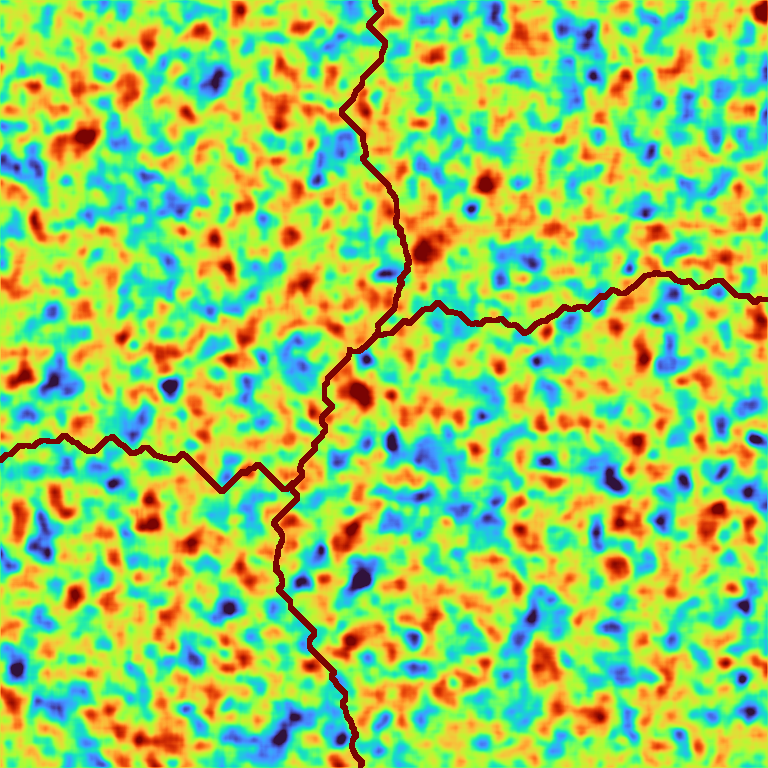}
        \caption*{L-shaped\\alignment.}
    \end{subfigure}
    \begin{subfigure}[t]{0.19\columnwidth}
        \centering
        \captionsetup{justification=centering}
        \includegraphics[width=\columnwidth]{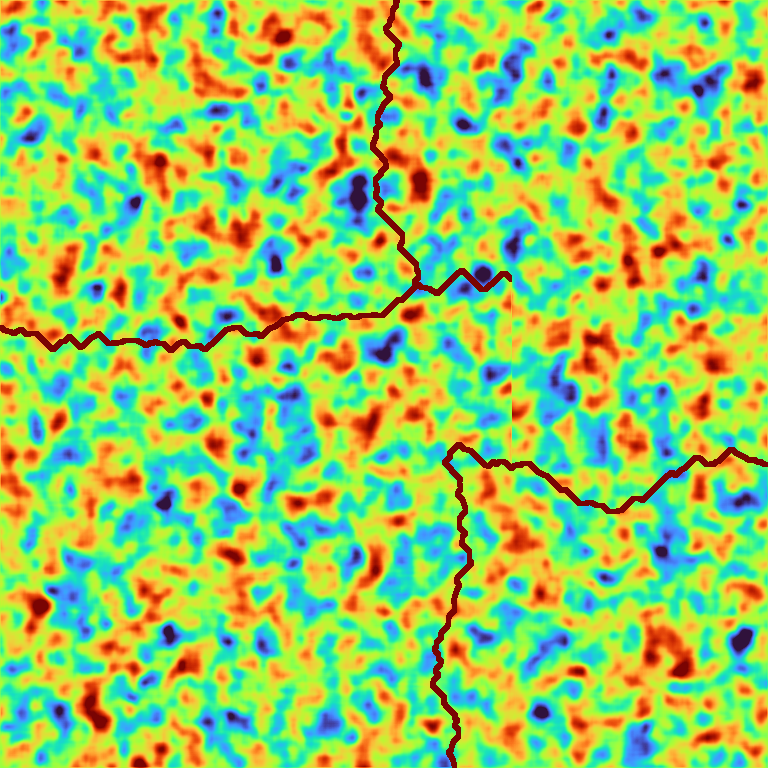}
        \caption*{Disconnected\\paths.}
    \end{subfigure}
	\caption{Illustration of original EF-stitching \cite{Efros2001EFOld,Raad2017EfrosFreeman} using patches of size $512\times 512$ and overlap width $\omicron=256$. The minimal paths are marked.}
	\label{fig:sand:EFstitching}
\end{figure*}

Assume quadratic patches $P_{i,j}$ of size $M_P\times M_P$ for $i,j=1,\dots,n$ and an overlap region of width $1<o<M_P$.
The resulting patchwork image $\mathcal{I}$ is of size $M_\mathcal{I}\times M_\mathcal{I}$ with $M_\mathcal{I}=(M_P-o)\cdot n+o$.
A new patch $P_{i,j}$ is inserted into $\mathcal{I}$ by updating the corresponding cut-out $\mathcal{I}_{i,j}$.
A mask $A_{i,j}\in\{0,1\}^{M_P\times M_P}$ decides whether the existing part or the new patch is taken. That is, $\mathcal{I}_{i,j}$  is updated to $A_{i,j}\cdot\mathcal{I}_{i,j} + \left(1-A_{i,j}\right)\cdot P_{i,j}$. 
Depending on the alignment of both patches, we distinguish between horizontal, vertical and L-shaped stitching, see Figure \ref{fig:sand:EFstitching}.
The corresponding masks are defined as follows
\begin{equation*}
	A_{i,j}=
	\begin{cases*}
		\textbf{0} &if $i=1$, $j=1$\\
		\text{horizontal}(P_{i,j},\mathcal{I}_{i,j}) &if $i>1$, $j=1$\\
		\text{vertical}(P_{i,j},\mathcal{I}_{i,j}) &if $i=1$, $j>1$\\
		\text{L-shaped}(P_{i,j},\mathcal{I}_{i,j}) &if $i>1$, $j>1$.
	\end{cases*}
\end{equation*} 

Given the position of the patch, the type of overlap region $\Omega^\bullet$, $\bullet\in\{h,v,L\}$ is determined.
For horizontal alignment, a vertically elongated rectangle serves as overlap region $\Omega^h = \{1,\dots,\omicron\}\times\{1,\dots,M_P\}$ while for vertical alignment we get $\Omega^v=\left(\Omega^h\right)^T$.
If the patch's alignment is none of both, the overlap region $\Omega^L=\Omega^h\cup\Omega^v$ is L-shaped.

To determine the mask, we compute the 8-connected path within the corresponding overlap region minimizing the sum of point-wise squared differences $e(p)=\left(\mathcal{I}_{i,j}(p)-P_{i,j}(p)\right)^2$ for $p\in\Omega^\bullet$. It separates the patches where they are most similar.

\begin{figure*}
	\centering
    \begin{subfigure}[b]{0.19\columnwidth}
        \centering
        \captionsetup{justification=centering}
        \includegraphics[width=1.5\columnwidth,angle=-90]{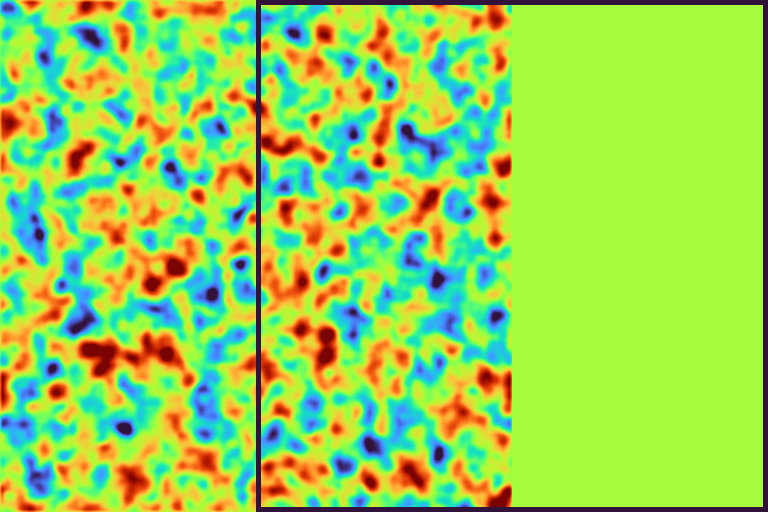}
        \caption*{Position of $\mathcal{I}_{1,2}$\\within $\mathcal{I}$.}
    \end{subfigure}
    \begin{subfigure}[b]{0.19\columnwidth}
        \centering
        \includegraphics[width=\columnwidth,angle=-90]{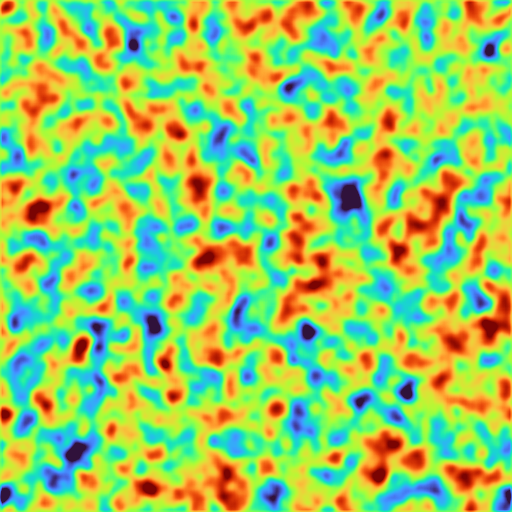}
        \caption*{$P_{1,2}$.\\ \phantom{place}}
    \end{subfigure}
    \begin{subfigure}[b]{0.19\columnwidth}
        \centering
        \includegraphics[width=0.5\columnwidth,angle=-90]{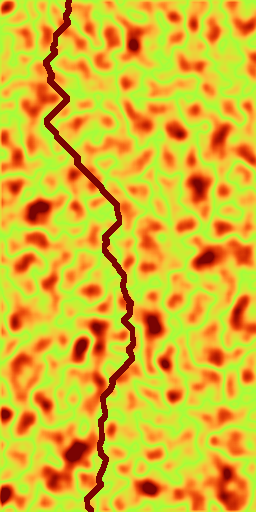}
        \vspace{0.485\columnwidth}
        \caption*{$e\left(\Omega^v\right)$ with $\zeta^h$.\\ \phantom{place}}
    \end{subfigure}
    \begin{subfigure}[b]{0.19\columnwidth}
        \centering
       \includegraphics[width=\columnwidth,angle=-90]{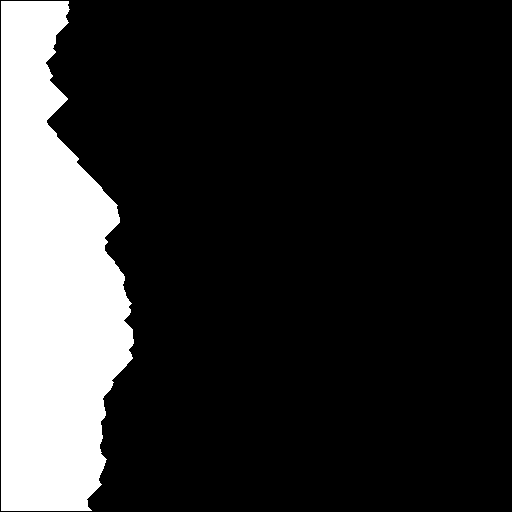}
        \caption*{$A_{1,2}$.\\ \phantom{place}}
    \end{subfigure}
    \begin{subfigure}[b]{0.19\columnwidth}
        \centering
        \captionsetup{justification=centering}
        \includegraphics[width=1.5\columnwidth,angle=-90]{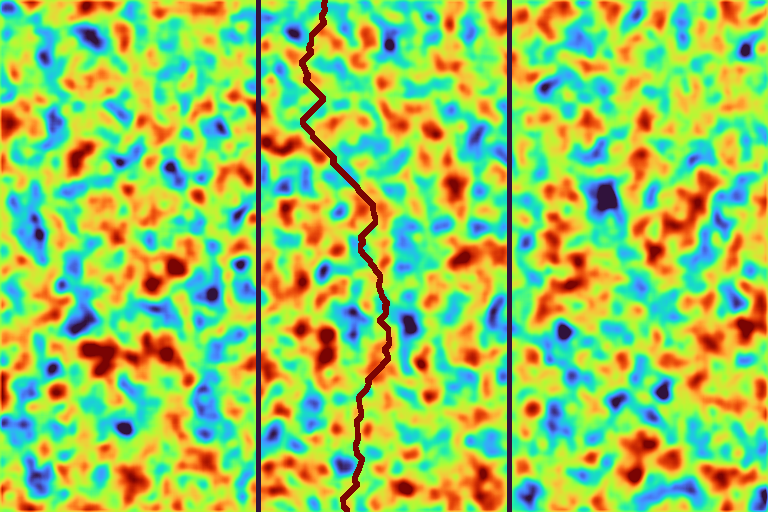}
        \caption*{Stitching based\\on $\zeta^h$.}
    \end{subfigure}
    \caption{Illustration of EF-stitching for vertical alignment using patches of size $512\times 512$ and overlap width $\omicron=256$.}
	\label{fig:sand:EF:vertical}
\end{figure*}

In case of vertical alignment, the overlap region is divided by a horizontal path $\zeta^h=\left(\zeta^h_1,\dots,\zeta^h_{M_P}\right)$.
The path is oriented from right to left such that for successive path elements $\left(\zeta^h_k\right)_1-\left(\zeta^h_{k-1}\right)_1=-1$ for $k=2,\dots,M_P$. 
The mask is then computed by 
\begin{equation*}
    \text{vertical}(P_{i,j},\mathcal{I}_{i,j})(p) = 
	\begin{cases*}
		0 &if $p_2\geq \left(\zeta_{p_1}^h\right)_2$\\
		1 &else 
	\end{cases*}
\end{equation*}
for $p\in\left\{1,\dots,M_P\right\}^2$. 
Figure \ref{fig:sand:EF:vertical} illustrates the stitching procedure for vertical alignment. The formulation for horizontal alignment is analogous.

\begin{figure*}
	\centering
    \begin{subfigure}[b]{0.24\columnwidth}
        \centering
        \captionsetup{justification=centering}
        \includegraphics[width=\columnwidth]{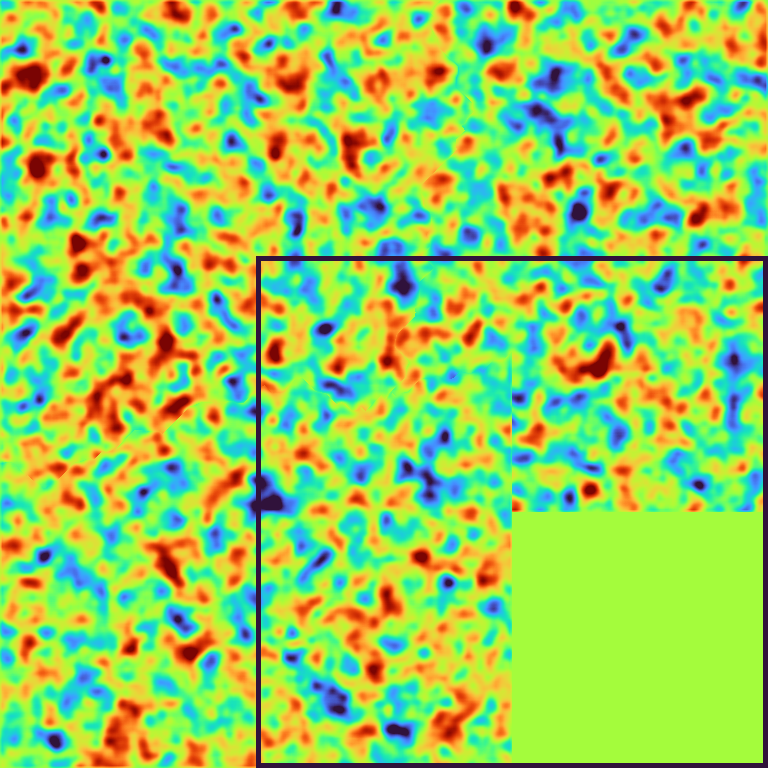}
        \caption*{Position of $\mathcal{I}_{2,2}$\\within $\mathcal{I}$.}
    \end{subfigure}
    \begin{subfigure}[b]{0.16\columnwidth}
        \centering
        \captionsetup{justification=centering}
        \includegraphics[width=\columnwidth]{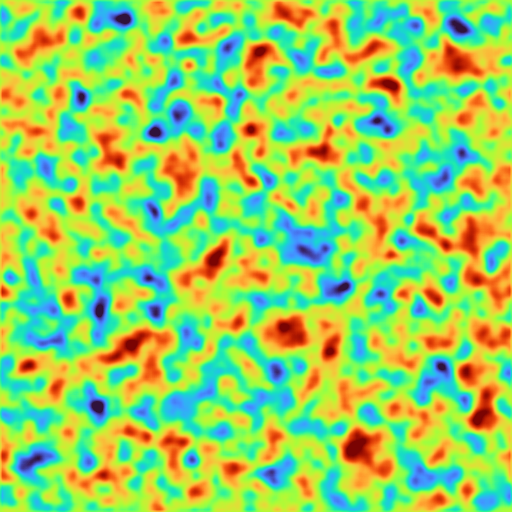}
        \caption*{$P_{2,2}$.\\ \phantom{space}}
    \end{subfigure}
    \begin{subfigure}[b]{0.16\columnwidth}
        \centering
        \captionsetup{justification=centering}
        \includegraphics[width=\columnwidth]{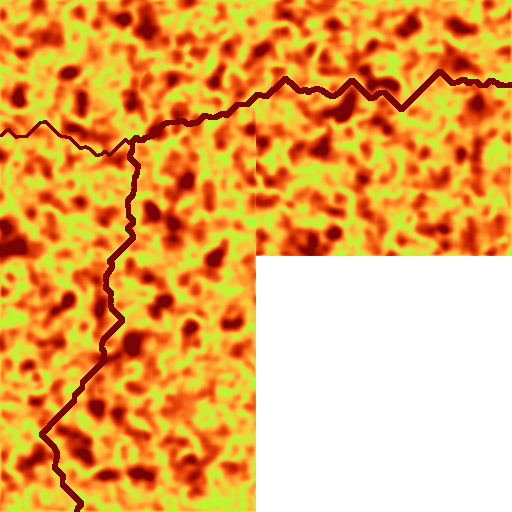}
        \caption*{$e\left(\Omega^L\right)$ with $\zeta^L$ and the old path.}
    \end{subfigure}
    \begin{subfigure}[b]{0.16\columnwidth}
        \centering
        \captionsetup{justification=centering}
        \includegraphics[width=\columnwidth]{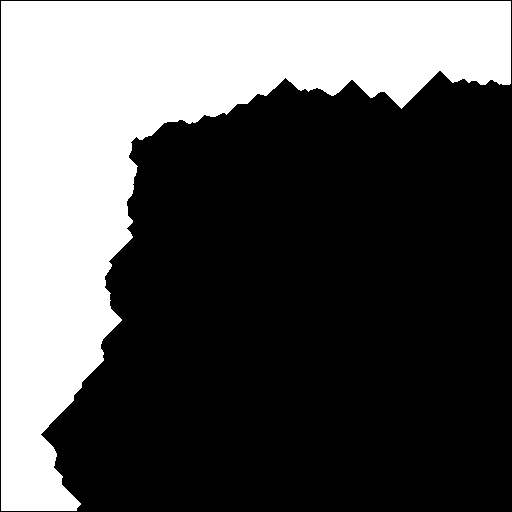}
        \caption*{$A_{2,2}$.\\ \phantom{space}}
    \end{subfigure}
    \begin{subfigure}[b]{0.24\columnwidth}
        \centering
        \captionsetup{justification=centering}
        \includegraphics[width=\columnwidth]{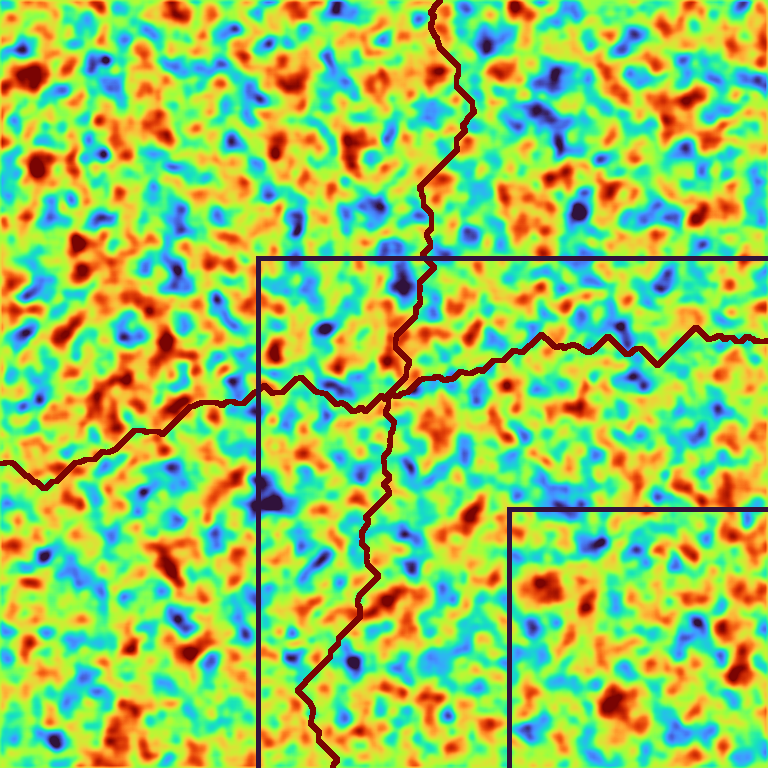}
        \caption*{Stitching based\\on $\zeta^L$.}
    \end{subfigure}
	\caption{Illustration of adapted EF-stitching for L-shaped alignment using patches of size $512\times 512$ and overlap width $\omicron=256$.}
	\label{fig:sand:EF:Lshape}
\end{figure*}

An L-shaped path $\zeta^L$ is a combination of a horizontal and a vertical path connected at a point located in $\{1\dots\omicron\}^2$.
Simply searching for the minimal L-shaped path in the overlap region as described in \cite{Raad2017EfrosFreeman} may result in a path being disconnected to the paths constructed in the previous stitching steps, see Figure~\ref{fig:sand:EFstitching}. This causes discontinuities in the stitched image. 
To avoid this problem, we elongate the existing horizontal path $\zeta^{L,h}$ from the previous stitching step optimally and find the minimal vertical path $\zeta^{L,v}$ connected to that elongation.
The mask for L-shaped overlap is defined by
\begin{equation*}
	\text{L-shaped}\left(\mathcal{P}_{i,j},\mathcal{I}_{i,j}\right)(p)=
	\begin{cases*}
		0 &if $p_1\geq\left(\zeta^{L,v}_{p_2}\right)_1$ and $p_2\geq\left(\zeta^{L,h}_{p_1}\right)_2$\\
		1 &else
	\end{cases*}
\end{equation*}
for $p\in\left\{1,\dots,M_P\right\}^2$. 
Figure \ref{fig:sand:EF:Lshape} illustrates the adapted EF-stitching for L-shaped alignment.

\section{Model for milled surfaces}
\label{sec:milling}

Milling produces significantly different textures than sandblasting, compare the measurements given in Figure \ref{fig:data:heightimage} and Figure \ref{fig:sand:measurements}.
In particular, face-milled surfaces are characterized by a non-stationary pattern of periodic ring-shaped indentations.
These structures cannot be reproduced by the methods introduced in Section \ref{sec:sand:methods} since they are  only suitable for stationary textures with short-range correlations \cite{Galerne2011AdsnRpnTheoretical,Raad2017Overview,Vacher2021PortillaSimoncelli}. 
Thus, a separate model to generate textures of milled surfaces is needed.

\begin{figure}
	\centering
	\includegraphics[height=2cm]{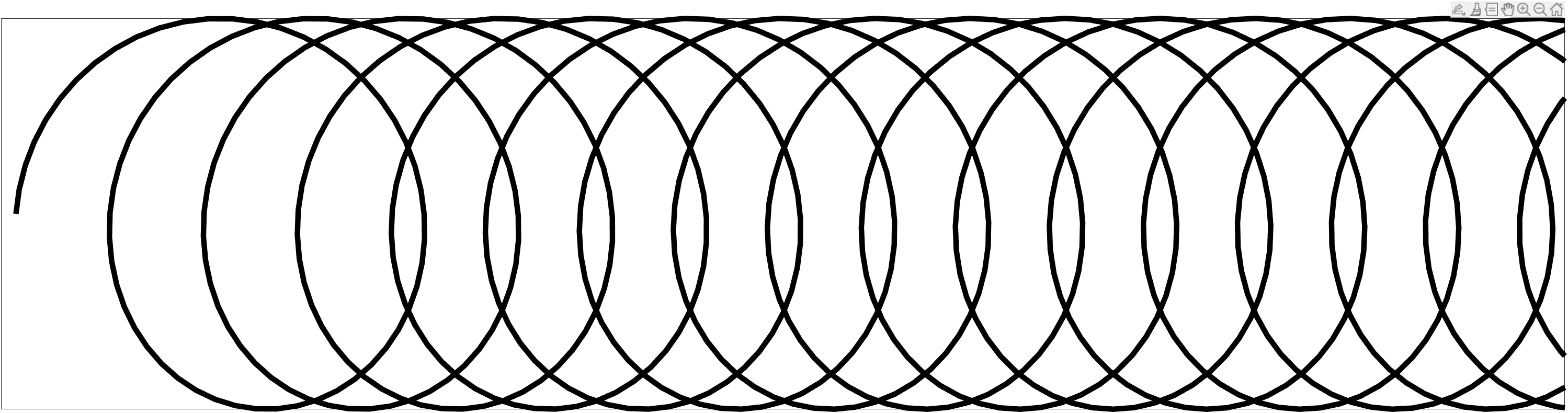}
	\hspace{10pt}
	\includegraphics[height=2cm]{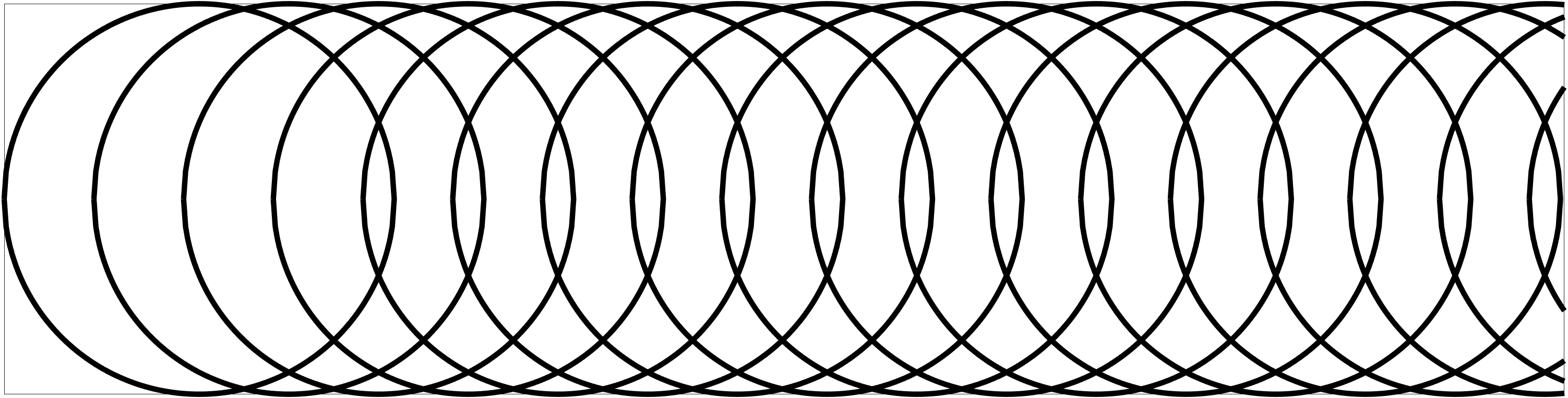}
	\caption{Path of one cutting edge (left) and its approximation with rings (right).}
	\label{fig:mill:edgepath}
\end{figure}

The characteristic ring-shaped indentations are formed by the milling tool when following its prescribed path along the material surface. 
During this motion, the cutting edge of the milling head follows a cycloidal path. 
In our model, this path will be simplified to a pattern of successive rings, see Figure \ref{fig:mill:edgepath}.

The milling model is introduced as parameterized function $f:\mathbb{R}^2\rightarrow \mathbb{R}$ on a continuous domain and is thus procedural.
It is then evaluated at given image grid points to obtain the texture image $\mathcal{T}$.
A model containing $n\in\mathbb{N}$ rings is given by $f\left(x\right) = f_n\left(R_1,\dots,R_n\right)(x)$.
The modeling procedure consists of three steps describing the appearance, interaction, and location of the rings as follows:
\begin{enumerate}
	\item the appearance of an individual ring is described by the function $R_k:\mathbb{R}^2\rightarrow\mathbb{R}$,
 	\item the interaction between $n$ rings is described by the mapping $f_n:\left(\mathbb{R}^2\rightarrow\mathbb{R}\right)^n\rightarrow\left(\mathbb{R}^2\rightarrow\mathbb{R}\right):\left(R_1, \ldots, R_n\right)\mapsto f_n\left(R_1, \ldots, R_n\right)$,
  	\item the tool-path providing the position and order of the rings by defining their center points $c_k\in\mathbb{R}^2$.
\end{enumerate}

These sub-models are explained in more detail in Sections \ref{sec:mill:ring}, \ref{sec:mill:inter} and \ref{sec:mill:path} below.
The milling process is controlled by several parameters both for the milling head, e.g. the head diameter and the blade length, and for the tool-path such as the feed rate.
The known real-world parameters are integrated into the model and the measurements are used to estimate the remaining unknown parameters.
The adaption of the model to given data is described in Section \ref{sec:mill:fit}.

\subsection{Sub-model for ring appearance}
\label{sec:mill:ring}

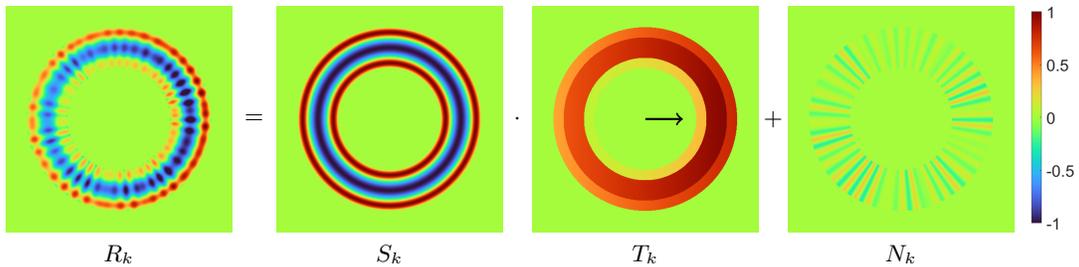
\begin{figure*}[b]
	\centering
	\begin{tikzpicture}
		\node at (0,1) {\includegraphics[height=3cm]{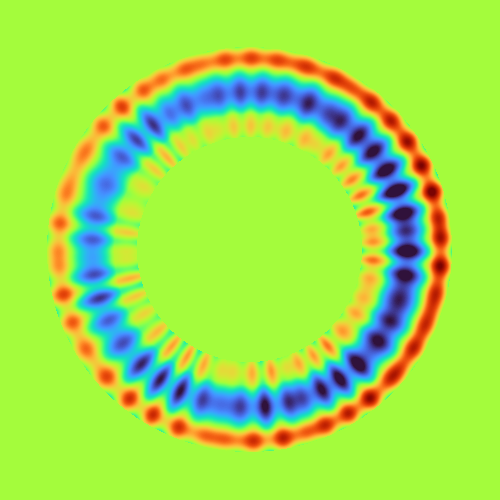}};
		\node at (3.6,1) {\includegraphics[height=3cm]{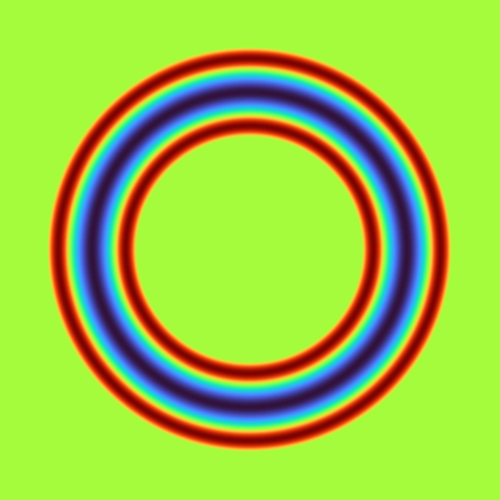}};
		\node at (7,1) {\includegraphics[height=3cm]{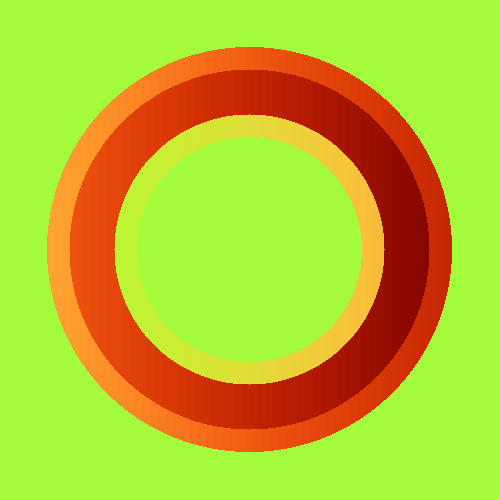}};
		\node at (10.4,1) {\includegraphics[height=3cm]{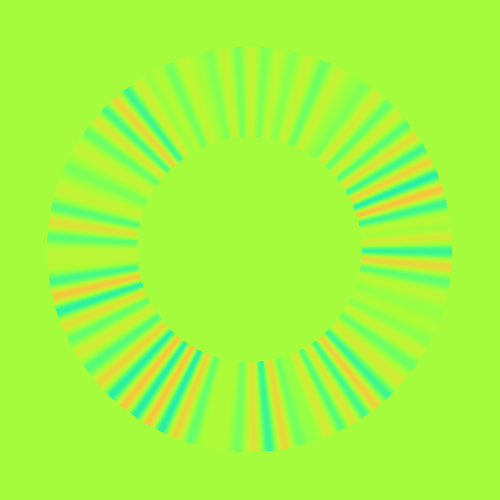}};
		\node at (12.4,1) {\includegraphics[height=3cm]{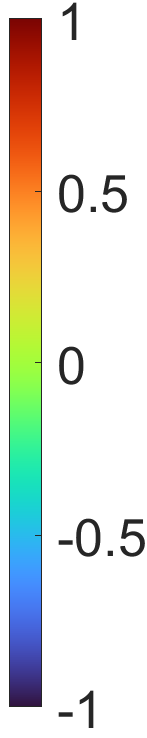}};

        \draw[->,thick] (7,1) -- (7.5,1);
		
		\node at (1.8,1) {\small$=$};
		\node at (5.3,1) {\small$\cdot$};
		\node at (8.7,1) {\small$+$};
		
		\node at (0,-0.8) {\small$R_k$};
		\node at (3.6,-0.8) {\small$S_k$};
		\node at (7,-0.8) {\small$T_k$};
		\node at (10.4,-0.8) {\small$N_k$};
	\end{tikzpicture}
\caption{Illustration of the composition of the sub-model for the ring appearance.}
\label{fig:mill:ring}
\end{figure*}

A ring $R_k:\mathbb{R}^2\rightarrow\mathbb{R}$ approximates the cut of one cutting edge during a complete rotation of the milling head. 
The shapes of indentation resulting from the cutting process and of possible material deposition are captured by the shape function $S_k:\mathbb{R}^2\rightarrow[-1,1]$.
In practice, the milling tool is often tilted forwards in direction of the tool-path to prevent re-cutting \cite{Hadad2016FaceMillingImages,2018KlockeMilling}. 
This circumstance is modeled by the tilting function $T_k:\mathbb{R}^2\rightarrow[-1,1]$ which is multiplied with the shape.
Moreover, irregularities can occur during the milling process due to vibrations or resistances in the material. These are added as noise $N_k:\mathbb{R}^2\rightarrow[-1,1]$ to the tilted ring.
Thus, the sub-model for a ring reads $R_k = S_k\cdot T_k+N_k$, see Figure \ref{fig:mill:ring}.

\subsubsection{Shape}
\begin{figure*}
	\centering
    \begin{subfigure}[t]{0.2\textwidth}
        \centering
        \begin{tikzpicture}
    		\node at (0,0) {\includegraphics[height=3cm]{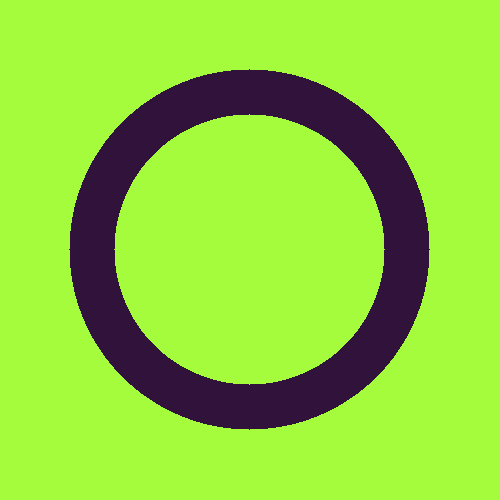}};
    		\draw[line width=1pt,color=red,-|] (0,0) -- (-1.08,0) node[pos=.5,anchor=south]{\small $r$};
    		\node at (0,0) {\small $\bullet$};	
    		\node at (0,-0.25) {\small $c_k$};		
    	\end{tikzpicture}
        \includegraphics[height=2.7cm]{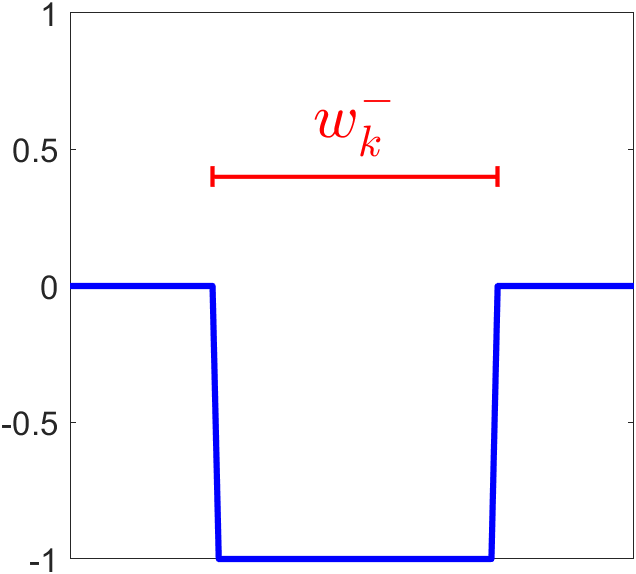}
        \caption*{$S_k^\text{Indicator}$}
    \end{subfigure}
    \begin{subfigure}[t]{0.2\columnwidth}
        \centering
        \begin{tikzpicture}
    		\node at (0,0) {\includegraphics[height=3cm]{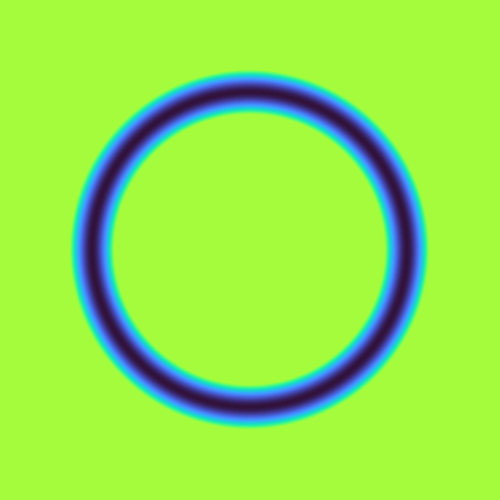}};	
    		\draw[line width=1pt,color=red,-|] (0,0) -- (-1.08,0) node[pos=.5,anchor=south]{\small $r$};
    		  \node at (0,0) {\small $\bullet$};	
    		\node at (0,-0.25) {\small $c_k$};			
    	\end{tikzpicture}
        \includegraphics[height=2.7cm]{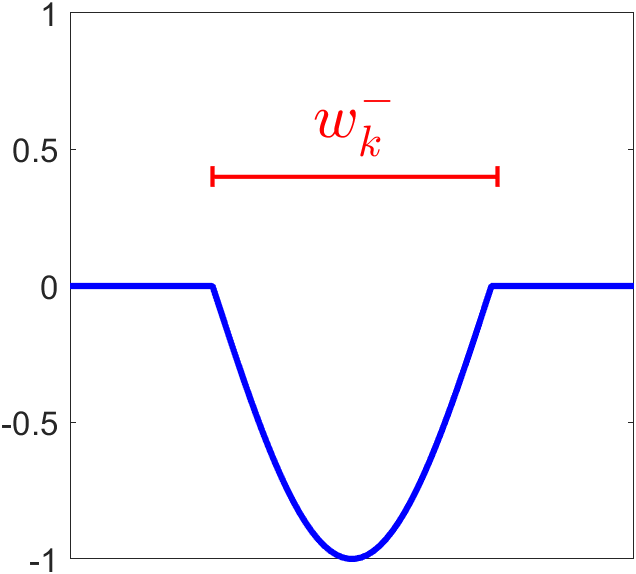}
        \caption*{$S_k^\text{Cosine}$}
    \end{subfigure}
    \begin{subfigure}[t]{0.2\columnwidth}
        \centering
        \begin{tikzpicture}
    		\node at (0,0) {\includegraphics[height=3cm]{images/mill/MillRingCosineShape.png}};
    		\draw[line width=1pt,color=red,-|] (0,0) -- (-1.08,0) node[pos=.5,anchor=south]{\small $r$};
    		  \node at (0,0) {\small $\bullet$};	
    		\node at (0,-0.25) {\small $c_k$};		
    	\end{tikzpicture}
        \includegraphics[height=2.7cm]{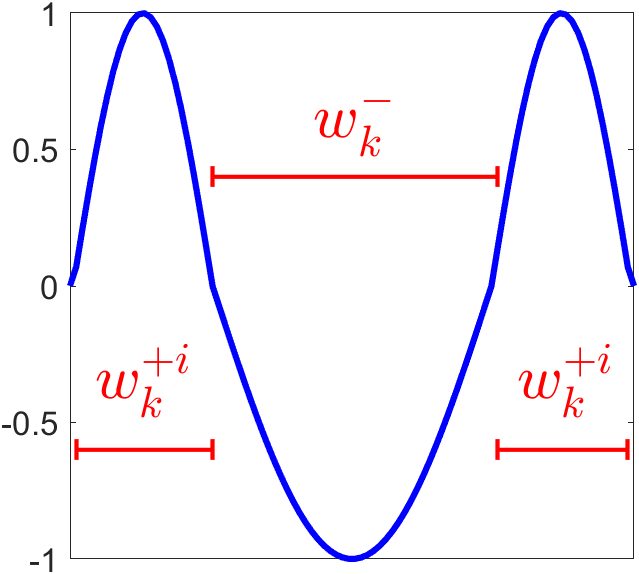}
        \caption*{$S_k^\text{Bump}$}
    \end{subfigure}
    \caption{Illustration of shape functions and their parameters: height images (top) and 1d-intersections thereof (bottom).}
	\label{fig:mill:shape}
\end{figure*}

To define a ring, the following spatial components are needed: the center point $c_k\in\mathbb{R}^2$, the radius $r\in\mathbb{R}_{>0}$ and the width $w^-_k$.
The location of the center point depends on the tool-path, described in Subsection \ref{sec:mill:path}.
The radius is directly given by the diameter $d$ of the milling tool as $r=\nicefrac{d}{2}$. 
The width is modeled by a random variable $w^-_k \sim \mathcal{N}\left(\mu_{w^-},\sigma_{w^-}\right)$ whose mean $\mu_{w^-}\in\left(0,r\right)$ is given by the length of the cutting edge.
The set of points belonging to the ring is denoted by $P_k^-=\left\{x\in\mathbb{R}^2:r-w_k^-\leq \Vert x-c_k \Vert_2\leq r \right\}$.
In the following, we define several indentation shapes which are illustrated in Figure \ref{fig:mill:shape}.

\begin{enumerate}
	\item \textbf{Indicator function}\\
	This shape function is constant on the set of points lying in the ring. Thus, there are no height differences within the ring.
		{\makeatletter
		\@fleqntrue
		\makeatother
		\begin{alignat*}{1}
			&S_k^\text{Indicator}(x) = 
			\begin{cases*}
				-1&if $x\in P_k^-$\\
				0 &else
			\end{cases*}
		\end{alignat*}}
	
	\item \textbf{Cosine function}\\
	The indicator shape function is extended to have a smooth transition and to resemble the shape of the cutting edge. 
    The function is modeled using the cosine function and is defined based on the normalized distance $$D_k^-:\mathbb{R}^2\rightarrow[-1,1]:x\mapsto\left(\Vert x-c_k \Vert_2-r+\frac{w^-_k}{2}\right)\cdot\frac{2}{w_k^-}\cdot\mathbb{1}_{P_k^-}(x)$$ between points in the ring and the ring's central circle.
    The resulting shape function is
		{\makeatletter
		\@fleqntrue
		\makeatother
		\begin{alignat*}{1}
			&S_k^\text{Cosine}(x) = 
			\begin{cases*}
				-\cos\left(\frac{\pi}{2}\cdot D_k^-(x)\right)&if $x\in P_k^-$\\
				0 &else.
			\end{cases*}
		\end{alignat*}}

	\item \textbf{Bump function}\\
    The cosine shape function is extended to include accumulations outside the indentation caused by material deposition.
    These are modeled by positive parts of a cosine function on two rings outside and inside $P_k^-$. 
    The inner and outer ring have widths $w_k^{+i}\sim\mathcal{N}(\mu_{w^{+i}},\sigma_{w^{+i}})$ and $w_k^{+o}\sim\mathcal{N}(\mu_{w^{+o}},\sigma_{w^{+o}})$.
    Denote the set of points lying in the inner and outer ring by 
    \begin{align*}
        P_k^{+i}&=\left\{x\in\mathbb{R}^2:r-w_k^--w_k^{+i}\leq \Vert x-c_k \Vert_2\leq r-w_k^-\right\}\\
        P_k^{+o}&=\left\{x\in\mathbb{R}^2:r\leq \Vert x-c_k \Vert_2\leq r+w_k^{+o}\right\}.
    \end{align*}
    The distance functions $D_k^{+i}(x)$ and $D_k^{+o}(x)$ are defined in analogy to $D_k^-(x)$. The complete shape function then reads
        {\makeatletter
		\@fleqntrue
		\makeatother
		\begin{alignat*}{1}
			&S_k^\text{Bump}(x) = 
			\begin{cases*}
				\cos\left(\frac{\pi}{2}\cdot D_k^{+i}(x)\right)&if $x\in P_k^{+i}$\\
				-\cos\left(\frac{\pi}{2}\cdot D_k^-(x)\right)&if $x\in P_k^-$\\
				\cos\left(\frac{\pi}{2}\cdot D_k^{+o}(x)\right)&if $x\in P_k^{+o}$\\
				0 &else.
			\end{cases*}
		\end{alignat*}}
\end{enumerate}

\subsubsection{Tilting}
So far, milling shapes were assumed to be aligned to the milled surface. Tilting of the milling head introduces a depth gradient within the rings with increasing depth towards the front, i.e., in motion direction of the milling tool. For a ring $R_k$, this direction is given by the angle $\theta_k\in(-\pi,\pi]$ between the planar tool-path at point $c_k$ and the x-axis.
It can directly be computed by $$\theta_k=\arccos\left(\frac{\left(c_{k+1}\right)_1-\left(c_{k}\right)_1}{\Vert c_{k+1}-c_{k}\Vert_2}\right).$$

\begin{figure*}[b]
	\centering
    \begin{subfigure}[b]{0.49\columnwidth}
        \centering
        \includegraphics[height=3cm]{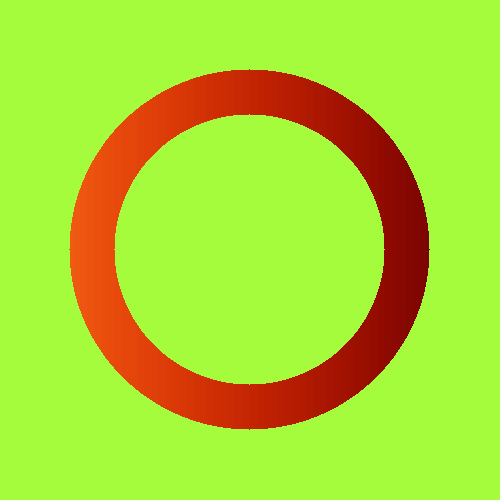}
	    \includegraphics[height=3cm]{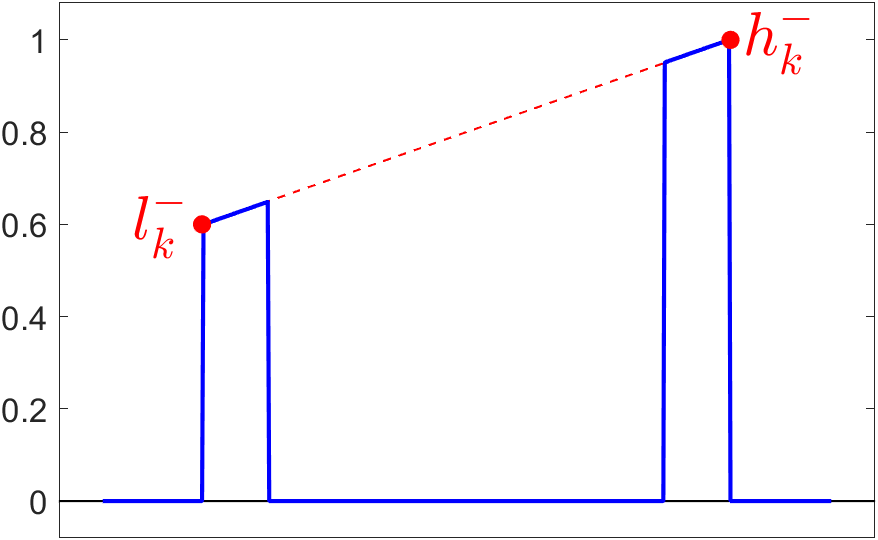}\\[4pt]
        \includegraphics[height=3cm]{images/mill/MillRingCosineWeights.png}
	    \includegraphics[height=3cm]{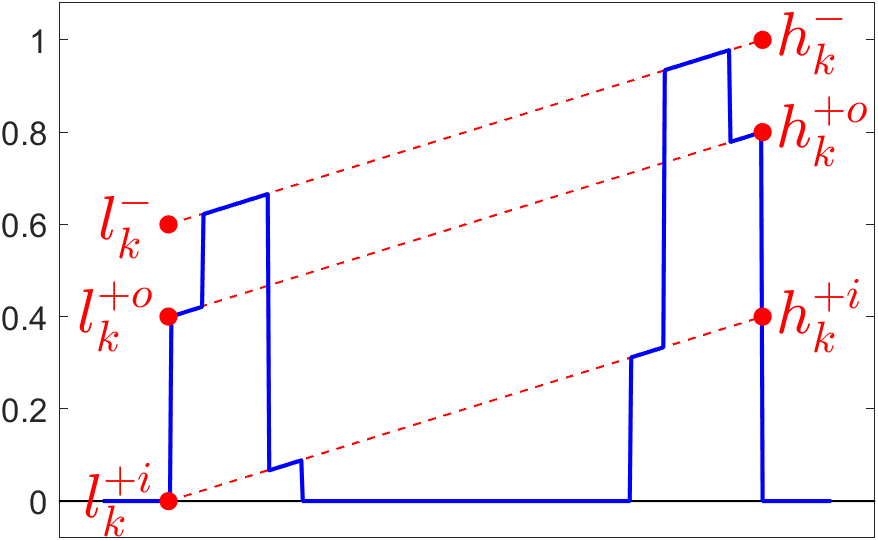}        
        \caption*{$T_k$.}
    \end{subfigure}
    \begin{subfigure}[b]{0.49\columnwidth}
        \centering
        \includegraphics[height=3cm]{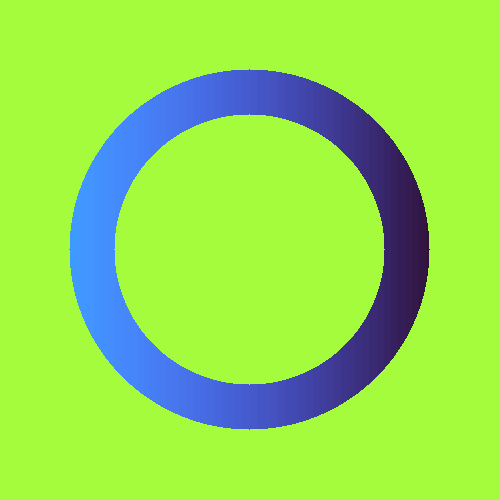}
        \includegraphics[height=3cm]{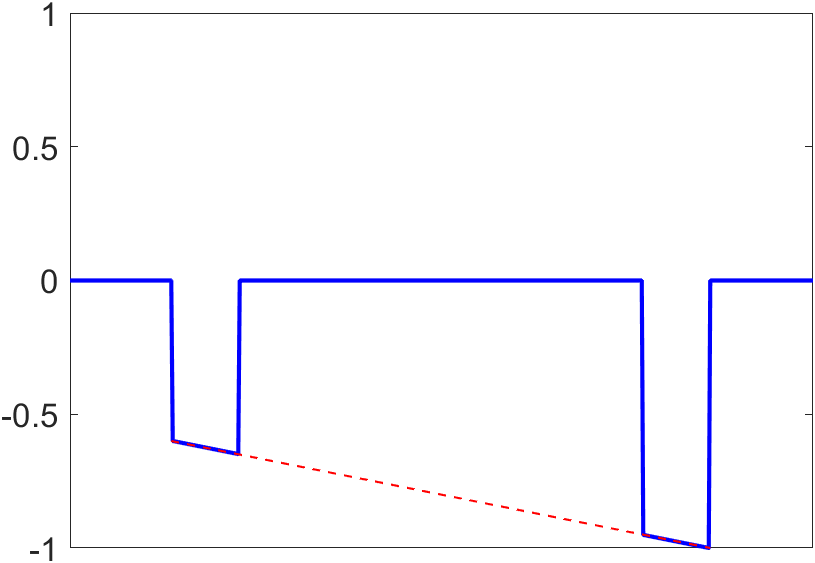}\\[4pt]
        \includegraphics[height=3cm]{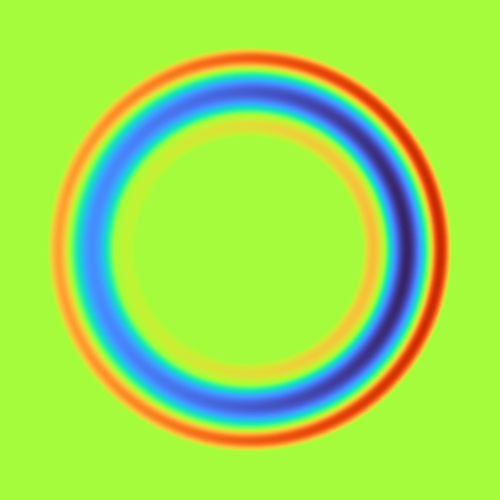}
		\includegraphics[height=3cm]{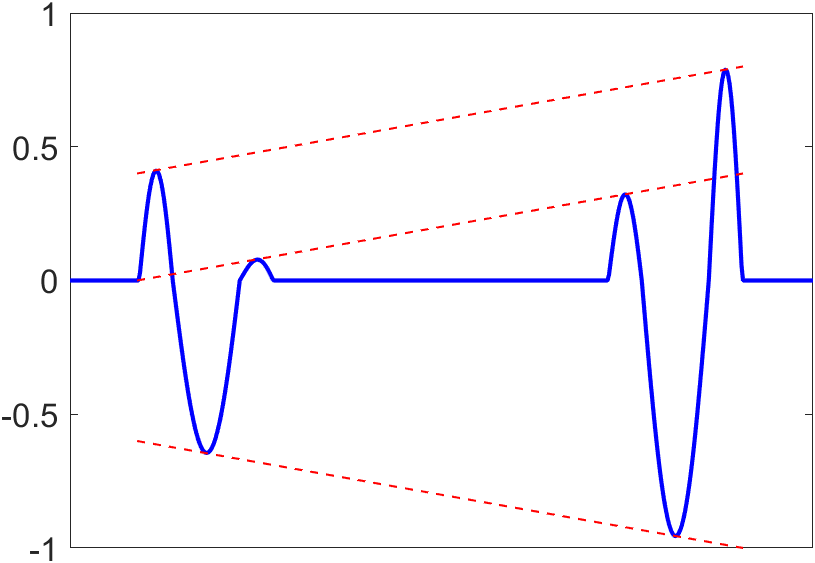}
        \caption*{$S_k\cdot T_k$.}
    \end{subfigure}
    \caption{Illustration of tilting with $\theta_k=0$, applied to indicator (top) and bump shape (bottom). Height images and their 1d-intersections.} 
	\label{fig:mill:tilting}
\end{figure*}

Tilting is introduced by location dependent scaling of the shape function $S_k$. The slope of the tilting plane is defined by the highest  and the lowest scaling factors $h^-_k\sim\mathcal{N}\left(\mu_{h^-},\sigma_{h^-}\right)$ and $l^-_k\sim\mathcal{N}\left(\mu_{l^-},\sigma_{l^-}\right)$, respectively, which are applied to the fore- and rearmost point of the indentation ring. 
The plane of contact between milling head and surface thus becomes
\begin{equation}
	\text{plane}_k^-(x)=\frac{h^-_k-l_k^-}{2r}\left(\cos(\theta_k)\cdot\left(x_1-\left(c_{k}\right)_1\right)+\sin(\theta_k)\cdot\left(x_2-\left(c_{k}\right)_2\right)\right) + \frac{h^-_k+l_k^-}{2} \label{Eq:plane}
\end{equation}
modeling the gradient of the indentation's height values, see Figure \ref{fig:mill:tilting} for an illustration.
Tilting is applied separately to indentation and accumulations by 
\begin{equation*}
    T_k(x) = \left(\text{plane}_k^-\cdot\mathbb{1}_{P_k^-} + \text{plane}_k^{+o}\cdot\mathbb{1}_{P_k^{+o}} + \text{plane}_k^{+i}\cdot\mathbb{1}_{P_k^{+i}}\right)(x).
\end{equation*}
The planes for the accumulation are defined similarly to Equation \eqref{Eq:plane} with the corresponding highest $h^\bullet_k\sim\mathcal{N}\left(\mu_{h^\bullet},\sigma_{h^\bullet}\right)$ and lowest scaling factors $l^\bullet_k\sim\mathcal{N}\left(\mu_{l^\bullet},\sigma_{l^\bullet}\right)$ for $\bullet\in\left\{+i,+o\right\}$.
To keep the model general, the highest and lowest scaling factors can be chosen independently. 
However, as real processing parameters serve the tilting angle $\varphi$ between the $x$-$y$-plane and the $z$-axis and the indentation depth $l$ in the center of the milling tool from which we obtain $\mu_{l^-}$ and $\mu_{h^-}$ by $l\pm r\cdot\sin(\varphi)$. 
None of the remaining tilting parameters can explicitly be derived from the machining parameters, but the planes are usually chosen to be parallel.
The outer accumulation needs higher scaling factors than the inner since the tool's presence prevents material from being deposited on the inside. 

\subsubsection{Noise}
\begin{figure*}
	\centering
    \begin{subfigure}[b]{0.49\columnwidth}
        \centering
        \includegraphics[height=3cm]{images/mill/MillRingCosineNoise.png}
	    \includegraphics[height=3cm]{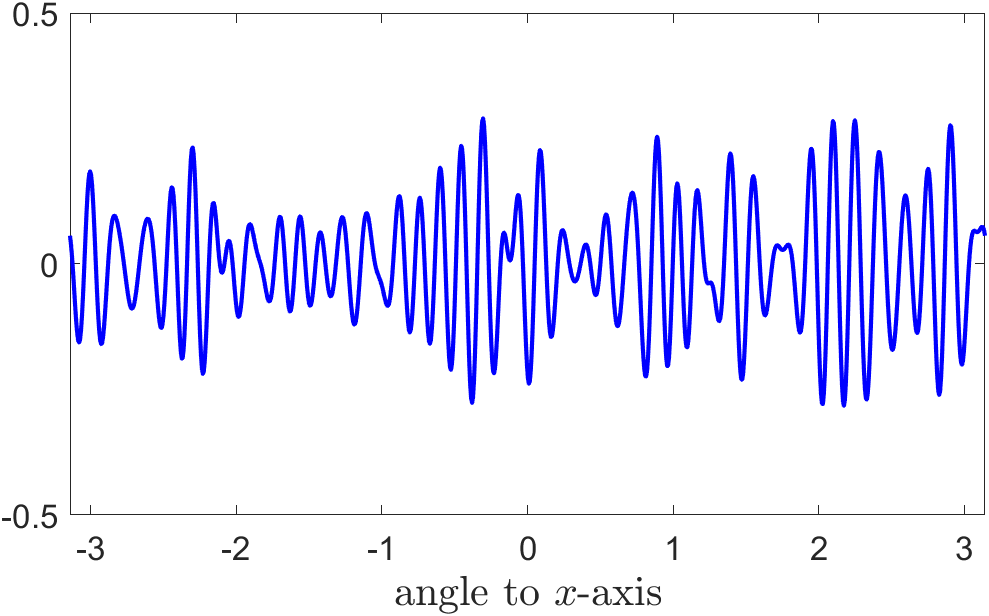}
        \caption*{$N_k$. Height image and its circular 1d-intersection}
    \end{subfigure}
    \begin{subfigure}[b]{0.21\columnwidth}
        \centering
        \includegraphics[height=3cm]{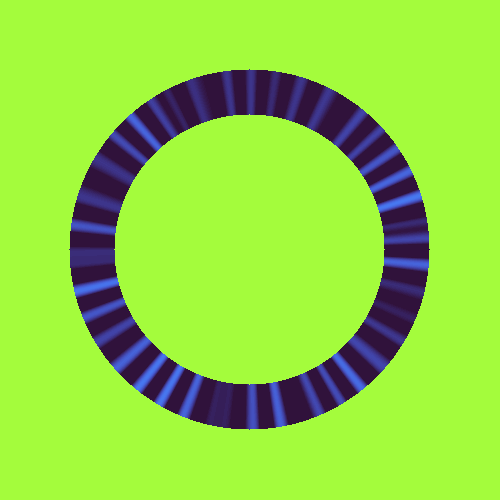}
	    \caption*{$S_k^\text{Indicator}+N_k$.}
    \end{subfigure}
    \begin{subfigure}[b]{0.19\columnwidth}
        \centering
        \includegraphics[height=3cm]{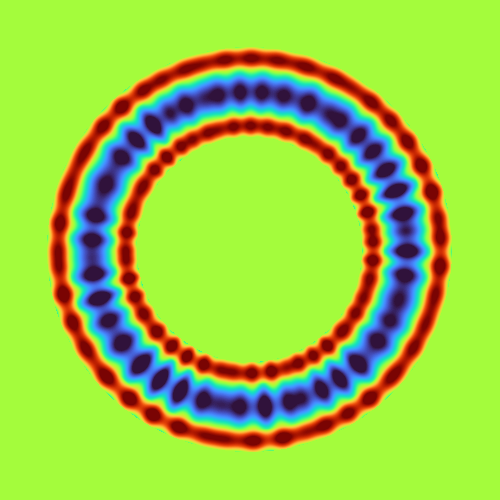}
        \caption*{$S_k^\text{Bump}+N_k$.}
    \end{subfigure}
    \caption{Illustration of noise using $\lambda=\tau=50$.}
	\label{fig:mill:noise}
\end{figure*}

The noise is modeled as a sum of $\lambda_k\sim\mathcal{P}(\lambda)$ sine curves, having frequencies $\tau_{k,j}\sim\mathcal{P}(\tau)$ and random shifts $\xi_{k,j}\sim\mathcal{U}((-\pi,\pi])$ for $j=1,\dots,\lambda_k$. 
Thus, the noise is given by
\begin{align*}
	N_k(x) = \left(\frac{1}{\lambda_k}\sum_{j=1}^{\lambda_k}\tau_{k,j}\cdot\sin\left(\arctantwo\left(x_2-\left(c_{k}\right)_2,x_1-\left(c_{k}\right)_1\right)+\xi_{k,j}\right)\right)\cdot\mathbb{1}_{P_k}(x)
\end{align*}
and illustrated in Figure \ref{fig:mill:noise}. 
Here, $P_k$ is defined as set of points to be considered, dependent on the used shape function.
For those having an indentation only, we have $P_k=P_k^-$ and those with additional accumulations we get $P_k=P_k^-\cup P_k^{+i}\cup P_k^{+o}$.

\subsection{Sub-model for ring interaction}
\label{sec:mill:inter}

\begin{figure*}
    \centering
    \begin{subfigure}[t]{0.24\columnwidth}
        \centering
        \includegraphics[width=\textwidth]{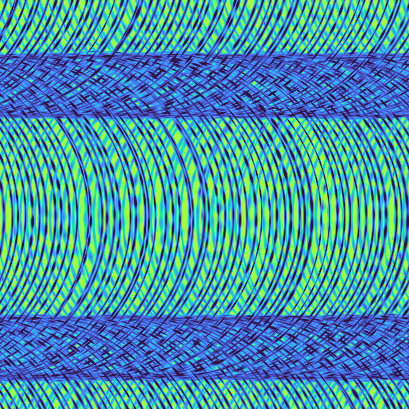}
        \caption*{$f^\text{Min}_n\left(R_1,\dots,R_{n}\right)$}
    \end{subfigure}
    \begin{subfigure}[t]{0.24\columnwidth}
        \centering
        \includegraphics[width=\textwidth]{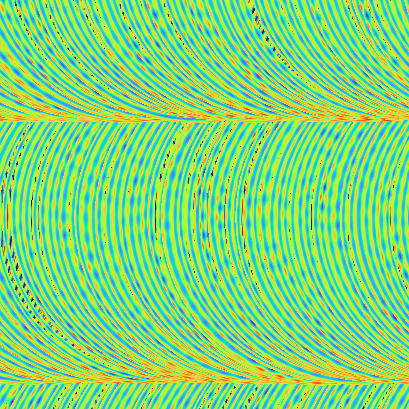}
        \caption*{$f^\text{Latest}_n\left(R_1,\dots,R_{n}\right)$}
    \end{subfigure}
    \begin{subfigure}[t]{0.24\columnwidth}
        \centering
        \includegraphics[width=\textwidth]{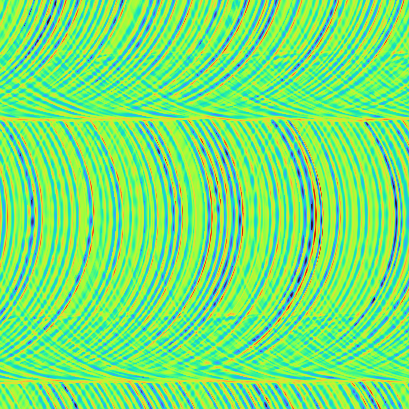}
        \caption*{$f^\text{Convex}_n\left(R_1,\dots,R_{n}\right)$}
    \end{subfigure}
    \includegraphics[height=3cm]{images/mill/Colorbar1}
    \caption{Illustration of interaction functions using bump shape. Imaged region is $5\text{ mm}\times 5\text{ mm}$.}
    \label{fig:mill:interaction}
\end{figure*}

\begin{figure*}
	\centering
	\begin{tikzpicture}
			\node at (0,1) {\includegraphics[height=2.9cm]{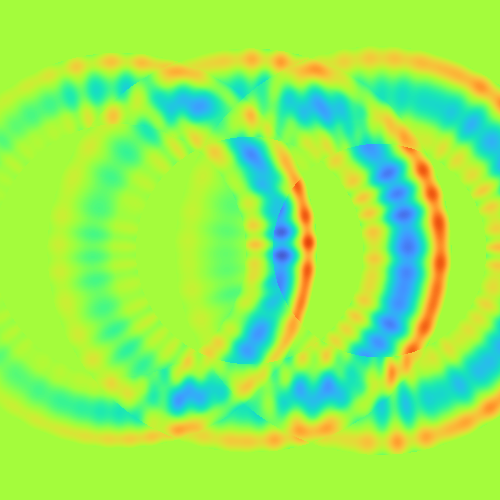}};
			\node at (3.4,1) {\includegraphics[height=2.9cm]{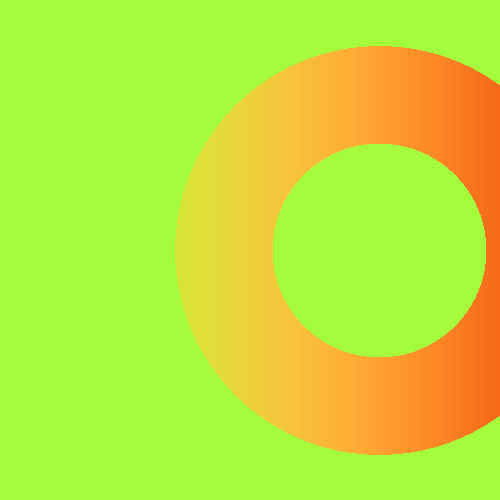}};
			\node at (6.6,1) {\includegraphics[height=2.9cm]{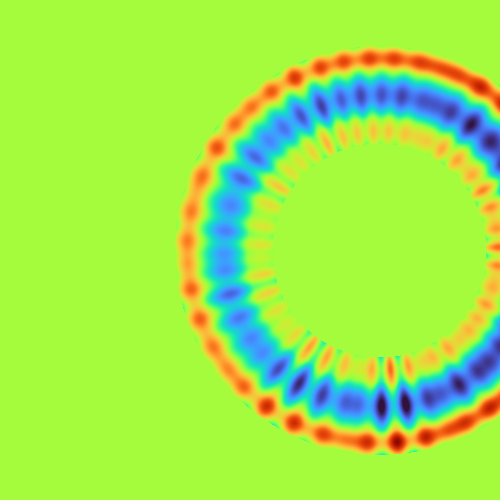}};
			\node at (9.8,1) {\includegraphics[height=2.9cm]{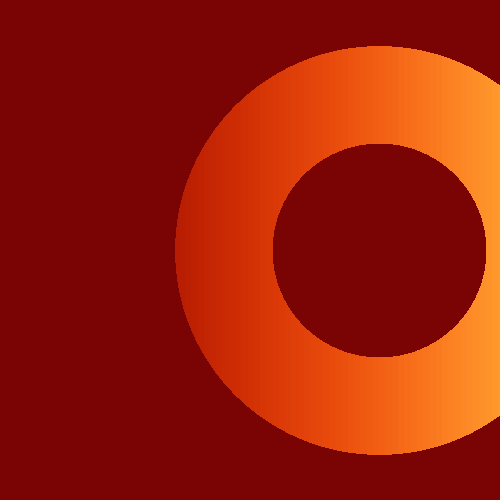}};
			\node at (13,1) {\includegraphics[height=2.9cm]{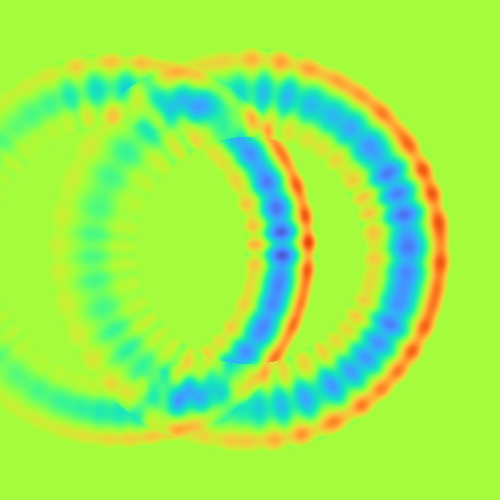}};
            \node at (14.8,1) {\includegraphics[height=2.9cm]{images/mill/Colorbar1.png}};
			
			\node at (1.7,1) {\small$=$};
			\node at (5,1) {\small$\cdot$};
			\node at (8.2,1) {\small$+$};
			\node at (11.4,1) {\small$\cdot$};
			
			\node at (0,-0.8) {\small$f_{3}^\text{Convex}\left(R_1,R_2,R_3\right)$};
			\node at (3.4,-0.8) {\small$A_3$};
			\node at (6.6,-0.8) {\small$R_3$};
			\node at (9.8,-0.8) {\small$\left(1-A_3\right)$};
			\node at (13,-0.8) {\small$f_{2}^\text{Convex}\left(R_1,R_2\right)$};
	\end{tikzpicture}
	\caption{Illustration of the convex combination as interaction function using tilting directions $\theta_k=0$ for $k=1,2,3$.}
	\label{fig:mill:inter:convex}
\end{figure*}

The interaction between $k\in\mathbb{N}$ rings is defined as mapping $f_k\left(R_1,\dots,R_k\right)$.
Assume $f_0(R_0)(x)=0$ for all $x\in\mathbb{R}^2$ and consider the following functions, which can be compared in Figure \ref{fig:mill:interaction}.
\begin{enumerate}
	\item \textbf{Minimum}\\
	Since material is removed during the milling process, the deepest indentation at any point should be decisive. 
    Note that the rings' order is not maintained when using the minimum. 
		{\makeatletter
		\@fleqntrue
		\makeatother
		\begin{alignat*}{1}
			&f_k^\text{Min}\left(R_1,\dots,R_k\right)(x) = \min\left(R_1(x),\dots,R_k(x)\right)
		\end{alignat*}}
	\item \textbf{Latest}\\
	Here, we take into account that milling is a consecutive process. 
    Hence, newly introduced rings overwrite older ones which leads to a dominant appearance of their rear parts.
	{\makeatletter
		\@fleqntrue
		\makeatother
		\begin{alignat*}{1}
			&f_{k}^\text{Latest}\left(R_1,\dots,R_k\right)(x) =
			\begin{cases*}
				R_k(x) &if $x\in P_k$\\
				f_{k-1}^\text{Latest}\left(R_1,\dots,R_{k-1}\right)(x) &else
			\end{cases*}
		\end{alignat*}}
	\item \textbf{Convex combination}\\
    In the previous approaches, the value in each point is determined by a single ring. 
    A more realistic assumption might be that the height value is obtained by a superposition of rings overlapping at that point.
    To this end, we use a convex combination of the rings with most weight on the last rings, see Figure \ref{fig:mill:inter:convex}.
    Since the surface is changed more by deeper cuts those should be weighted more.
    Thus, we use a tilted plane $A_k:\mathbb{R}^2\rightarrow\mathbb{R}$ instead of a global parameter.
    It is defined similarly to Equation \eqref{Eq:plane} but with $l_k=a_k\sim\mathcal{U}([a^\text{min},a^\text{max}])$ and $h_k=b_k\sim\mathcal{U}([b^\text{min},b^\text{max}])$ for $a^\text{min},a^\text{max},b^\text{min},b^\text{max}\in[0,1]$ fulfilling $a^\text{min}\leq a^\text{max}$ and $b^\text{min}\leq b^\text{max}$. 
    {\makeatletter
		\@fleqntrue
		\makeatother
		\begin{alignat*}{1}
			&f_{k}^\text{Convex}\left(R_1,\dots,R_k\right)(x) = \left(A_k\cdot R_k + \left(1-A_k\right)\cdot f_{k-1}^\text{Convex}\left(R_1,\dots,R_{k-1}\right)\right)(x)
		\end{alignat*}}
\end{enumerate}

\subsection{Sub-model for tool-path} 
\label{sec:mill:path}

In this section, a parametric model is formulated for both, the parallel and the spiral tool-path. 
Ring center points $c_{(k)}$ are modeled by random displacements of points $\bar{c}_{(k)}$ on the tool-path. 
That is, $c_{(k)}\sim\mathcal{N}\left(\bar{c}_{(k)},\Sigma_c\right)$. 
To account for irregularities in the visibility of rings, the deterministic ring order is changed by a certain amount $\epsilon\in[0,1]$. To do so, choose $\mathcal{E}\subset\left\{1,\dots,n\right\}$ with $\vert\mathcal{E}\vert=\lceil\epsilon\cdot n\rceil$ randomly. 
The order of rings with index in $\mathcal{E}$ is then permuted randomly to obtain $c_k$.

\begin{figure*}
	\centering
    \begin{subfigure}[t]{0.24\columnwidth}
        \centering
        \captionsetup{justification=centering}
        \includegraphics[width=\textwidth]{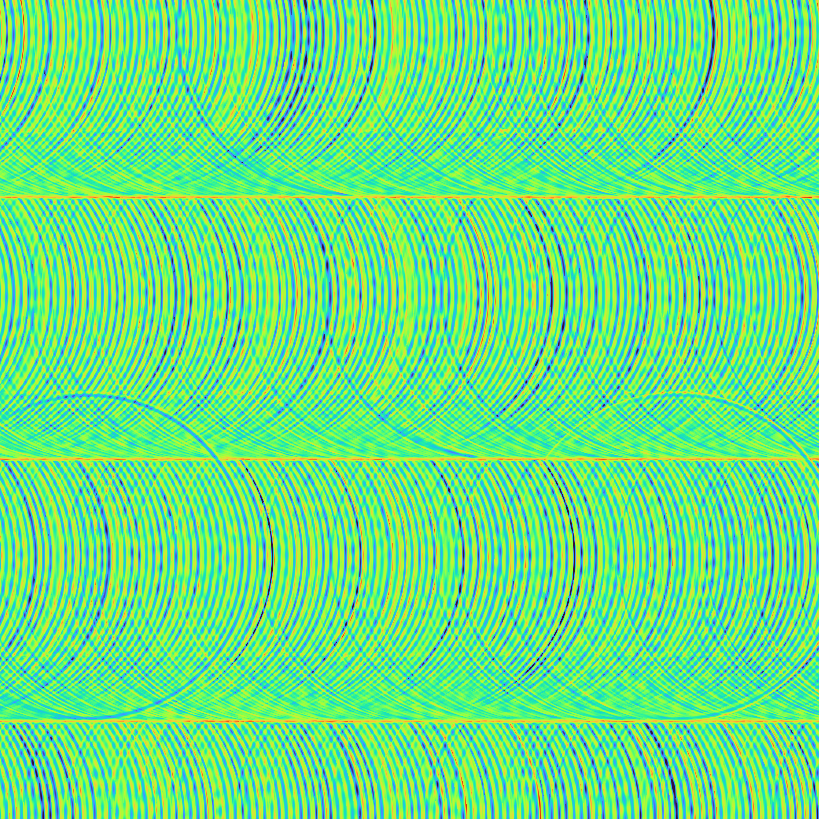}
        \caption*{$\alpha = 0.2$, $d=4\text{ mm}$, $\delta = 0.09\text{ mm}$.}
    \end{subfigure}
    \begin{subfigure}[t]{0.24\columnwidth}
        \centering
        \captionsetup{justification=centering}
        \includegraphics[width=\textwidth]{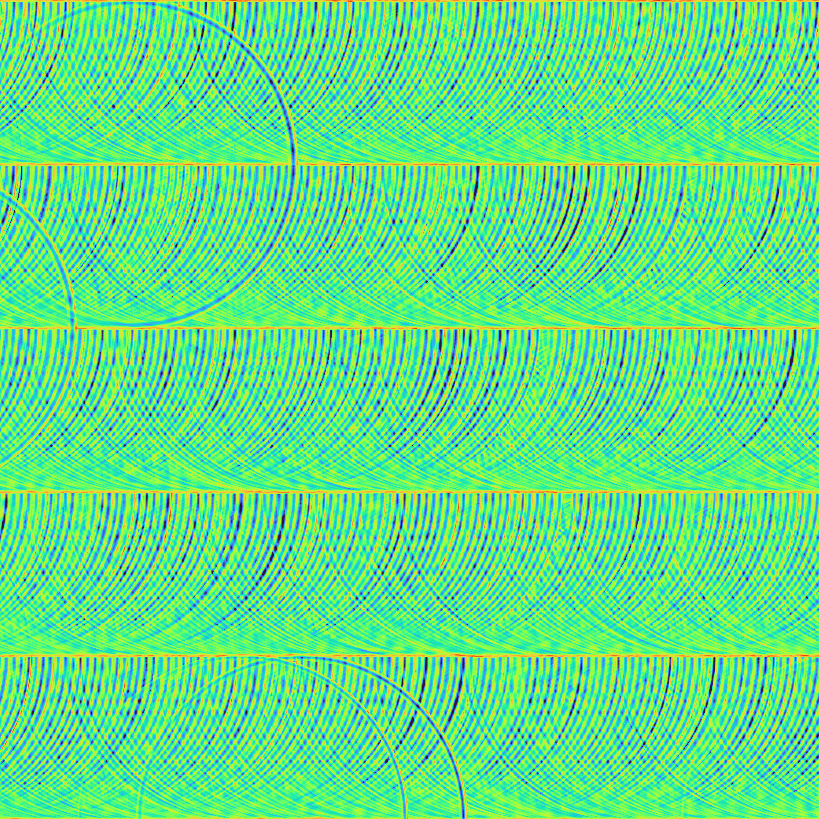}
        \caption*{$\alpha = 0.5$, $d=4\text{ mm}$, $\delta = 0.09\text{ mm}$.}
    \end{subfigure}
    \begin{subfigure}[t]{0.24\columnwidth}
        \centering
        \captionsetup{justification=centering}
        \includegraphics[width=\textwidth]{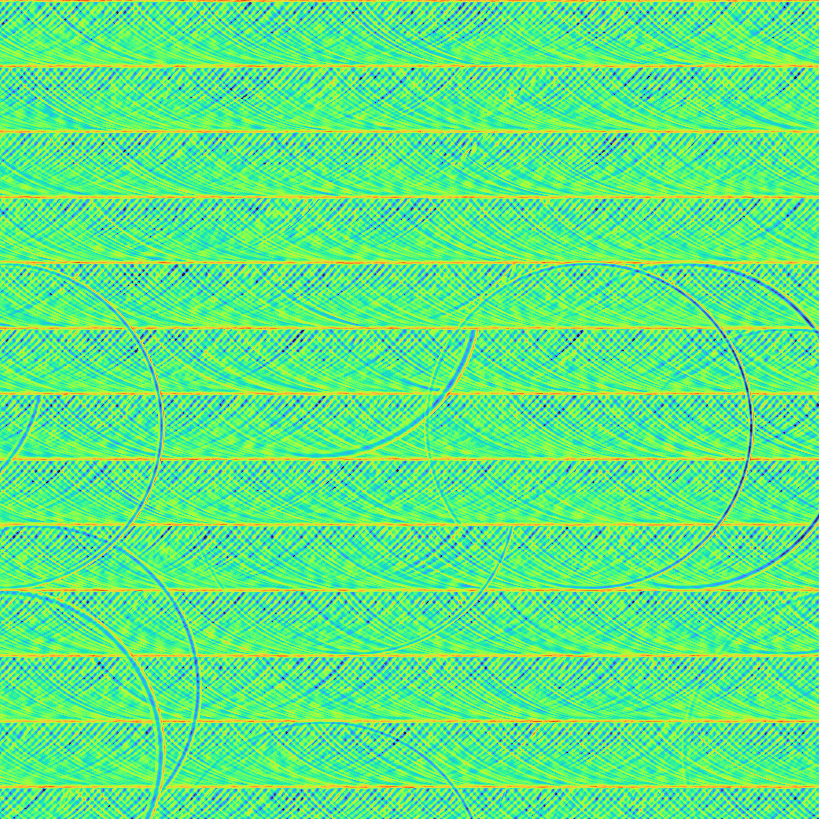}
        \caption*{$\alpha = 0.8$, $d=4\text{ mm}$, $\delta = 0.09\text{ mm}$.}
    \end{subfigure}
    \includegraphics[height=3cm]{images/mill/Colorbar1}\\[8pt]
    \begin{subfigure}[t]{0.24\columnwidth}
        \centering
        \captionsetup{justification=centering}
        \includegraphics[width=\textwidth]{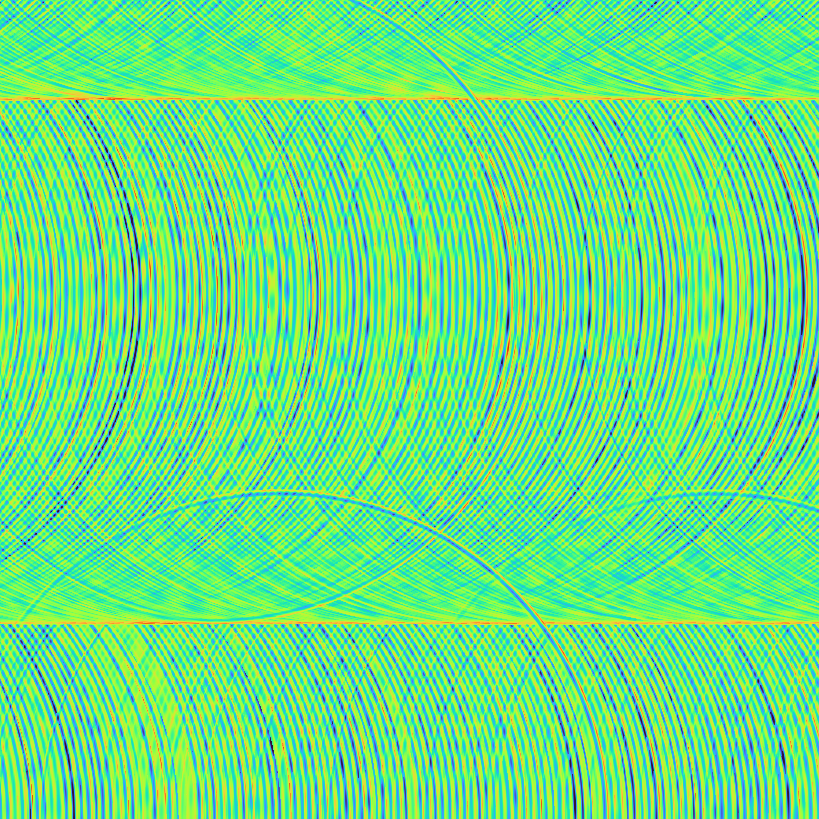}
        \caption*{$\alpha = 0.2$, $d=8\text{ mm}$, $\delta = 0.09\text{ mm}$.}
    \end{subfigure}
    \begin{subfigure}[t]{0.24\columnwidth}
        \centering
        \captionsetup{justification=centering}
        \includegraphics[width=\textwidth]{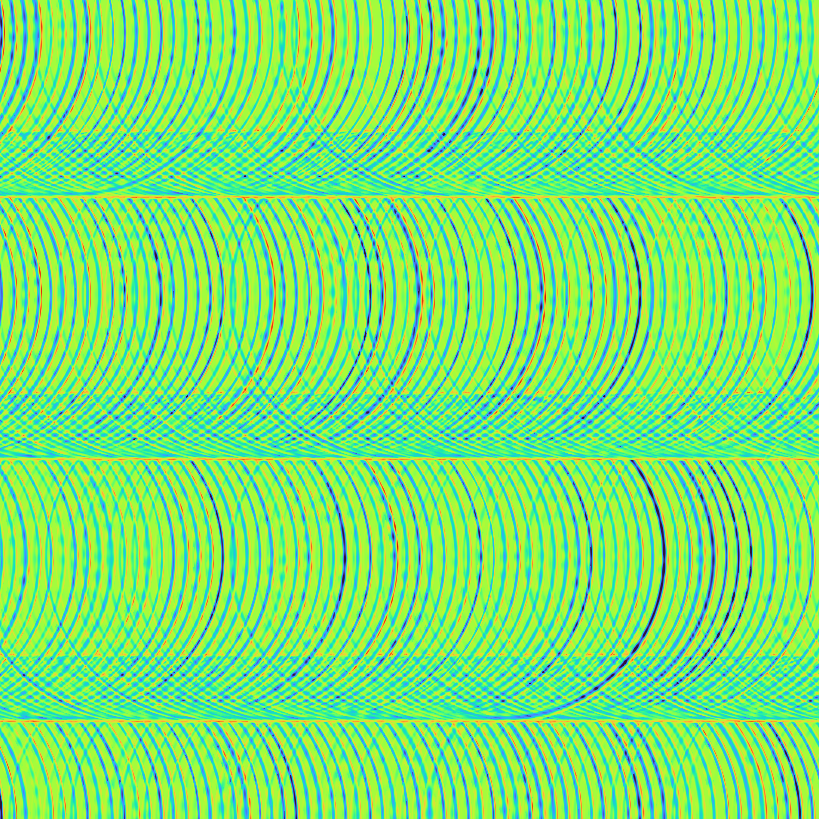}
        \caption*{$\alpha = 0.2$, $d=4\text{ mm}$, $\delta = 0.15\text{ mm}$.}
    \end{subfigure}
    \begin{subfigure}[t]{0.24\columnwidth}
        \centering
        \captionsetup{justification=centering}
        \includegraphics[width=\textwidth]{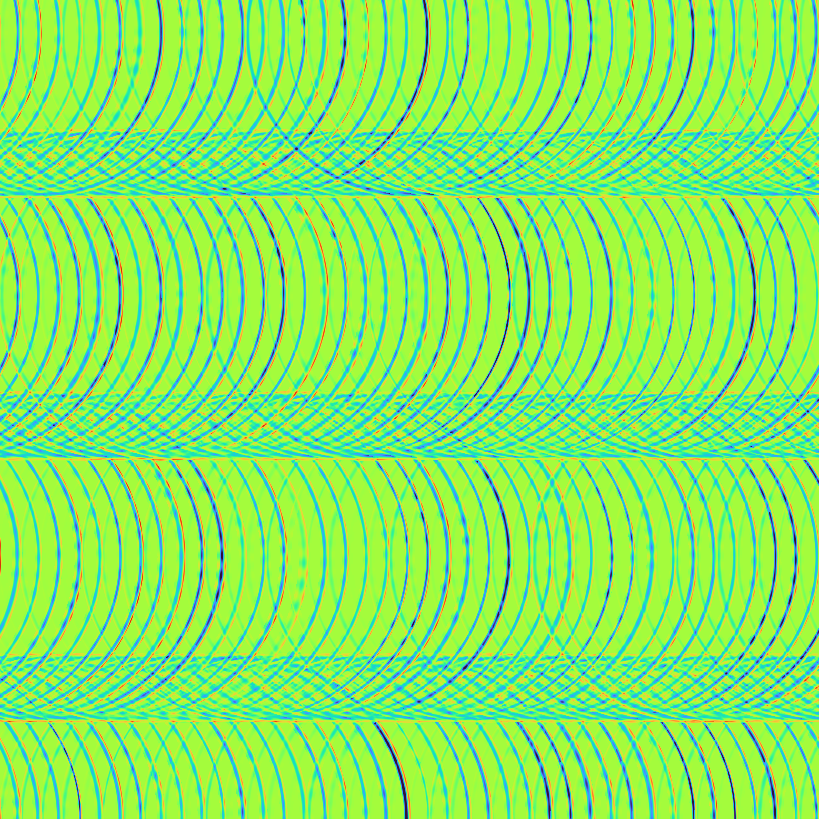}
        \caption*{$\alpha = 0.2$, $d=4\text{ mm}$, $\delta = 0.25\text{ mm}$.}
    \end{subfigure}
    \phantom{\includegraphics[height=3cm]{images/mill/Colorbar1}}
	\caption{Influence of parameters for the tool-path. Imaged region is $10\text{ mm}\times 10\text{ mm}$.}
	\label{fig:mill:path}
\end{figure*}

We denote the distance between neighboring tool-paths by $\rho=(1-\alpha)\cdot d\in(0,d)$, where $\alpha\in(0,1)$ specifies the amount of ring overlap.
It depends on the radial width of cut via $\alpha=1-a_e$.
Moreover, the distance between successive center points along the tool-path is $\delta\in\mathbb{R}_{>0}$, which depends on the tool's feed rate and the rotational speed of the milling head.
All parameters are process-induced and directly modifiable in the machine setting. 
Figure \ref{fig:mill:path} illustrates the influence of those parameters.

\subsubsection{Parallel pattern}
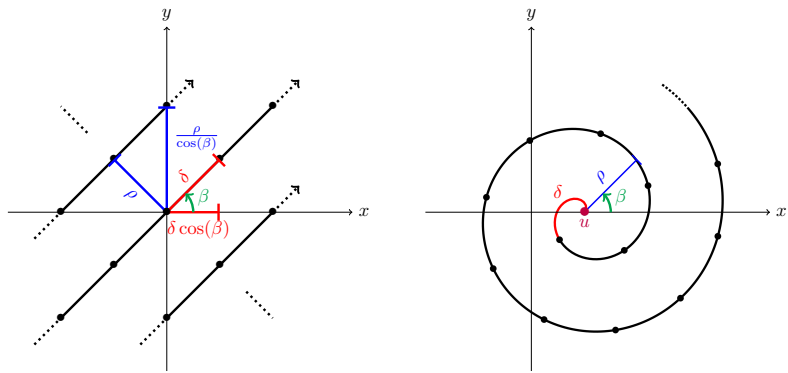
\begin{figure*}
	\centering
    \resizebox{0.3\columnwidth}{!}{
    \begin{tikzpicture}
		\coordinate (o) at (0,0);
		\coordinate (xx) at (1,0);
		\coordinate (yy) at (1,1);
		
		\draw[->,line width = 0.3pt] (-3,0) -- (3.5,0) node[right] {$x$};
		\draw[->,line width = 0.3pt] (0,-3) -- (0,3.5) node[above] {$y$};
		
		\draw[dotted,line width = 1.3pt] (-0.5,-2.5) -- ( 0,-2);
		\draw[dotted,line width = 1.3pt] (-2.5,-2.5) -- (-2,-2);
		\draw[dotted,line width = 1.3pt] (-2.5,-0.5) -- (-2, 0);
		
		\draw[line width = 1.3pt] ( 0,-2) -- (2,0);
		\draw[line width = 1.3pt] (-2,-2) -- (2,2);
		\draw[line width = 1.3pt] (-2, 0) -- (0,2);
		
		\draw[dotted,->,line width = 1.3pt] (2,0) -- (2.5,0.5);
		\draw[dotted,->,line width = 1.3pt] (2,2) -- (2.5,2.5);
		\draw[dotted,->,line width = 1.3pt] (0,2) -- (0.5,2.5);
		
		\draw[dotted,line width = 1.3pt] ( 1.5,-1.5) -- ( 2,-2);
		\draw[dotted,line width = 1.3pt] (-1.5, 1.5) -- (-2, 2);
		
		\node at (-2, 0) {$\bullet$};
		\node at (-1, 1) {$\bullet$};
		\node at ( 0, 2) {$\bullet$};
		
		\node at (-2,-2) {$\bullet$};
		\node at (-1,-1) {$\bullet$};
		\node at ( 0, 0) {$\bullet$};
		\node at ( 1, 1) {$\bullet$};
		\node at ( 2, 2) {$\bullet$};
		
		\node at (0,-2) {$\bullet$};
		\node at (1,-1) {$\bullet$};
		\node at (2, 0) {$\bullet$};
		
		\draw[-|,color=blue,line width = 1.3pt] (0,0) -- (-1,1) node[midway,below,sloped] {$\rho$};
		\draw[-|,color=blue,line width = 1.3pt] (0,0) -- (0,2) node[pos=0.7,right] {$\frac{\rho}{\cos(\beta)}$};
		
		\draw[-|,color=red,line width = 1.3pt] (0,0) -- (1,1) node[midway,above,sloped] {$\delta$};
		\draw[-|,color=red,line width = 1.3pt] (0,0) -- (1,0) node[pos=0.6,below] {$\delta\cos(\beta)$};	
		
		\node at (0, 0) {$\bullet$};	
        \pic [draw,->,color=Green,line width = 1.3pt,"$\beta$",angle eccentricity=1.5] {angle = xx--o--yy};
	\end{tikzpicture}}
    \hspace{12pt}
    \resizebox{0.3\columnwidth}{!}{
    \begin{tikzpicture}
		\coordinate (o) at (0,0);
		\coordinate (xx) at (1,0);
		\coordinate (yy) at (1,1);
		\node at (0,0) {\includegraphics[height=5.5cm]{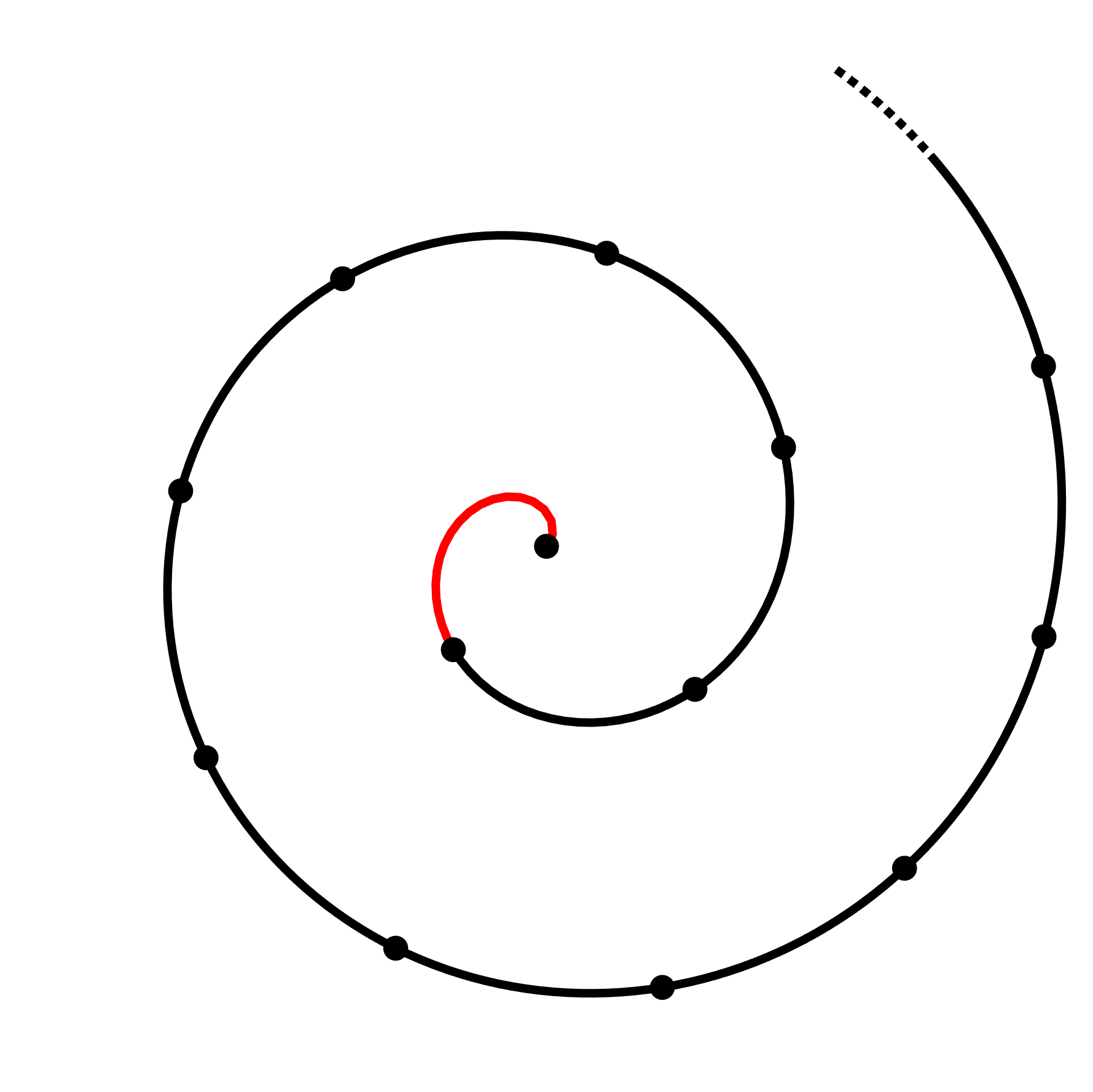}};
		\draw[->,line width=0.3pt] (-3,0) -- (3.5,0) node[right] {$x$};
		\draw[->,line width=0.3pt] (-1,-3) -- (-1,3.5) node[above] {$y$};
		\draw[-|,line width=0.8pt,color=blue] (0,0) -- (1,1) node[midway,above,sloped] {$\rho$};	
		\node at (0.005,-0.005) {\textcolor{purple}{\large{$\bullet$}}};
		\node at (-0.5,0.4) {\textcolor{red}{$\delta$}};	
		\node[below] at (0.005,-0.005) {\textcolor{purple}{$u$}};	
		\pic [draw,->,color=Green,line width = 1.3pt,"$\beta$",angle eccentricity=1.5] {angle = xx--o--yy};
	\end{tikzpicture}}
	\caption{Illustration of parallel (left) and spiral tool-path (right), their parameters and preliminary center points lying thereon.}
    \label{fig:mill:toolpath}
\end{figure*}

For parallel milling, the center points $\bar{c}_{(i,j)}$ are located on a parallel system of lines with distance $\rho$. Here, index $j$ is the index of the line while index $i$ represents a set of equidistant points at distance $\delta$ along each line.
Depending on the angle $\beta\in\left(-\pi,\pi\right]$ between the lines and the $x$-axis, we get the set of center points as $C^\text{parallel} = \left\{\bar{c}_{(i,j)}\right\}_{i,j\in\mathbb{Z}}$ for
\begin{equation*}
	\bar{c}_{(i,j)} =
	\begin{cases*}
		\left(x_i,\tan(\beta)x_i+\frac{j\rho}{\cos(\beta)}\right):x_i=i\delta\cos(\beta) &if $\beta\in\left(-\pi,\pi\right]\setminus\left\{-\frac{\pi}{2},\frac{\pi}{2}\right\}$\\
		\left(j\rho,i\delta\right) &if $\beta\in\left\{-\frac{\pi}{2},\frac{\pi}{2}\right\}$\\
	\end{cases*}
\end{equation*}
as illustrated in Figure \ref{fig:mill:toolpath}.

\begin{figure*}
	\centering
    \begin{subfigure}[t]{0.24\columnwidth}
        \centering
        \captionsetup{justification=centering}
        \includegraphics[width=\textwidth]{images/mill/Milling_parallel_equal_equal_1orientation_10mm_8192resolution_4mm_02_100seed_1range}
        \caption*{Milling in the same direction, bottom to top.}
    \end{subfigure}
    \begin{subfigure}[t]{0.24\columnwidth}
        \centering
        \captionsetup{justification=centering}
        \includegraphics[width=\textwidth]{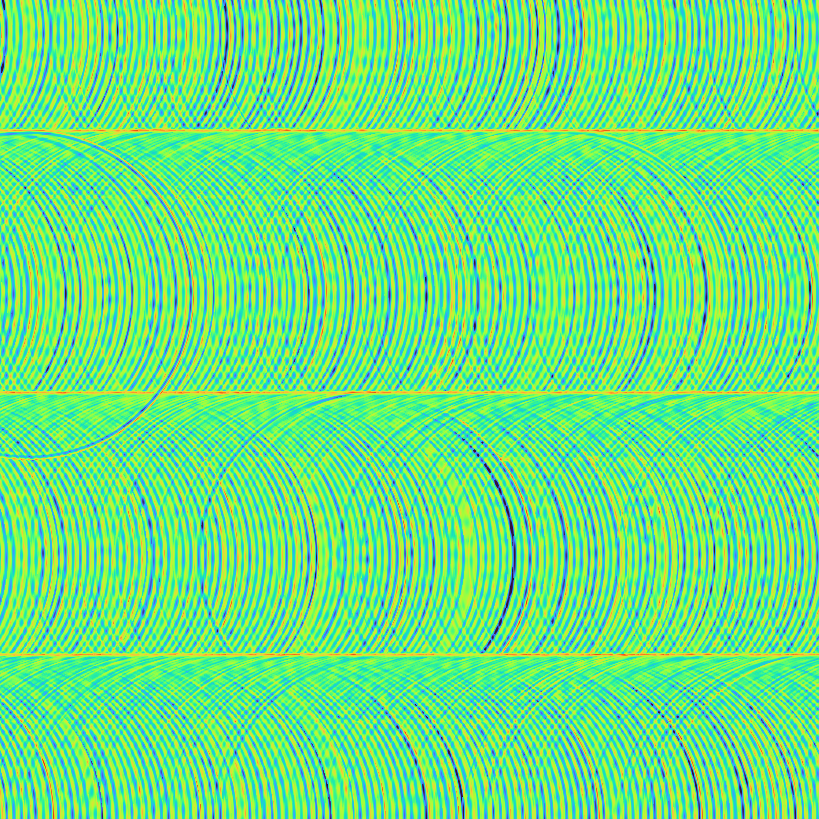}
        \caption*{Milling in the same direction, top to bottom.}
    \end{subfigure}
    \begin{subfigure}[t]{0.24\columnwidth}
        \centering
        \captionsetup{justification=centering}
        \includegraphics[width=\textwidth]{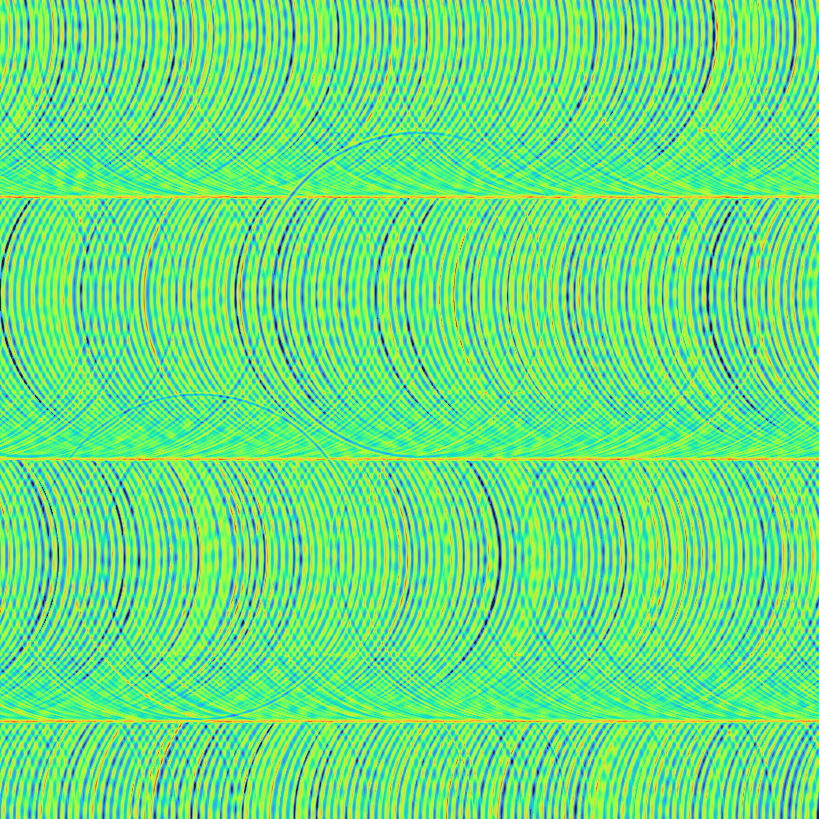}
        \caption*{Milling in alternating directions.}
    \end{subfigure}
    \begin{subfigure}[t]{0.24\columnwidth}
        \centering
        \captionsetup{justification=centering}
        \includegraphics[width=\textwidth]{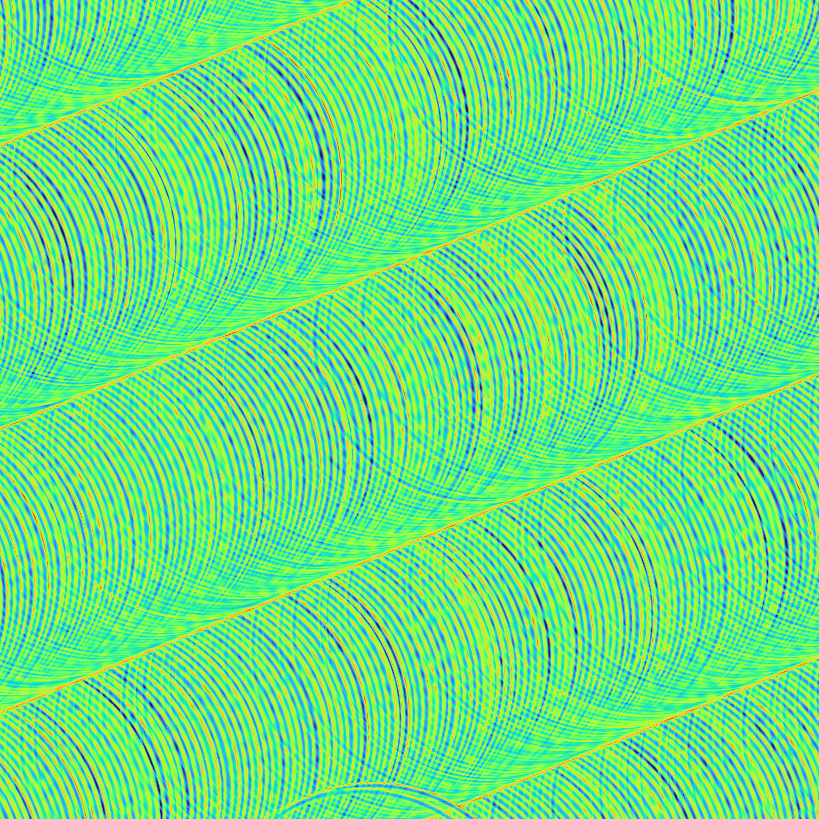}
        \caption*{Milling for $\beta=\nicefrac{\pi}{8}$.}
    \end{subfigure}
	\caption{Several parallel milled patterns. Imaged region is $10\text{ mm}\times 10\text{ mm}$.}
	\label{fig:mill:parallel}
\end{figure*}

The indices $k$ required for the interaction function are derived from the temporal order of milling the lines. 
We start by the first visible point on the line that is milled first. 
Then, we visit all indices $i$ on this line following the milling direction. 
After all visible rings on the line have been captured, we proceed to the line that is milled next. 
Figure \ref{fig:mill:parallel} shows patterns that are produced by different orderings.
Finally, to obtain $\left\{\bar{c}_{(k)}\right\}_{k\in\mathbb{N}}$, restrict the computation of $\bar{c}_{(i,j)}$ such that the resulting rings are visible in the image.

\subsubsection{Spiral pattern}
\begin{figure*}
	\centering
    \begin{subfigure}[t]{0.24\columnwidth}
        \centering
        \captionsetup{justification=centering}
        \includegraphics[width=\textwidth]{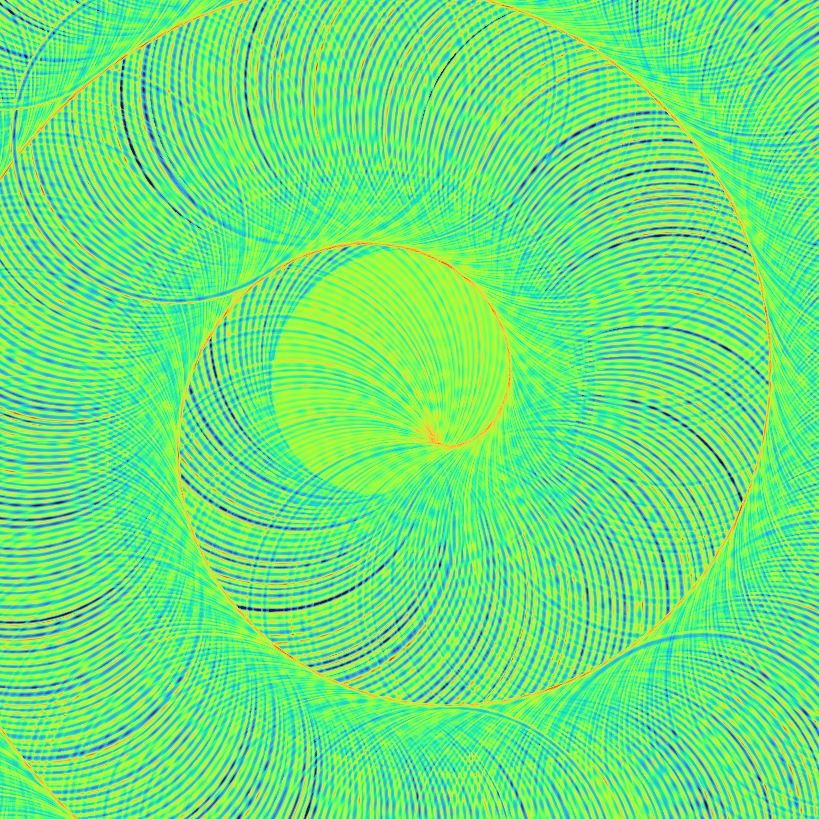}
        \caption*{Outward milling.}
    \end{subfigure}
    \begin{subfigure}[t]{0.24\columnwidth}
        \centering
        \captionsetup{justification=centering}
        \includegraphics[width=\textwidth]{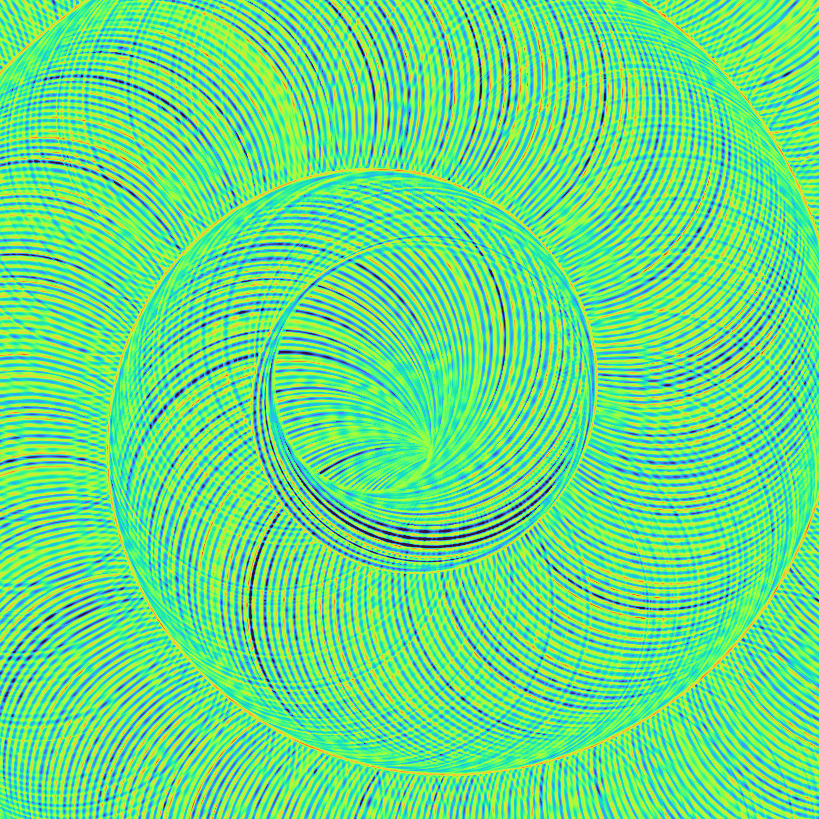}
        \caption*{Inward milling.}
    \end{subfigure}
    \begin{subfigure}[t]{0.24\columnwidth}
        \centering
        \captionsetup{justification=centering}
        \includegraphics[width=\textwidth]{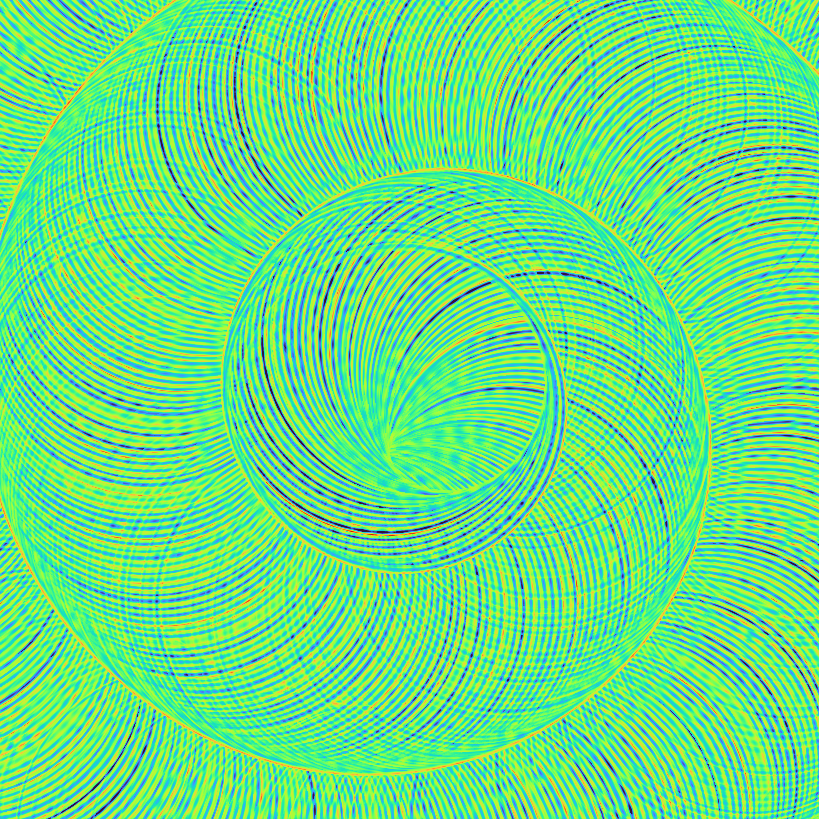}
        \caption*{Counter-clockwise inward milling.}
    \end{subfigure}
    \begin{subfigure}[t]{0.24\columnwidth}
        \centering
        \captionsetup{justification=centering}
        \includegraphics[width=\textwidth]{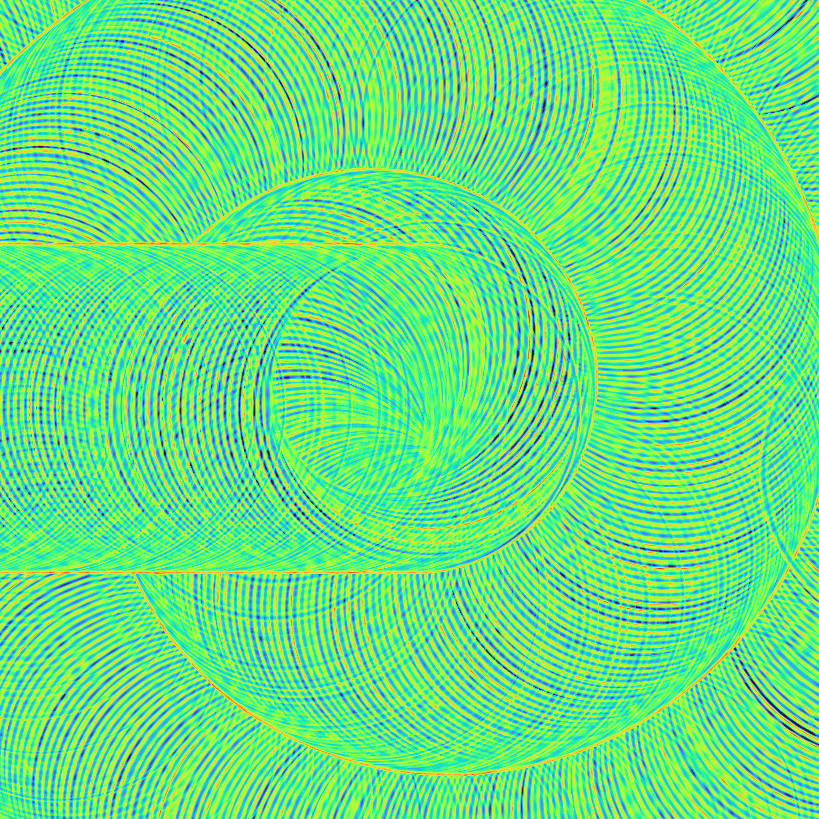}
        \caption*{Inward milling\\with exit line.}
    \end{subfigure}
	\caption{Several spiral milled patterns. Imaged region is $10\text{ mm}\times 10\text{ mm}$.}
	\label{fig:mill:spiral}
\end{figure*}

The spiral tool-path is modeled using the Archimedean spiral which is defined such that the radius grows proportionally to the angle.
Its parametric representation in polar coordinates is
\begin{equation*}
	\text{spiral}:\varphi\mapsto
	\left(\begin{array}{r}
		\omicron\cdot a\varphi\cdot\cos\left(\varphi+\beta\right)\\
		a\varphi\cdot\sin\left(\varphi+\beta\right)
	\end{array}\right)+u, \quad \varphi \ge 0.
\end{equation*} 
Here, $u\in\mathbb{R}^2$ is the origin of the spiral, $\beta\in(-\pi,\pi]$ the angle to the $x$-axis at the origin and $\omicron\in\{-1,1\}$ its orientation. 
If $\omicron=1$, then the spiral turns counter-clockwise. 
The distance of neighboring arcs is given by $\Vert \text{spiral}(\varphi)-\text{spiral}(\varphi+2\pi)\Vert_2=2\pi a$. 
To achieve a distance of $\rho$, we have to choose $a=\frac{\rho}{2\pi}$. 
Figure \ref{fig:mill:toolpath} shows an exemplary spiral tool-path explaining all required parameters.
For simplicity, we only consider $\omicron=1$, $\beta=0$ and $u=(0,0)$ in the following.
According to our definition, the tool-path starts at the spiral's origin and moves to the outside (outward milling). 
Alternatively, the tool can also move in the opposite direction (inward milling). 
In this case, the tool can produce an exit line when it is not lifted from the surface before removing it. 
Exemplary textures of spiral milled surfaces are shown in Figure \ref{fig:mill:spiral}.

\begin{figure*}
	\centering
    \begin{subfigure}[t]{0.45\columnwidth}
        \centering
        \includegraphics[height=3.5cm]{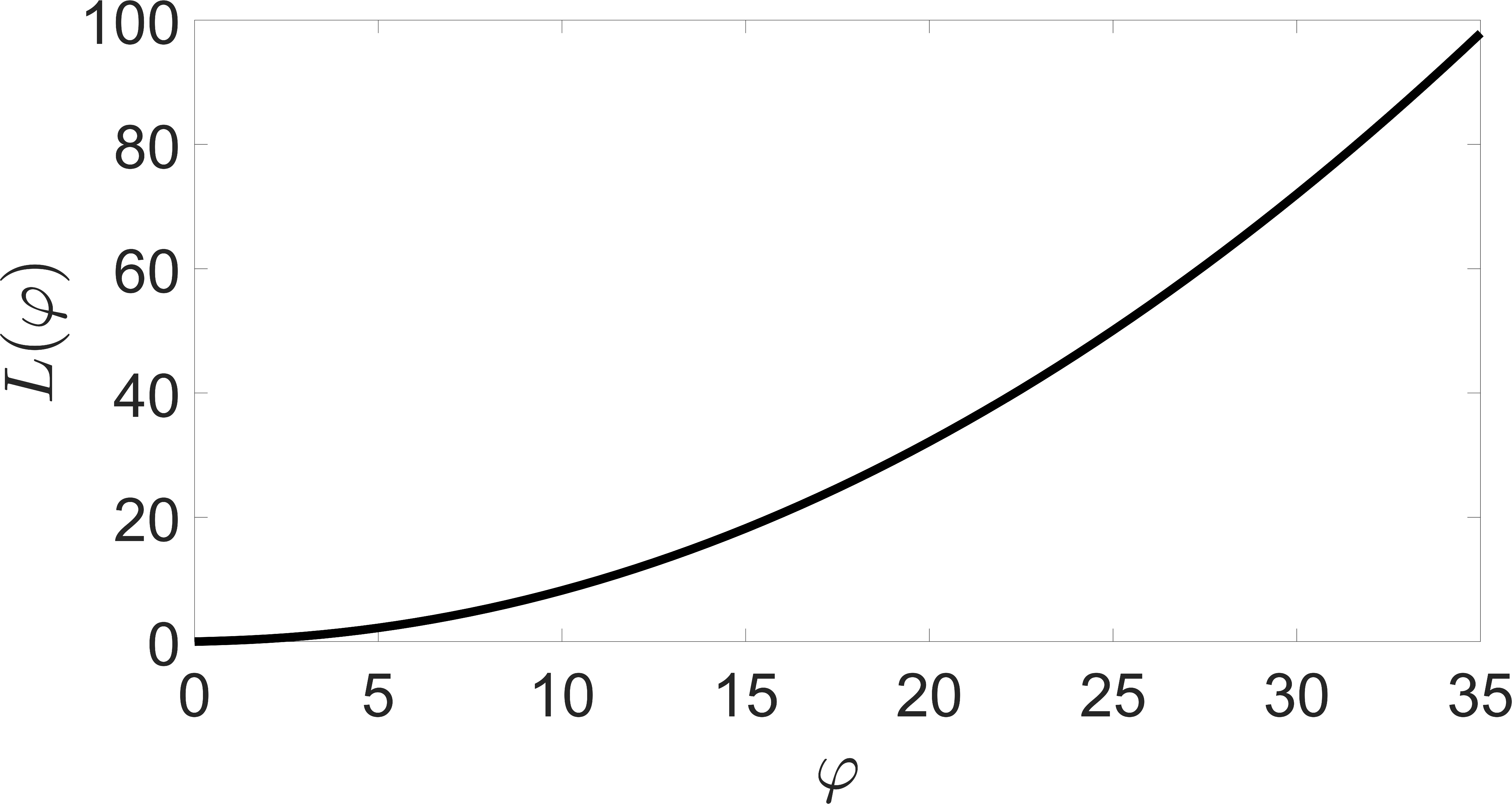}
        \caption{Arc length $L(\varphi$).}
    \end{subfigure}
    \begin{subfigure}[t]{0.49\columnwidth}
        \centering
        \includegraphics[height=3.5cm]{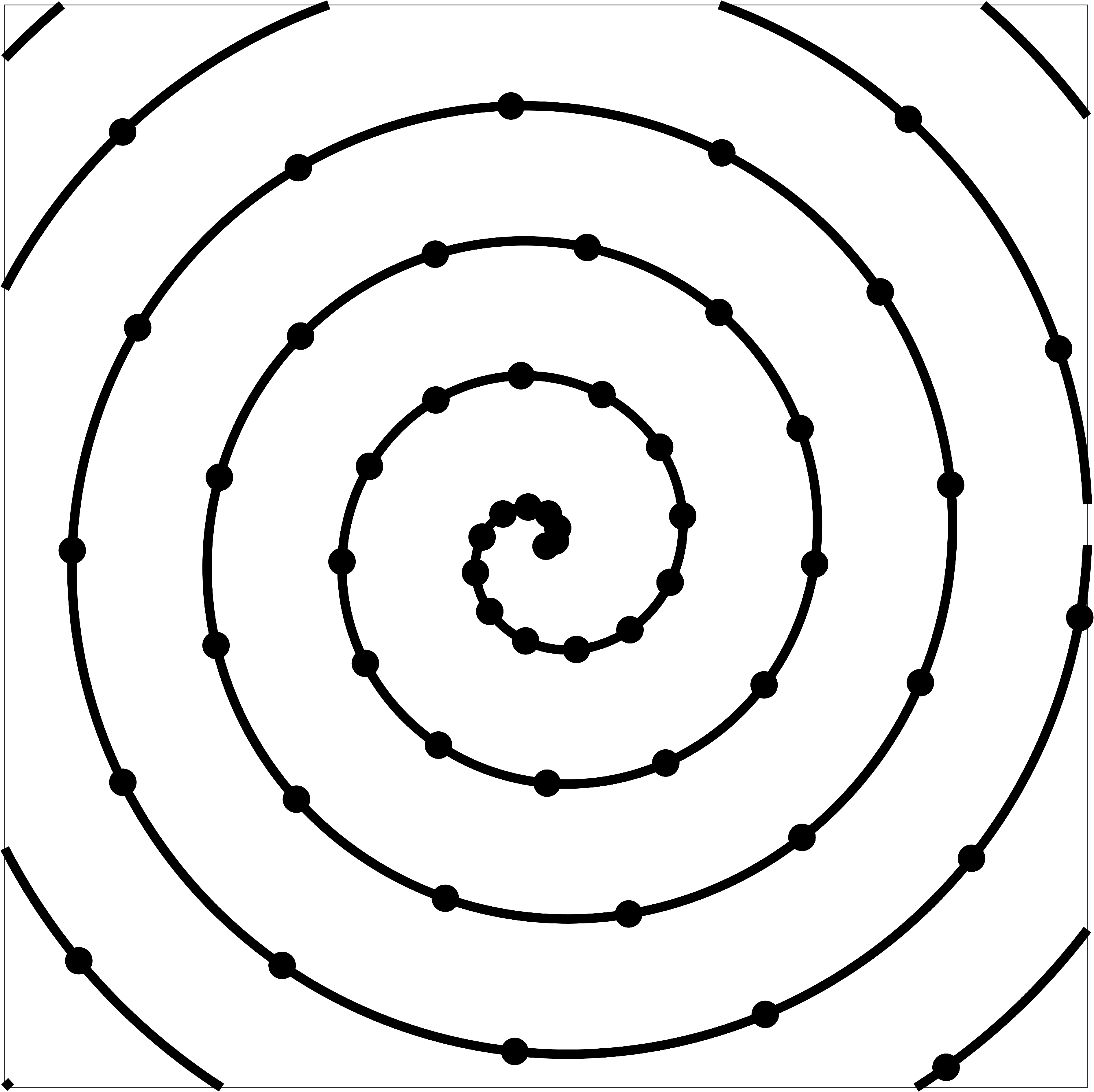}
	    \includegraphics[height=3.55cm]{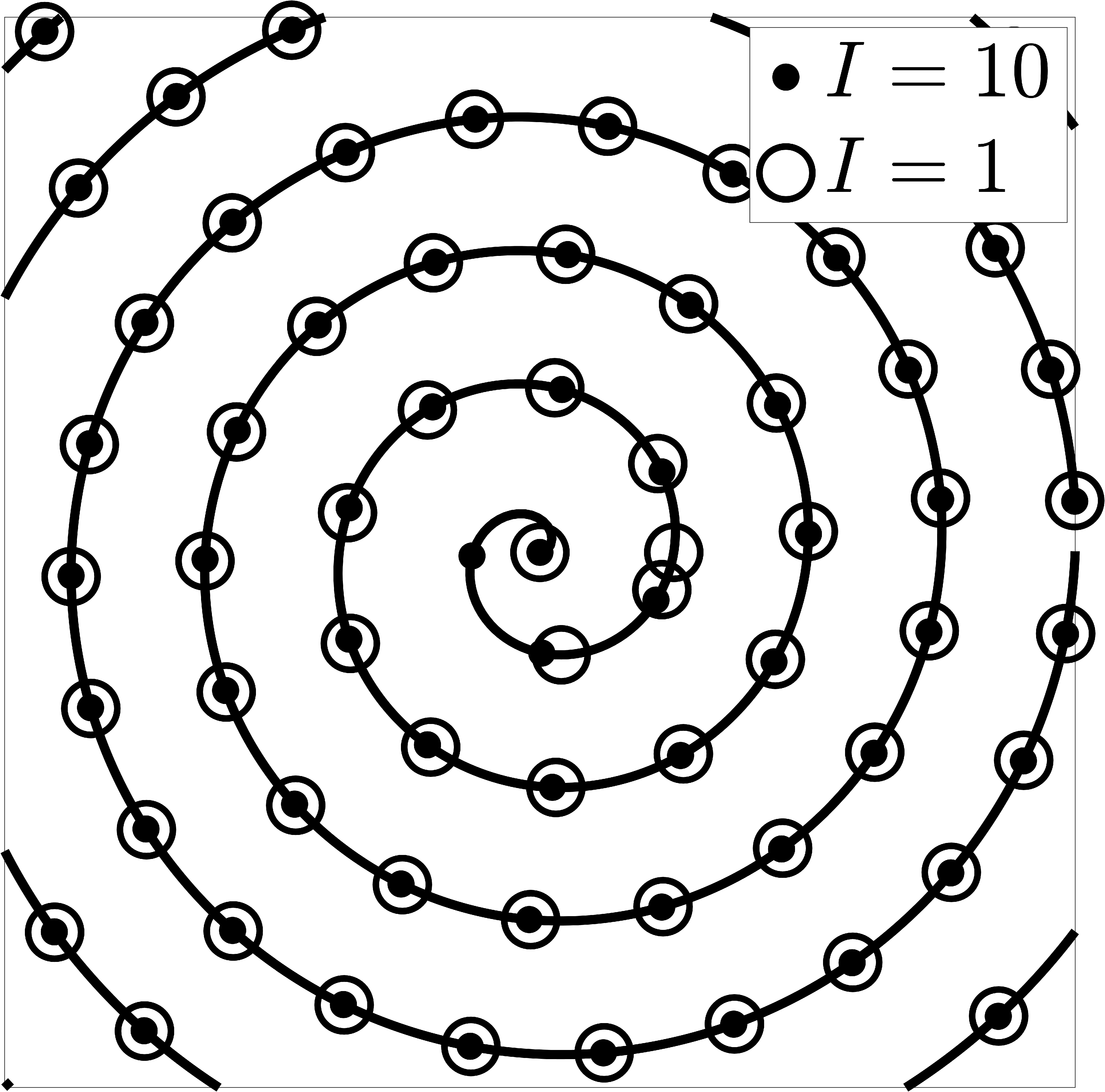}
        \caption{Comparison of equidistant $\varphi_{(k)}$ (left) and those computed using Newton's method (right) with different $I$. }
    \end{subfigure}
	\caption{Explanation to use Newton's method for the computation of $\varphi_{(k)}$ considering $a=\frac{1}{2\pi}$.}
	\label{fig:mill:spiral:points}
\end{figure*}

The center points are given by $C^\text{spiral}=\left\{\bar{c}_{(k)}=\text{spiral}\left(\varphi_{(k)}\right)\right\}_{k\in\mathbb{N}}$ for suitable $\varphi_{(k)}$.
To ensure that consecutive center points have equal distance $\delta$, the angles $\varphi_{(k)}$ have to be chosen such that they fulfill $L\left(\varphi_{(k)}\right)=k\cdot\delta$, where $L\left(\varphi\right)=\frac{a}{2}\left(\varphi\sqrt{1+\varphi^2}+\ln\left(\varphi+\sqrt{1+\varphi^2}\right)\right)$ is the spiral's arc length.
Equidistant $\varphi_{(k)}$ are not possible since $L(\cdot)$ is not linear, see Figure \ref{fig:mill:spiral:points}.
Moreover, no analytic solution of $L\left(\varphi_{(k)}\right)-k\cdot\delta=0$ exists. 
Therefore, we use Newton's method 
\begin{alignat*}{2}
    \varphi_{(k,0)} = \varphi_{(k-1)} &\hspace{20pt}&
    \varphi_{(k,j+1)} = \varphi_{(k,j)} - \frac{L\left(\varphi_{(k,j)}\right)-k\cdot\delta}{L^\prime\left(\varphi_{(k,j)}\right)},\, j\in\mathbb{N}
\end{alignat*}
with $\varphi_{(0)}=0$. The derivative of $L(\cdot)$ is given by $L'(\varphi) = \frac{a}{2}\left(\frac{2\varphi^2+1}{\sqrt{1+\varphi^2}} + \frac{1}{\sqrt{1+\varphi^2}}\right) = a\sqrt{1+\varphi^2}$.
Hence, approximate $\varphi_{(k)}\approx\varphi_{(k,I)}$ for $I\in\mathbb{N}$. 
We decided for $I=1$ because negligible mistakes are just visible for small $k$, see Figure \ref{fig:mill:spiral:points}. 
Hence, use
\begin{alignat*}{2}
	\varphi_{(0)} = 0 &\hspace{20pt}&
	\varphi_{(k+1)} = \varphi_{(k)} - \frac{L\left(\varphi_{(k)}\right)-k\cdot\delta}{L^\prime\left(\varphi_{(k)}\right)}.
\end{alignat*}

\subsection{Fitting the model to data}
\label{sec:mill:fit}

The modular multi-step model for generating texture images of milled surfaces has components for modeling the shape of an individual ring and the interaction between rings which are compared in Table \ref{tab:mill:improvements}.
It shows the necessity of randomizing the ring structures and of all components for the sub-model of the rings' appearance. 
We decided for the bump shape and the convex combination as interaction function since these two variants reflect the corresponding processes most realistically.
Although this is the most complex choice, most of the parameters are related to the milling process. 
Some parameters are real-world parameters and thus can be taken immediately from those of the machining process. 
The remaining parameters are estimated by visual comparison of the resulting patterns to the measurements. 
All parameters are summarized in Table \ref{tab:mill:parameter}.

\begin{table*}
    \caption{Comparison of interaction functions using indicator and bump shape. The model's improvements are added successively: variability  in the ring width ($\sigma_{w^\bullet},\left(\Sigma_c\right)_{ii}\neq 0$), integration of tilting and noise and change of the rings' ordering. Imaged region is $5\text{ mm}\times 5\text { mm}$.}
	\label{tab:mill:improvements}
	\centering
	\resizebox{\textwidth}{!}{
	\begin{tabular}{|@{\hskip 5pt}c@{\hskip 5pt}|@{\hskip 5pt}c@{\hskip 1.5pt}c@{\hskip 7pt}|c@{\hskip 3pt}c@{\hskip 3pt}c@{\hskip 3pt}c@{\hskip 3pt}c|}
		\hline
		\multicolumn{3}{|c|}{$\sigma\neq 0$} & \xmark & \cmark & \cmark & \cmark & \cmark \\
		\multicolumn{3}{|c|}{Tilting} & \xmark & \xmark & \cmark & \cmark & \cmark \\
		\multicolumn{3}{|c|}{Noise} & \xmark & \xmark & \xmark & \cmark & \cmark \\
		\multicolumn{3}{|c|}{$\epsilon\neq 0$} & \xmark & \xmark & \xmark & \xmark & \cmark \\
		\hline
		\hline
		
		\multirow{8}{*}{\begin{turn}{90} Indicator shape \end{turn}} &
		\multicolumn{2}{@{\hskip -1.5pt}c|}{\begin{turn}{90} \phantom{iiiiiiii} Minimum \phantom{iiiiiiii} \end{turn}} &
		\includegraphics[width=0.19\columnwidth]{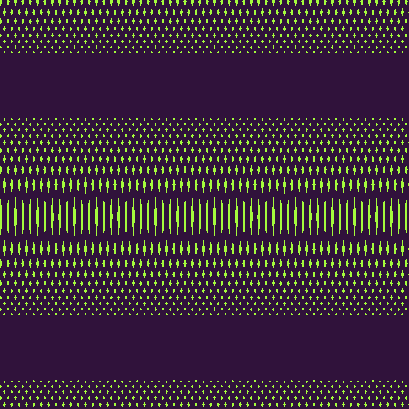} &
		\includegraphics[width=0.19\columnwidth]{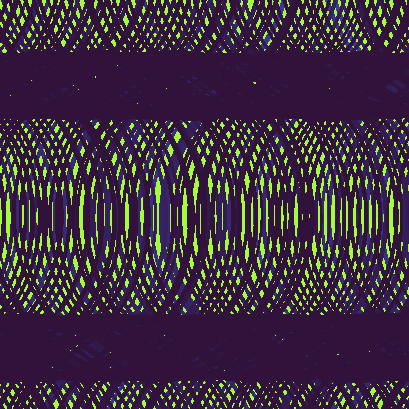} &
		\includegraphics[width=0.19\columnwidth]{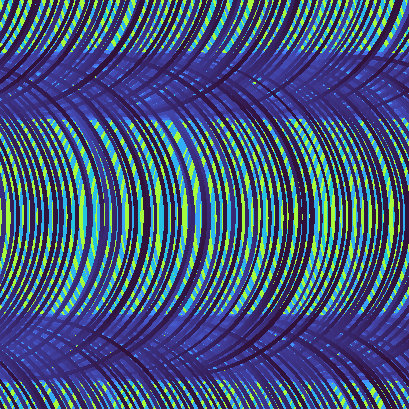} &
		\includegraphics[width=0.19\columnwidth]{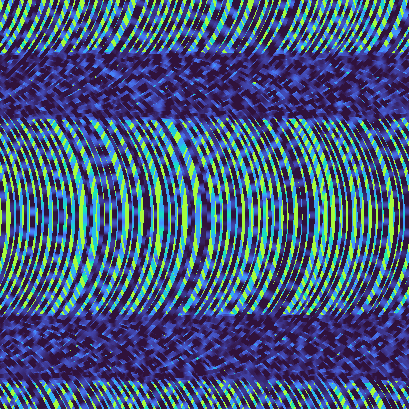} &
		\includegraphics[width=0.19\columnwidth]{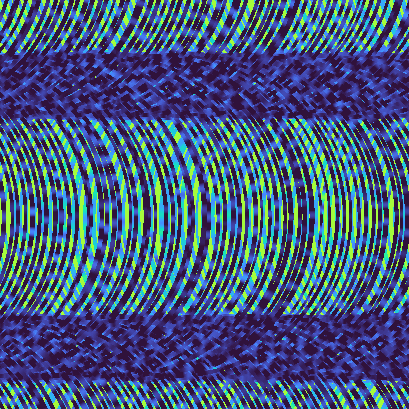} \rule[93pt]{0pt}{0pt} \\
		
		&\multicolumn{2}{@{\hskip -1.5pt}c|}{\begin{turn}{90} \phantom{iiiiiiiiiiii}  Latest \phantom{iiiiiiii}  \end{turn}} &
		\includegraphics[width=0.19\columnwidth]{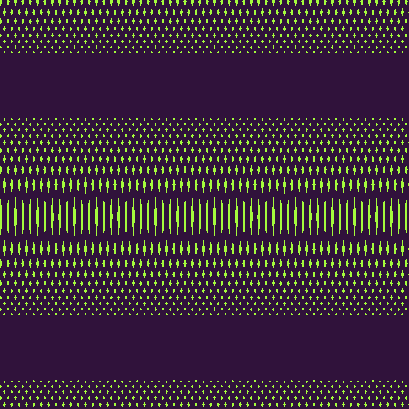} &
		\includegraphics[width=0.19\columnwidth]{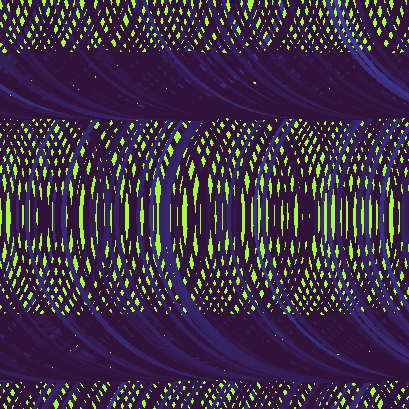} &
		\includegraphics[width=0.19\columnwidth]{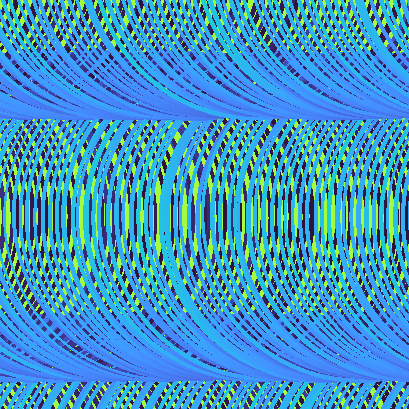} &
		\includegraphics[width=0.19\columnwidth]{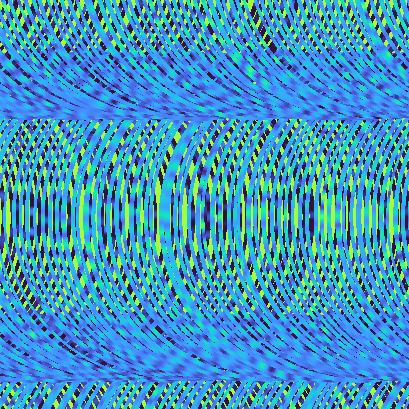} &
		\includegraphics[width=0.19\columnwidth]{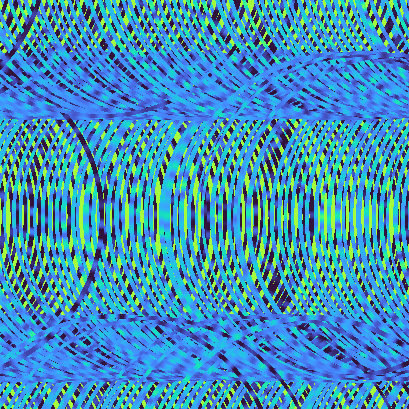} \rule[0pt]{0pt}{0pt} \\
		
		&\begin{turn}{90} \phantom{iiiiiiiii} Convex \phantom{iiiiiii} \end{turn} &
		\begin{turn}{90} \phantom{iiiiii} combination \phantom{iii} \end{turn} &
		\includegraphics[width=0.19\columnwidth]{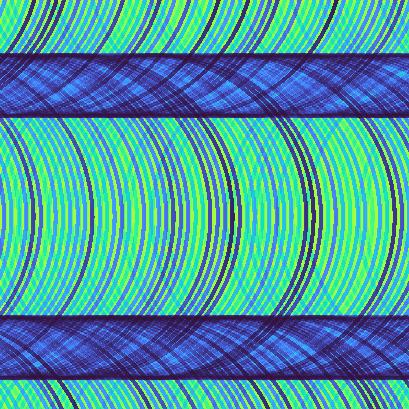} &
		\includegraphics[width=0.19\columnwidth]{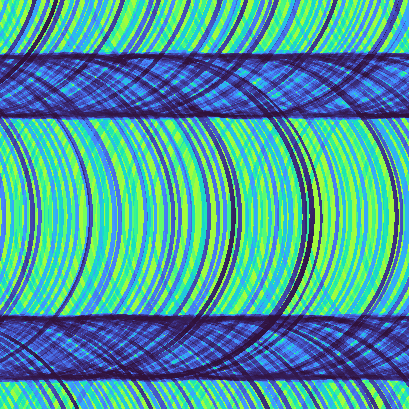} &
		\includegraphics[width=0.19\columnwidth]{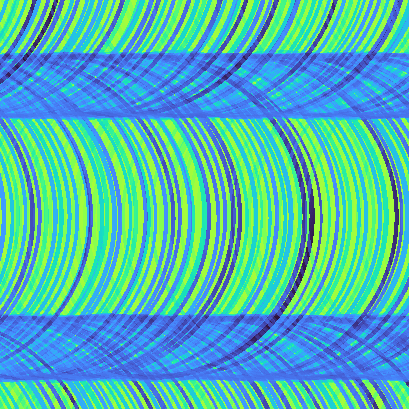} &
		\includegraphics[width=0.19\columnwidth]{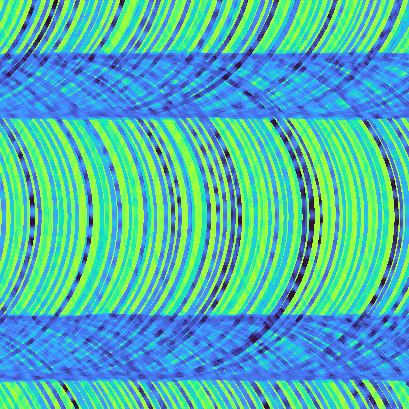} &
		\includegraphics[width=0.19\columnwidth]{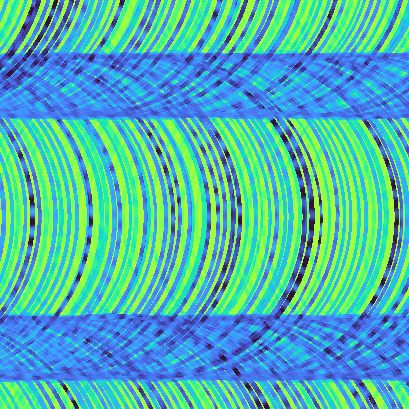} \rule[0pt]{0pt}{0pt} \\
		
		\hline
		\hline
		
		\multirow{8}{*}{\begin{turn}{90} Bump shape \end{turn}} &
		\multicolumn{2}{@{\hskip -1.5pt}c|}{\begin{turn}{90} \phantom{iiiiiiii} Minimum \phantom{iiiiiiii} \end{turn}} &
		\includegraphics[width=0.19\columnwidth]{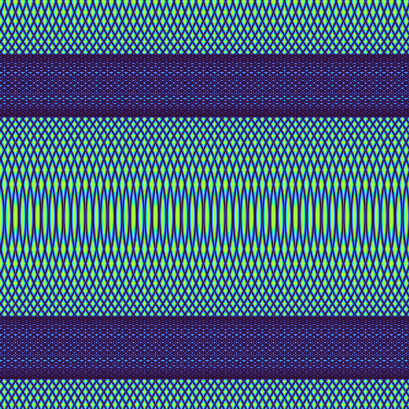} &
		\includegraphics[width=0.19\columnwidth]{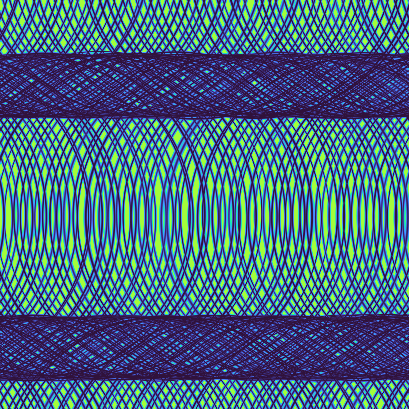} &
		\includegraphics[width=0.19\columnwidth]{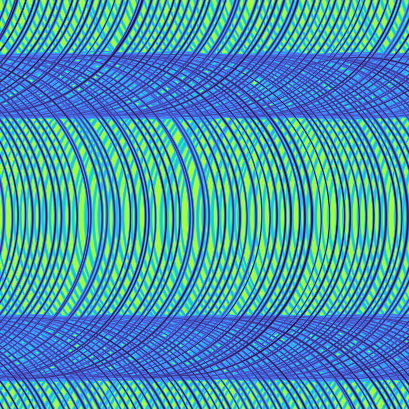} &
		\includegraphics[width=0.19\columnwidth]{images/mill/cosine3_minimum_withNoise_withTilting_withSineNoise_noChange_Size819} &
		\includegraphics[width=0.19\columnwidth]{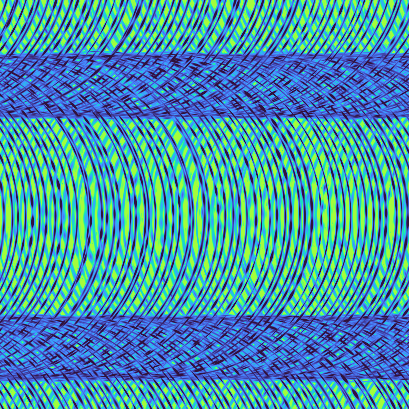} \rule[93pt]{0pt}{0pt} \\
		
		&\multicolumn{2}{@{\hskip -1.5pt}c|}{\begin{turn}{90} \phantom{iiiiiiiiiiii}  Latest \phantom{iiiiiiii}  \end{turn}} &
		\includegraphics[width=0.19\columnwidth]{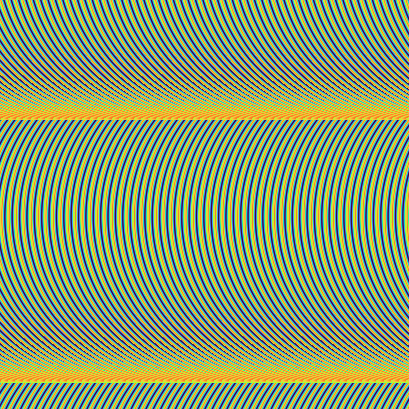} &
		\includegraphics[width=0.19\columnwidth]{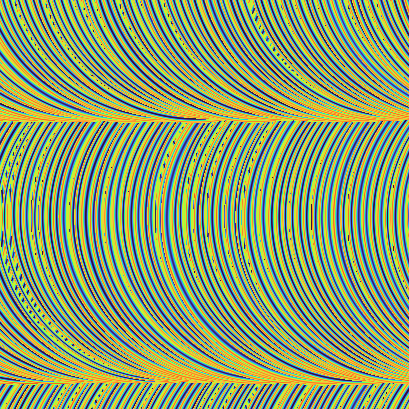} &
		\includegraphics[width=0.19\columnwidth]{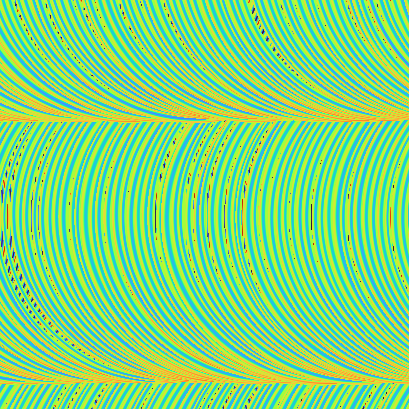} &
		\includegraphics[width=0.19\columnwidth]{images/mill/cosine3_latest_withNoise_withTilting_withSineNoise_noChange_Size819} &
		\includegraphics[width=0.19\columnwidth]{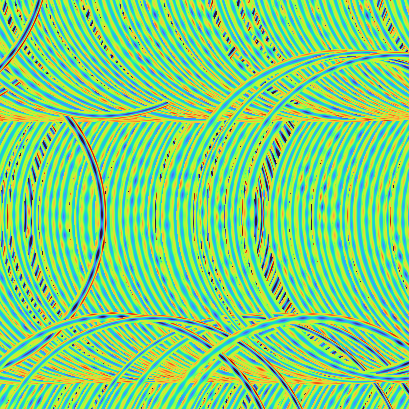} \rule[0pt]{0pt}{0pt} \\
		
		&\begin{turn}{90} \phantom{iiiiiiiii} Convex \phantom{iiiiiii} \end{turn} &
		\begin{turn}{90} \phantom{iiiiii} combination \phantom{iii} \end{turn} &
		\includegraphics[width=0.19\columnwidth]{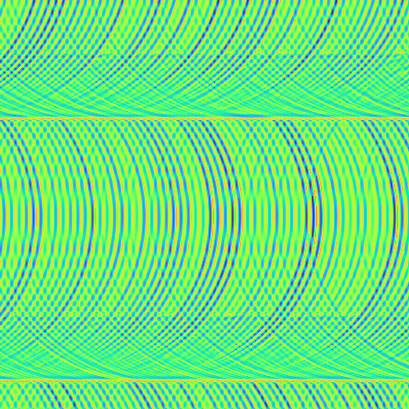} &
		\includegraphics[width=0.19\columnwidth]{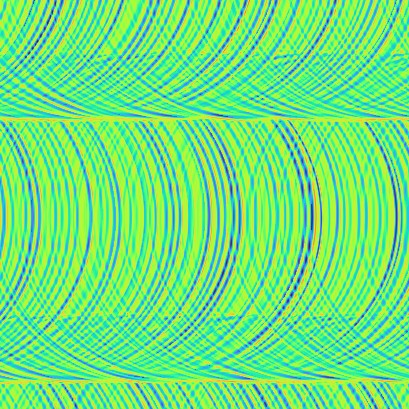} &
		\includegraphics[width=0.19\columnwidth]{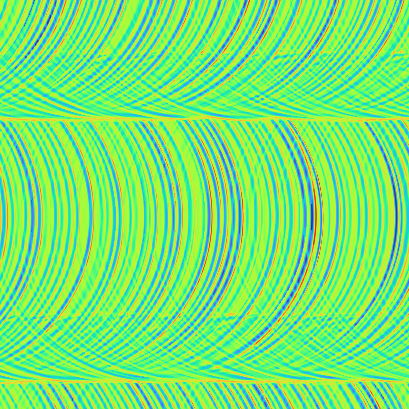} &
		\includegraphics[width=0.19\columnwidth]{images/mill/cosine3_averaged_withNoise_withTilting_withSineNoise_noChange_Size819} &
		\includegraphics[width=0.19\columnwidth]{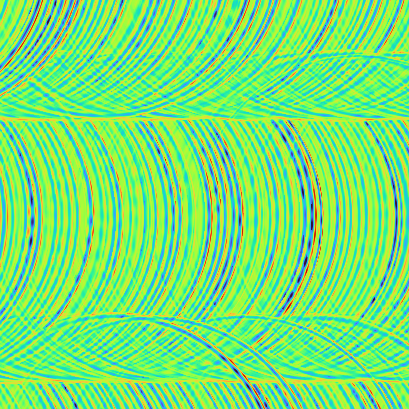} \rule[0pt]{0pt}{0pt} \\
		
		\hline
	\end{tabular}}
\end{table*}

\begin{table*}
    \caption{Overview of parameters needed for the model to simulate milled surfaces. Directly known process parameters are highlighted.} 
	\label{tab:mill:parameter}
	\centering
	\resizebox{\linewidth}{!}{
	\begin{tabular}{|c|c||c|c|c|c|c|}
		\hline
		\multicolumn{2}{|c||}{}& notation & definition & relation to process & distribution & parameters \\
		\hline
		\hline

        \multirow{17}*[-18pt]{\rotatebox[origin=c]{90}{appearance}}
        & \multirow{7}*[-9pt]{\rotatebox[origin=c]{90}{shape}}
   
		& $d$ & diameter of ring & \textbf{diameter of milling head} && $d\in\mathbb{R}_{>0}$\\[4pt]
   
		&& \multirow{2}{*}{$w^-_k$} & width of & \multirow{2}{*}{\textbf{width of cutting edge}} & \multirow{2}{*}{$\mathcal{N}\left(\mu_{w^-},\sigma_{w^-}\right)$}
        & $\mu_{w^-}\in\left(0,\nicefrac{d}{2}\right)$\\
        &&& indentation &&& $\sigma_{w^-}\in\mathbb{R}_{>0}$\\[4pt]

        && \multirow{2}{*}{$w^{+i}_k$} & width of & depends on & \multirow{2}{*}{$\mathcal{N}\left(\mu_{w^{+i}},\sigma_{w^{+i}}\right)$}
        & $\mu_{w^{+i}}\in\left[0,\nicefrac{d}{2}-\mu_{w^-}\right)$\\
        &&& inner accumulation & edges' sharpness && $\sigma_{w^{+i}}\in\mathbb{R}_{>0}$\\[4pt]

        && \multirow{2}{*}{$w^{+o}_k$} & width of & depends on & \multirow{2}{*}{$\mathcal{N}\left(\mu_{w^{+o}},\sigma_{w^{+o}}\right)$}
        & $\mu_{w^{+o}}\in\mathbb{R}_{\geq 0}$ \\
        &&& outer accumulation & edges' sharpness && $\sigma_{w^{+o}}\in\mathbb{R}_{>0}$ \\

        \cline{2-7}

        & \multirow{7}*[-9pt]{\rotatebox[origin=c]{90}{tilting}}
        & $\theta_k$ & tilting direction & depends on tool-path && $\theta_k\in(-\pi,\pi]$\\[2pt]

        && \multirow{2}{*}{$l^-_k$, $h^-_k$} & minimal/maximal scaling & \multirow{2}{*}{\textbf{cutting depth with tilting}} & \multirow{2}{*}{$\mathcal{N}\left(\mu_{\bullet^-},\sigma_{\bullet^-}\right)$}
        & $\mu_{\bullet^-}\in\mathbb{R}_{\geq 0}$\\
        &&& of indentation depth &&& $\sigma_{\bullet^-}\in\mathbb{R}_{>0}$ \\[4pt]

        && \multirow{2}{*}{$l^{+i}_k$, $h^{+i}_k$} & minimal/maximal scaling & depends on & \multirow{2}{*}{$\mathcal{N}\left(\mu_{\bullet^{+i}},\sigma_{\bullet^{+i}}\right)$}
        & $\mu_{\bullet^{+i}}\in\mathbb{R}_{\geq 0}$ \\
        &&& of inner accumulation height & edges' sharpness && $\sigma_{\bullet^{+i}}\in\mathbb{R}_{>0}$ \\[4pt]

        && \multirow{2}{*}{$l^{+o}_k$, $h^{+o}_k$} & minimal/maximal scaling & depends on & \multirow{2}{*}{$\mathcal{N}\left(\mu_{\bullet^{+o}},\sigma_{\bullet^{+o}}\right)$}
        & $\mu_{\bullet^{+o}}\in\mathbb{R}_{\geq 0}$\\
        &&& of outer accumulation height & edges' sharpness && $\sigma_{\bullet^{+o}}\in\mathbb{R}_{>0}$ \\[4pt]

        \cline{2-7}

        & \multirow{3}*[-4pt]{\rotatebox[origin=c]{90}{noise}}
        & $\lambda_k$ & number of sine curves && $\mathcal{P}(\lambda)$ & $\lambda\in\mathbb{N}$\\[4pt]

        && $\tau_{k_j}$ & frequency of sine curves && $\mathcal{P}(\tau)$ & $\tau\in\mathbb{N}$ \\[4pt]

        && $\xi_{k_j}$ & shift of sine curves && $\mathcal{U}\left((-\pi,\pi]\right)$ & \\

        \hline

        \multirow{4}*[-2pt]{\rotatebox[origin=c]{90}{interaction}}

        && \multirow{2}{*}{$a_k$} & front value for && \multirow{2}{*}{$\mathcal{U}\left([a^{\text{min}},a^{\text{max}}]\right)$}
        & $a^{\text{min}}\in[0,1]$\\
        &&& convex combination &&& $a^{\text{max}}\in[a^{\text{min}},1]$ \\[4pt]

        && \multirow{2}{*}{$b_k$} & rear value for && \multirow{2}{*}{$\mathcal{U}\left([b^{\text{min}},b^{\text{max}}]\right)$}
        & $b^{\text{min}}\in[0,1]$\\
        &&& convex combination &&& $b^{\text{max}}\in[b^{\text{min}},1]$ \\

        \hline
        
        \multirow{11}*[-6pt]{\rotatebox[origin=c]{90}{tool-path}}
        & \multirow{7}{*}{}

        & $c_k$ & ring center points & determined by tool-path & $\mathcal{N}\left(\bar{c}_k,\Sigma_c\right)$ & $\Sigma_c\in\mathbb{R}^{2\times 2}_{>0}$ \\[4pt]

        && \multirow{2}{*}{$\alpha$} & defines $\rho$ (distance between & \textbf{amount of overlap of} && \multirow{2}{*}{$\alpha\in(0,1)$} \\
        &&& neighboring tool-paths) & \textbf{neighboring tool-paths} &&\\[4pt]
            
        && \multirow{2}{*}{$\delta$} & distance between & \textbf{depends on feed rate and} && \multirow{2}{*}{$\delta\in\mathbb{R}_{>0}$}\\
        &&& center points & \textbf{tool rotational speed} &&\\[4pt]

        && \multirow{2}{*}{$\epsilon$} & amount of rings &&& \multirow{2}{*}{$\epsilon\in[0,1]$}\\
        &&& with changed order &&&\\[4pt]

        && $\beta$ & angle to $x$-axis &&& $\beta\in(-\pi,\pi]$\\[4pt]
        && $u$ & spiral origin &&& $u\in\mathbb{R}^2$\\[4pt]
        && $\omicron$ & spiral orientation & \textbf{(counter-)clockwise} && $\omicron\in\{-1,1\}$\\[4pt]

        \hline
	\end{tabular}}
\end{table*}

\begin{figure*}
	\centering
    \begin{subfigure}[t]{0.4\columnwidth}
        \centering
        \includegraphics[height=3.5cm]{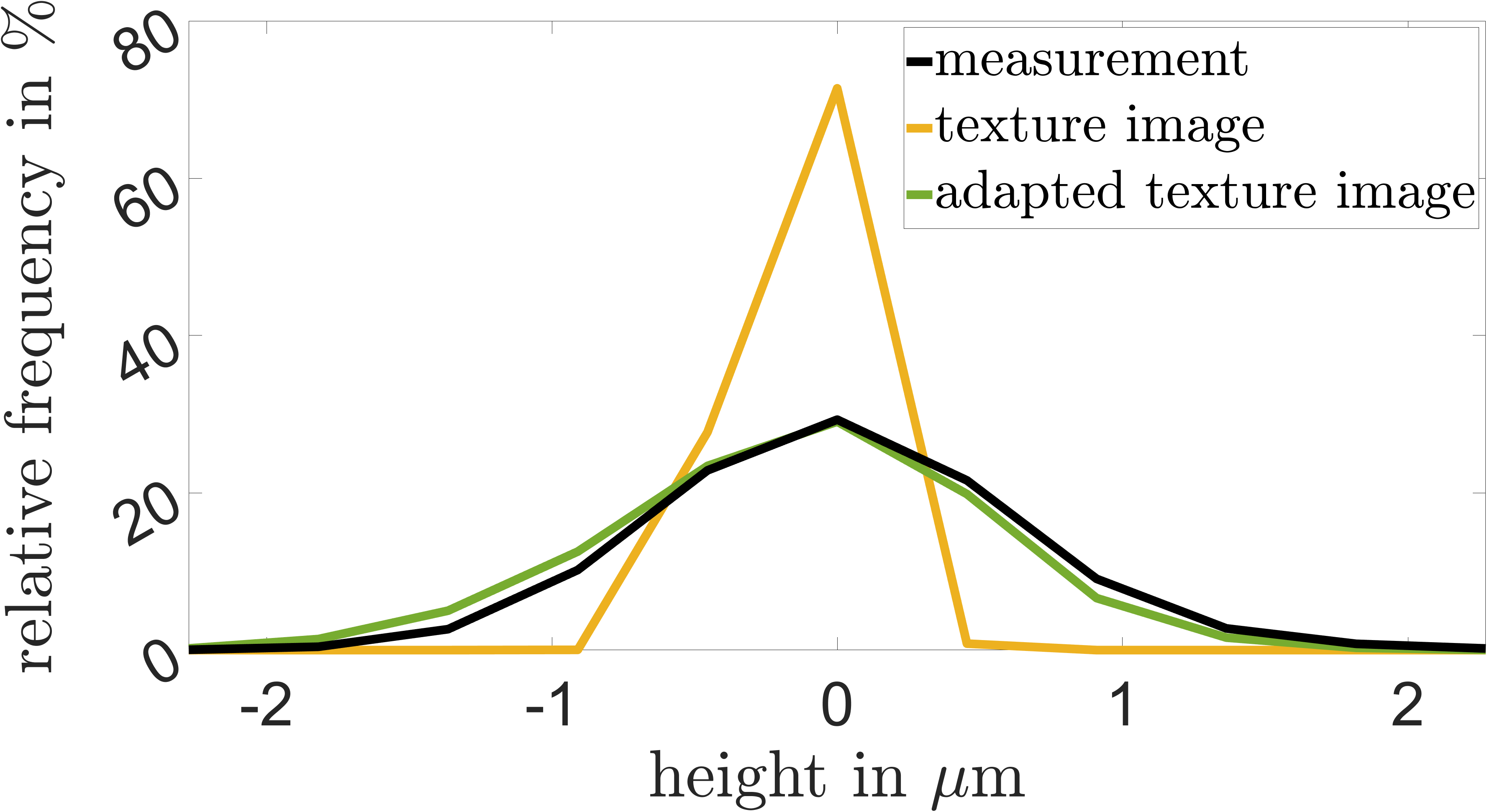}
        \caption{Height value distribution of the measurement compared to the simulation with and without height value adaption.}
    \end{subfigure}
    \hspace{4pt}
    \begin{subfigure}[t]{0.5\columnwidth}
        \centering
	    \includegraphics[height=3.5cm]{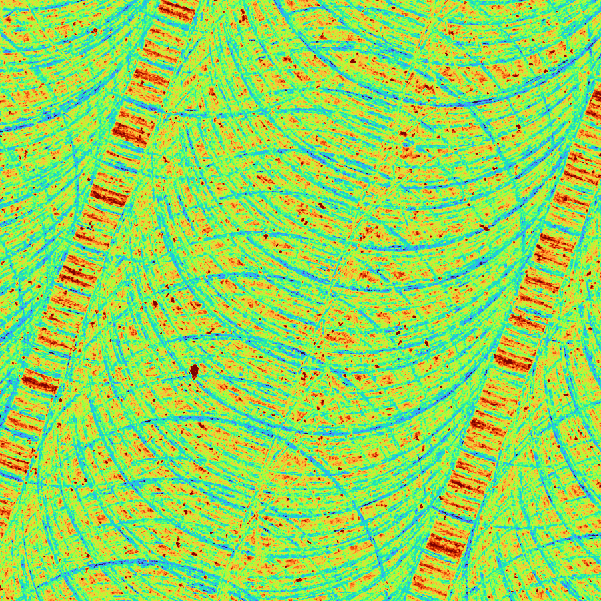}
        \includegraphics[height=3.5cm]{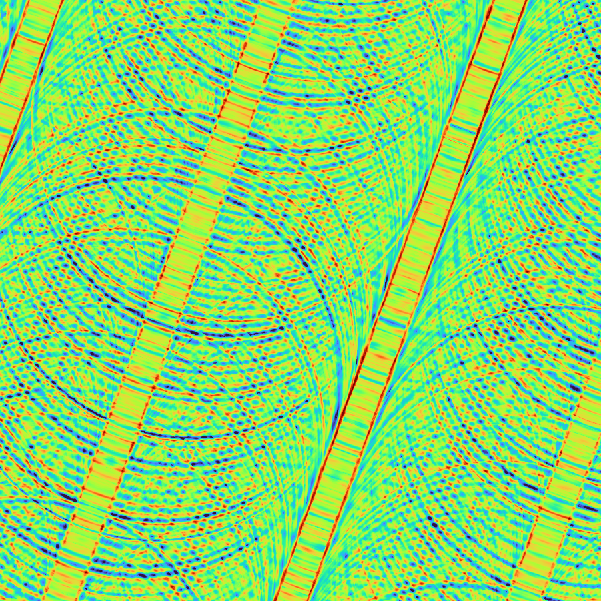}
	    \includegraphics[height=3cm]{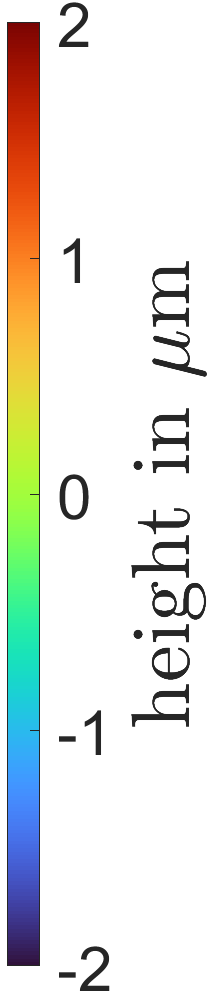}
        \caption{Height images of the measurement (left) and the adapted simulation (right).}
    \end{subfigure}
	\caption{Comparison of measurement and simulation of parallel milling with milling head diameter $4$ mm and ring overlap $0.46$. Imaged region is $5.5\text{ mm}\times 5.5\text{ mm}$. }
	\label{fig:mill:adaption}
\end{figure*}

Due to the choice of convex combination as interaction function, the height values of a simulated texture image $\mathcal{T}$ may not resemble those of the corresponding measurement $\mathcal{M}$, see Figure \ref{fig:mill:adaption}.
Hence, we adapt the height values of the simulated image such that their sample mean $\hat{\mu}_{\mathcal{T}}$ and sample variance $\hat{\sigma}_{\mathcal{T}}$ coincide with the sample mean $\hat{\mu}_\mathcal{M}$ and the sample variance $\hat{\sigma}_\mathcal{M}$ of the measurement.

\begin{figure*}
	\centering
	\includegraphics[height=3.5cm]{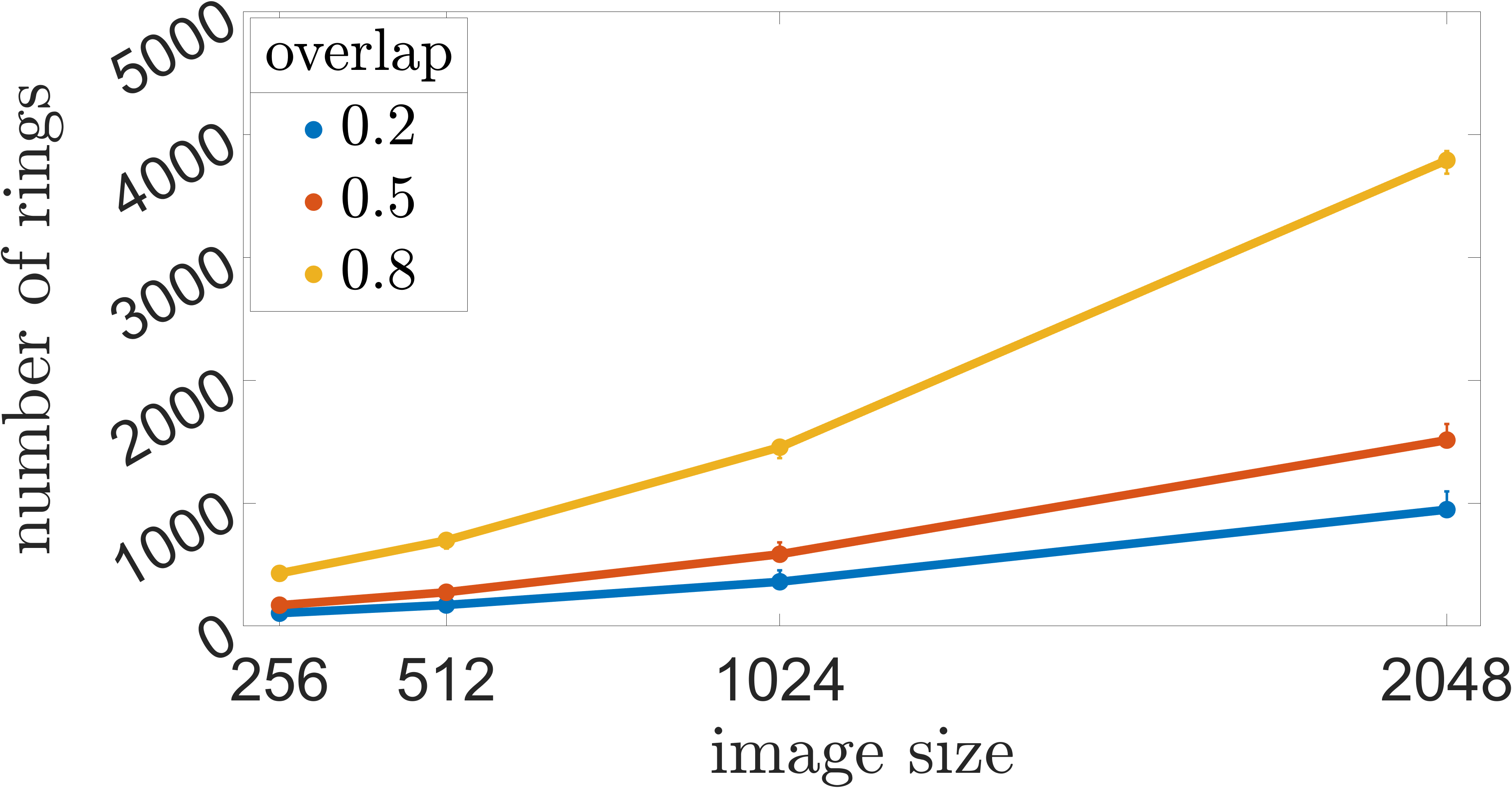}\hspace{8pt}
	\includegraphics[height=3.5cm]{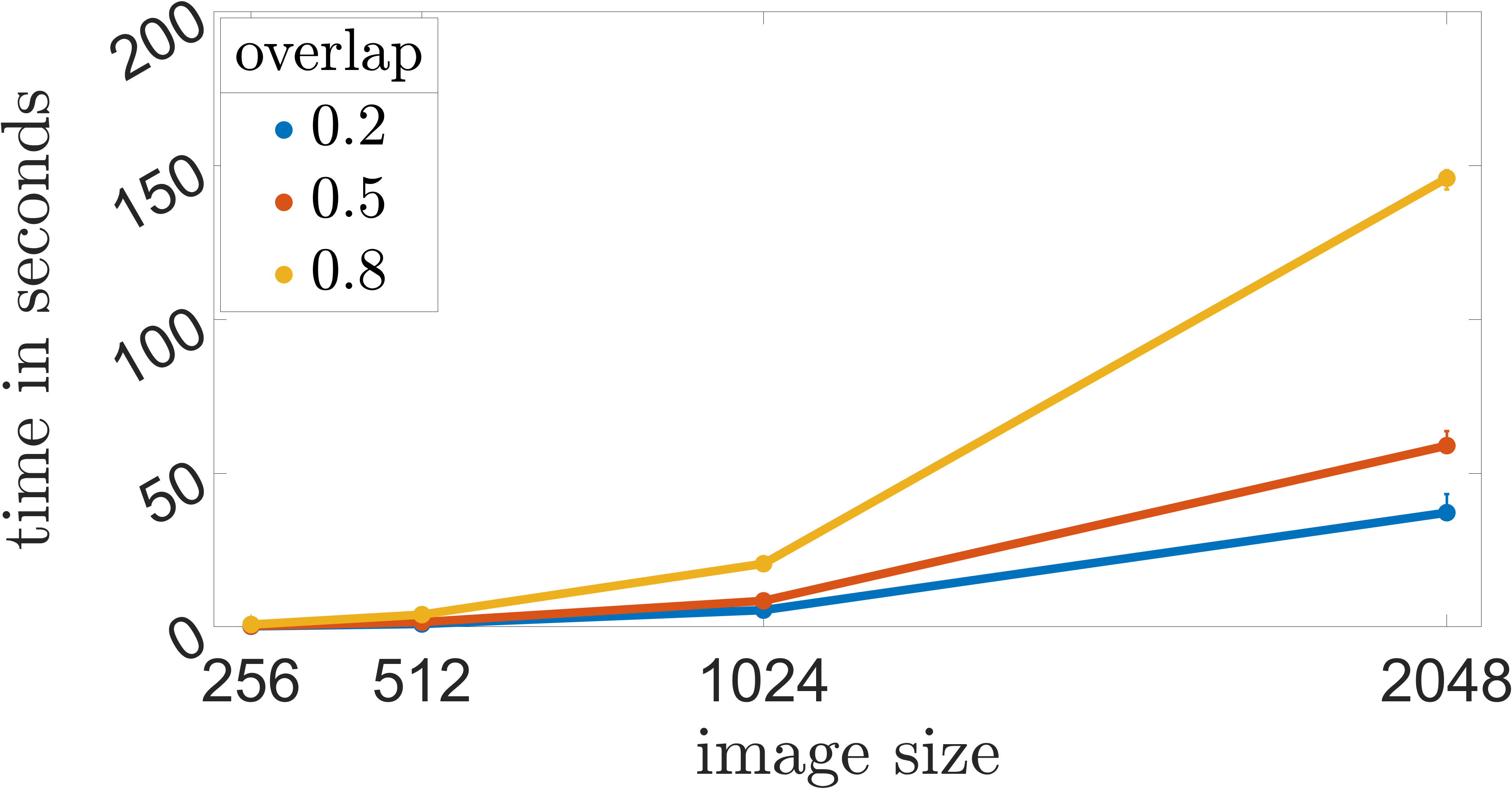}
	\caption{Mean computation time (left) and mean number of relevant rings (right) for milling pattern using pixel spacing $\nu=6.1\,\mu$m, milling tool diameter $d=4\text{ mm}$ and several choices for the overlap. Means of $100$ realizations. Error bars indicate minimal and maximal values. Run-times of the \textsc{matlab}-implementations were measured on a HP Z240 Tower Workstation equipped with Intel Core i7-3700 CPU at 3.40GHz $\times$ 8. }
	\label{fig:mill:runtime}
\end{figure*}

The computation time of the texture synthesis method grows non-linearly with increasing image size, see Figure \ref{fig:mill:runtime}.
It depends on both, the pixel spacing and the number of rings.
The number of rings increases with the size of the overlap.
Thus, run-time for large high-resolution images can get critical.
Parallelization is possible by precomputing and storing the required parameters for each ring. Then, smaller sub-images can be computed simultaneously and merged afterwards.

\section{Conclusion}
\label{sec:conclusion}

We developed  texture synthesis models for sandblasted and milled surfaces.
A common model was not feasible, as the surface patterns of the two methods have no common properties.
The models are based on topography measurements of real surfaces.

There are only few parameters that determine the sandblasting process.
Thus, it is difficult to develop a procedural model that only refers to these parameters.
Therefore, we use a combination of existing data-based texture models that rely entirely on the measurement.
Sandblasted texture images are created in two steps.
First, image patches are generated and stitched afterwards.
The second step can be neglected if the output texture is to be smaller than the measurement.

We examined and compared several methods for the patch generation step, namely the ADSN, the RPN, the method of Heeger and Bergen and the method of Portilla and Simoncelli.
For our purpose, RPN turned out to be the best method, as it has a short computation time and similar output texture images compared to the input measurement.
In particular, the height value distribution and the auto-correlation resemble those of the measurement.

To achieve a coarser pixel spacing than the measurement, the input is down-sampled beforehand using nearest neighbor interpolation.
The case that the output texture should have a higher resolution than the measurement was not investigated.
Images larger than the input can be calculated directly by padding the image before applying the RPN.
However, the resulting texture image is highly dependent on the given input patch.
This can be avoided by computing smaller patches with inputs taken from random positions and stitching them using the method of Efros and Freeman.

Texture images of arbitrary size and with a pixel spacing coarser or equal to that of the measurement can be generated.
For future research, a parametric texture synthesis model would be beneficial that is independent of input images.
Ideally, the model parameters should rely on the real machining parameters.
Moreover, we aim to develop a procedural method which is not feasible with the current model.

The texture synthesis model for milled surfaces is parameterized including the knowledge of process parameters.
A wide variety of texture images can be generated simply by using different parameter settings.
The variability of textures generated with the same parameter configuration is ensured, as their values are selected from suitable probability distributions.

The model for milled surfaces is procedural since it is defined as function on a continuous domain.
Texture images can be obtained by evaluating the function at given grid points. 
For efficient use of the model in rendering, fast approaches for computing the height value at a given surface point have to be developed. 
This could avoid the intermediate image-generating step.
The crucial point is a fast searching algorithm to find all rings affecting one surface point.

\vspace{0.5cm}
\textbf{Acknowledgment} 
This work was supported by the German Federal Ministry of Education and Research (BMBF) [grant number 01IS21058 (SynosIs)]. 
We thank Fraunhofer IOF for the design, production and topography measurements of test objects and Fraunhofer ITWM for the acquisition and rendering images.

\bibliographystyle{unsrt}  
\bibliography{literature}

\end{document}